\documentclass[twocolumn,secnumarabic,amssymb, nobibnotes, aps, prd]{revtex4}
\usepackage{graphicx}

\begin{document}
\title{Why Vortex Lattice Melting Theory is Science Fiction}%
\author{Alexey Nikulov}%
\email[]{nikulov@ipmt-hpm.ac.ru}
\affiliation{Institute of Microelectronics Technology and High Purity Materials, Russian Academy of Sciences, 142432 Chernogolovka, Moscow District, RUSSIA. }
\begin{abstract}The distinguished work by A.A. Abrikosov awarded of the Nobel Prize in Physics for 2003 not only has determined the orientation of all investigations of the mixed state of type II superconductors during forty years but also has provoked a mass delusion. Most scientists consider the Abrikosov state first of all as a vortex lattice. But Abrikosov has found a periodic lattice structure with crystalline long-rang order only because that he searched it. It is impossible to obtain any other result besides a periodical structure for the case of homogeneous, symmetric, infinite space. The long-rang order of superconducting state is phase coherence. It is strange that most scientists have lightly admitted that the Abrikosov state can have two long-rang order although only phase transition is assumed always on the way from this state in the normal state. Moreover, the vortex lattice is considered by many scientists as only long-rang order 
of the Abrikosov state and therefore vortex lattice melting is one of the 
most popular themes in physics of superconductors during last 15 years. It 
is explained in the present paper why the only first order phase transition 
observed on the way from the Abrikosov state in the normal state can not be 
interpreted as vortex lattice melting. History of the problem is considered 
and it is analysed why some delusions about the Abrikosov state became 
popular. It is emphasized that taking into account thermal fluctuations 
changes in essence the habitual notion about the mixed state of type II 
superconductor built because of the Abrikosov result. First of all it shows 
that the Abrikosov solution is not valid just in the ideal case for which it 
was obtained. The false concept of vortex lattice melting appeared because 
of some causes main of them are erroneous interpretation of direct 
observation of the Abrikosov state and the use by theorists the habitual 
determination of phase coherence invalid for multi-connected superconducting state.
\end{abstract}
 
\maketitle

\narrowtext

\noindent
\textsf{\textbf{CONTENTTS}}

\noindent
\textbf{1. Introduction}

\noindent
\textbf{2. History of the mixed state investigations}

\noindent
\textit{2.1. The Shubnikov phase.}

\noindent
\textit{2.2. The Mendelssohn model.}

\noindent
\textit{2.3. The Ginzburg-Landau theory}

\noindent
\textit{2.4. The Abrikosov solution.}

\noindent
\textit{2.5. Experimental corroboration of the Abrikosov prediction.}

\noindent
\textit{2.6. Absence of the crystalline long-range order of vortices in any real sample. The Larkin's result.}

\noindent
\textit{2.7. Invalidity of the Abrikosov solution for the ideal case. The result by Maki and Takayama.}

\noindent
\textit{2.8. Absence of the second order phase transition at the second critical field.}

\noindent
\textit{2.9. The discovery of the lost transition into the Abrikosov state.} 

\noindent
\textit{2.10. The expansion of attention to fluctuation phenomena in the mixed state after the discovery of the high-Tc superconductors. The concept of vortex lattice melting.}

\noindent
\textit{2.11. Experimental evidence of first order nature of the transition in the Abrikosov state.}

\noindent
\textbf{3. Two long rang order predicted by Abrikosov's solution.}

\noindent
\textit{3.1. Long rang phase coherence.} 

\noindent
\textit{3.2. The Abrikosov vortex is singularity in the mixed state with long-range phase coherence.}

\noindent
\textit{3.3. Crystalline long-rang order of vortex lattice.}

\noindent
\textbf{4. Delusions provoked by the direct observation of the Abrikosov state.}

\noindent
\textit{4.1. Abrikosov vortices are not magnetic flux lines.}

\noindent
\textit{4.2. Flux flow resistance is not induced by flux flow.}

\noindent
\textit{4.3. Long range phase coherence but no vortex pinning is main reason of zero resistance in the Abrikosov state.}

\noindent
\textit{4.4. Fundamental differences between vortex ``lattice'' and crystal lattice.}

\noindent
\textbf{5. Qualitative changes because of taking into account of thermal fluctuation.}

\noindent
\textit{5.1. The lowest Landau level approximation.}

\noindent
\textit{5.2. Reduction of the effective dimensionality near $H_{c2}$.}

\noindent
\textit{5.3. Scaling law.}

\noindent
\textit{5.4. Density of superconducting pairs and phase coherence.}

\noindent
\textbf{6. Transition into the Abrikosov state.} 

\noindent
\textit{6.1. The phase transition into the Abrikosov state of an ideal superconductor must be first order.}

\noindent
\textit{6.2. Transition into the Abrikosov state in bulk superconductors with weak pinning disorders.}

\noindent
\textit{6.3. Vortex creep induced by thermally activated depinning of vortices just below $H_{c4}$.}

\noindent
\textit{6.4. Sharp change of the vortex flow resistance at the transition into the Abrikosov state.}

\noindent
\textit{6.5. Absence of the transition into the Abrikosov state in two-dimensional superconductors with weak disorder down to very low magnetic field.}

\noindent
\textit{6.6. Experimental confirmation of phase coherence absence below $H_{c2}$ of thin films with weak pinning disorders.}

\noindent
\textbf{7. Influence of pinning disorders on the nature and position of the transition into the Abrikosov state.}

\noindent
\textit{7.1. Pinning disorders smooth out the sharp transition observed only in high quality samples.} 

\noindent
\textit{7.2. Abrikosov's model and Mendelssohn's model.} 

\noindent
\textit{7.3. Pinning disorders stabilize the Abrikosov state.} 

\noindent
\textbf{8. What can be in the ideal case considered by Abrikosov?}

\noindent
\textbf{9. Why could the erroneous concept of vortex lattice melting appear and become popular?}

\noindent
\textit{9.1. The conception of vortex lattice melting could appear because an incorrect interpretation of direct observation of the Abrikosov state.}

\noindent
\textit{9.2. Could vortices be in the vortex liquid?}

\noindent
\textit{9.3. Lack of information about absence of any phase transition at $H_{c2}$.}

\noindent
\textit{9.4. Definition of phase coherence.}

\noindent
\bigskip

\noindent
\textbf{1. Introduction}

The awarding of the Nobel Prize in Physics for 2003 is the well-deserved appreciation of the distinguished work by A.A.Abrikosov published in 1957 [1]. Results of this work determined the orientation of all following investigations of the mixed state of type II superconductors. But it is necessary to note to draw attention on a contradiction of the results [1] which has provoked some delusions. The famous Abrikosov solutions [1], as 
well as all following solutions [2] obtained in the mean field approximation 
for an ideal superconductor (i.e. without pinning disorders), predict two 
long-rang orders in the Abrikosov state: crystalline long-rang order of 
vortex lattice and long-rang phase coherence. Whereas only second order 
phase transition was assumed from the normal state to the Abrikosov state at 
the second critical field, $H_{c2}$. The situation became more 
dramatic when the consideration of the thermal fluctuations in the seventies 
showed that the second order phase transition, assumed during a long time, 
can not be at  $H_{c2}$ because of the reduction of the effective 
dimensionality of thermal fluctuations on two near  $H_{c2}$ [3-7]: two long-rang orders but no transition. The investigations of the 
thermal fluctuations changed habitual notion about the mixed state formed by the Abrikosov's work.

The lost transition into the Abrikosov state was found first in a bulk 
superconductor in the early eighties [8]. It was found that this sharp transition is observed below  $H_{c2}$ and its position was 
denoted  $H_{c4}$ in [9]. This result was repeated in ten years 
at investigation of high-Tc superconductors (HTSC) [10-12] and was shown that it is first order phase transition [13,14], but not the second order phase transition predicted by theory [1] obtained in the mean field approximation. But only transition is observed on the way from the normal state into the Abrikosov state in all experimental works, in both conventional and HTSC. What long-rang order does at this only transition appear: crystalline long-rang order, long-rang phase coherence or both?

Only few experts comprehended before the HTSC discovery that taking into account of thermal fluctuations changes qualitatively the notion about the mixed state built by Abrikosov and that the second order phase transition into the Abrikosov state is not observed at  $H_{c2}$. But the situation became even worse after 1986 when many scientists turned their attention to the fluctuation phenomena in the mixed state. The visual model of the Abrikosov state as vortex lattice was more acceptable for they than less visual long-rang phase coherence. Therefore when the first order phase transition from the Abrikosov state was found also in HTSC it was interpreted by most people as vortex lattice melting. This concept is very popular and predominated up to now. But it can not be correct. It assumes 
that vortex liquid (i.e. a state with vortices and consequently with long-rang phase coherence) exists above the only phase transition observed at  $H_{c4}$. In this case an additional transition should be observed above vortex lattice melting, i.e. above  $H_{c4}$, where the long-rang phase coherence can disappear. As it will be shown below this disappearance should be first order phase transition in superconductor with weak pinning disorders. But because only first order phase transition is observed in all known observation then it is obvious that it is rather long-rang phase coherence disappearance than vortex lattice melting.

The history of the investigation of the Abrikosov state can be used as a confirmation of the wisdom by Richard Feynman that an understanding is a habit. It may be useful for scientists concerned oneself with psychology of the scientific research. This history shows that images and even denomination can influence on a orientation of physical researches. Therefore it is very important to use correct interpretation of images and 
correct denomination. In order to find what interpretation is correct the 
history of the problem should be considered. It will be made in the Second 
Section. Main results obtained during 70 years are considered in this 
section briefly. Some most important results are considered in the next 
Sections. Two long-rang orders predicted by Abrikosov are considered in the 
Third Section. Some images, such as flux line, flux flow resistance and 
others, appear because of no quite correct interpretation of direct 
observations of the Abrikosov state. It is explained in the Fourth Section 
why these images and denominations are not correct. The lowest Landau level 
approximation is considered in the Fifth Section. According to this 
approximation of the fluctuation theory valid near  $H_{c2}$ the ``jump'' of the specific heat and the fracture of the magnetization 
dependence are not connected with the second order phase transition into the Abrikosov state. The problem of the transition into the Abrikosov state in 
an ideal superconductor and real superconductors with weak pinning disorders 
is considered in the Sixth Section. It is shown that there is an important 
difference between tree- and two-dimensional superconductors and that the shape transition into the Abrikosov state is observed only in bulk superconductors with enough weak pinning disorders. It is explained in the Seventh Section why the transition into the Abrikosov state is smooth in 
most samples in which pinning disorders are not enough weak and it is 
emphasized that pinning disorders stabilize the Abrikosov state in thin 
films and layered superconductors. The stabilization of the Abrikosov state 
by pinning disorders (including superconductor boundaries) and the influence on it the crystal lattice of the superconductor raise a question about a 
state in an homogeneous, symmetric, infinite space. It is stated in the 
Eighth Section that a mixed state without phase coherence should be expected in this ideal case considered by Abrikosov. Some motives of the appearance of the erroneous concept of vortex lattice melting are analysed in the Ninth Section. 

\bigskip

\noindent
\textbf{2. History of the Mixed State Investigations}

The superconductivity is one of the most marvelous phenomenon in physics. A flow without friction, without energy dissipation, is at variance with our everyday experience. But there is a more wonderful effect. Following the discovery of the disappearance of electrical resistance in quicksilver at low temperatures by Kamerlingh Onnes in 1911, a more fundamental aspect of superconductivity has been discovered by Meissner and Ochsenfeld in 1933, when they observed that a superconductor, placed in a weak magnetic field, completely expels the field from the superconducting material except for a thin layer at the surface [15]. This expulsion of a magnetic field generally 
referred to as the Meissner effect is first and most obvious evidence that superconductivity is macroscopic quantum phenomenon. A profound comprehension this effect allows to understand many peculiarities of superconductor behaviour including the one considered in this paper. 

Complete flux expulsion in the simple form of the Meissner effect takes place only in a weak magnetic field. If the applied field is sufficiently strong and if demagnetization becomes appreciable then magnetic flux penetrates through the superconductor. This penetration is qualitative difference in two types of superconductors having various sign of the wall energy associated with the interface between the normal and superconducting domains. In type-I superconductors the wall energy is positive and therefore the amount of flux contained in a single normal domain is many flux quanta. This domain configuration is called the intermediate state. Type-II superconductors considered in this paper are characterized by a negative 
wall energy. Therefore in this case magnetic flux can be dispersed through the superconductor in the form of single flux quanta $\Phi _{0} = \pi \hbar/e$, the smallest possible quantum unit of magnetic flux. Type-II superconductivity occurs preferentially in alloys, or more generally, in impurity systems. Pure metals usually are expected to display type-I superconductivity.

\bigskip

\noindent
\textit{2.1. The Shubnikov phase.}

Type-II superconductors was encountered first in experiment by Shubnikov et al. 70 years ago. These authors investigated the superconducting properties of alloys and discovery persistent of superconductivity up to unusual high magnetic fields. They found that in contrast to pure superconductors where magnetic flux is completely expelled in the superconducting state and the Meissner effect [15] is established the magnetic flux can penetrate in the alloys without complete destruction of superconductivity [16]. After the Abrikosov's work [17] this unusual behaviour is interpreted in terms of mixed state of type-II superconductors. Because of Shubnikov's early experiments, often the mixed state is also referred to as the Shubnikov phase.

\bigskip

\noindent
\textit{2.2. The Mendelssohn model}. 

The investigations showed that the magnetization dependence of alloys is rather like to the one of an ideal conductor than of superconductor with its Meissner effect [16]. In order to explain the difference between pure superconductors and alloys Mendelssohn assumed [18] that superconducting region is not continuous in alloys and it is multi-connected structure like superconducting sponge. This model described enough well some properties of type-II superconductors such as penetration of the magnetic field, irreversible magnetization curves, resistive peculiarities and others. But it was neglected when the Abrikosov model became popular. Nevertheless Mendelssohn sponge may be considered as limit case of strong pinning disorder and it is more suitable for the description of the mixed state of real type-II superconductor with strong pinning disorder than the Abrikosov model made for an ideal superconductor without pinning.

\bigskip

\noindent
\textit{2.3. The Ginzburg-Landau theory}

The following progress in the investigation of the mixed state is connected 
with the Ginzburg-Landau (GL) theory [19] awarded the Nobel Prize in Physics for 2003. This theory is based on Landau's theory [20] of second-order phase 
transition, in which Landau introduced the important concept of the order 
parameter. In the GL theory the order parameter is a complex quantity, namely a wave function $\Psi (r) = \vert \Psi \vert exp(i\varphi )$. The absolute value $\vert \Psi (r)\vert $ is connected with the local density of superconducting electrons, $\vert \Psi (r)\vert ^{2} = n_{s}(r)$. The phase $\varphi $ of the order parameter is needed for describing supercurrents. The free-energy density $f_{GL}$ is then expanded in powers of $\vert \Psi (r)\vert ^{2}$ and $\vert \nabla \Psi (r)\vert ^{2}$, assuming that $\Psi $ and $\nabla \Psi $ are small
$$f_{GL} = \alpha (T)\vert \Psi \vert ^{2} + {\frac{{\beta} }{{2}}}\vert \Psi \vert ^{4} + {\frac{{1}}{{2m}}}\vert ({\frac{{\hbar} }{{i}}}\nabla - 2eA)\Psi \vert ^{2} \eqno{(1)}$$
Here $\alpha (T) = \alpha _{0}(T-T_{c})$ and $\beta $ are coefficients of the GL theory; $\alpha_{0}$ and $\beta $ are constants $ > 0$. The minimum energy is found from a variational method leading to a pair of coupled differential equations for $\Psi (r)$ and the vector potential $A(r)$. Following a standard variational procedure, one finds the two Ginzburg-Landau differential equations 
$$\alpha (T)\Psi + \beta \vert \Psi \vert ^{2}\Psi + {\frac{{1}}{{2m}}}({\frac{{\hbar} }{{i}}}\nabla - 2eA)^{2}\Psi = 0 \eqno{(2a)}$$

\noindent
and 
$$ j_{s} = {\frac{{2e\hbar} }{{2mi}}}(\Psi^{*} \nabla \Psi - \Psi \nabla \Psi^{*} ) - {\frac{{4e^{2}}}{{m}}}\vert \Psi \vert ^{2}A \eqno{(2b)}$$

\noindent
Equation (2) has the form of the Schrodinger equation with the energy eigenvalue $-\alpha $, the term $\beta \vert \Psi \vert ^{2}\Psi $ acting like a repulsive potential. Equation (3) represent the quantum-mechanics description of a superconducting current $j_{s}$. 

The two characteristic lengths: the coherence length $\xi (T) = \hbar /(2m\vert \alpha \vert )^{1 / 2}$ and the penetration depth $\lambda (T) = (m/4e^{2}\vert \Psi \vert ^{2})^{1 / 2}$ can be derived from the GL theory. These lengths $\xi (T)$ and $\lambda (T)$ show the same proportionality to $(1-T/T_{c})^{ - 1 / 2}$ near the critical temperature $T_{c}$. Therefore, it was useful to introduce the Ginzburg-Landau parameter $\kappa  = \lambda (T)/\xi (T)$. This parameter is very important: at $\kappa  < 1/\surd 2$ the wall energy of a normal-superconducting interface is positive and the superconductor is type-I and at $\kappa  > 1/\surd 2$ the wall energy is negative and the superconductor is type II [21].

\

\noindent
\textit{2.4. The Abrikosov solution}.

Since the wall energy of type II superconductor is negative the mixed state can be possible in a magnetic field region. First it was noted by Abrikosov in 1952 [17]. Abrikosov wrote in this work that the normal state becomes unstable at the magnetic field $H < \surd 2 \kappa H_{c}$ and the mixed state can be observed between the first $H_{c1}$ and second $H_{c2}$ critical fields, where $H_{c}$ is the thermodynamic critical field. 

During the first appearance of superconductivity the density of superconducting pairs $\vert \Psi \vert ^{2}$ will be small. Therefore, the GL equation can be linearized neglecting the term $\beta \vert \Psi \vert ^{2}\Psi $ in (2). The linear equation
$${\frac{{1}}{{2m}}}({\frac{{\hbar} }{{i}}}\nabla - 2eA)^{2}\Psi = - \alpha \Psi \eqno{(3)}$$

\noindent
resembles the Schrodinger equation of a free particle in a uniform magnetic field, with the energy eigenvalue given by $-\alpha $. 

The energy eigenvalues $E_{n}$ of a free particle in a uniform magnetic field correspond to quantizide states of the Landau levels
$$E_{n} = m\frac{v_{z}^{2}}{2} + (n+0.5)\frac{\hbar 2eH}{m} \eqno{(4)}$$
$v_{z}$ is the velocity component along magnetic field; n is a positive integer number. The non-zero solution for $\Psi $ can be at $E_{n} = -\alpha $. Therefore the highest field $H$ satisfying this equation $-\alpha  = \hbar eH/m$ is the second critical field H$_{c2}$. This highest value is attained for $v_{z} = 0$ and $n = 0$. Therefore Abrikosov considered only state corresponded the lowest Landau level $n = 0$ and without any gradient along magnetic field $\nabla _{z}\Psi  = 0$. 

Developing of this work Abrikosov has obtained in 1957 the famous solution of the GL equations according to which the mixed state of type II superconductor is a periodical vortex structure [1]. He treated the regime with magnetic field $H$ only slightly less than $H_{c2}$, where the solution of the complete GL equations (2) must have strong similarity to a certain solution of the linear equation (2). The equation (2) has many degenerate eigenvalues having the form
$$\Psi _{\kappa}  = \exp i\kappa y\exp [ - {\textstyle{{1} \over {2}}}({\frac{{x - x_{0}} }{{\xi (T)}}})^{2}] \eqno{(5)}$$

\noindent
for $A_{x} = A_{z} = 0$ and $A_{y} = H_{c2}x$. Where $x_{0} = \hbar \kappa /2eH_{c2}$ and $\kappa $ is an arbitrary parameter. A general solution must be a linear combination of the $\Psi _{\kappa}$. Abrikosov was interested in a solution which is periodic both in x- and y-direction. Periodicity in y-direction is achieved by setting $\kappa  = \kappa_{k} = 2\pi k/l_{y}$ yielding the period $l_{y}$, where $k$ is a integer number. In this case the general solution will have the form 
$$\Psi = {\sum\limits_{k} {C_{k}} } \exp iky\exp [ - {\textstyle{{1} \over {2}}}({\frac{{x - x_{k}} }{{\xi (T)}}})^{2}] \eqno{(6)}$$

Periodicity in x-direction can be established, if the coefficients $C_{k}$ are periodic functions of $k$, such that $C_{k + l} = C_{k}$, where $l$ is some integer. The particular choice of l determines the type of periodic lattice structure: at $l = 1$ -- square lattice, at $l = 2$ -- triangular lattice. The periodicity in x-direction $l_{x} = \hbar 2\pi /l_{y}2eH_{c2}$ yields $l_{x}l_{y}H_{c2} = \Phi_{0}$, i.e., each unit cell of the periodic array contains one flux quantum. 

Thus, Abrikosov has found a periodic lattice structure since he searched it. It is impossible to obtain any other result besides a periodical structure for the case of homogeneous, symmetric, infinite space. The long-rang order of superconducting state is phase coherence. Because of it superconductor has zero resistance, the Meissner effect is observed and other macroscopic quantum phenomena take place. The famous Abrikosov result [1] states that an additional long-range order can exist in type-II superconductor. Can this statement be correct? Two long-rang order in an one state it is too many, the more so, as single phase transition should be on the way from the normal to the Abrikosov state according to [1]. Non-zero density of superconducting pairs $n_{s} > 0$, the long-range phase coherence and the crystalline long-range order of vortex lattice appear simultaneously at $H = H_{c2}$ according to [1] obtained in the mean field approximation. It is assumed that second order phase transition takes place at $H_{c2}$ as well as at $T_{c}$. Therefore it would be natural to think that long-rang phase coherence appears at $H_{c2}$ as well as at $T_{c}$. Nevertheless other long-rang order - the vortex lattice engrossed attention of all scientists during posterior forty years. The scientists were deluded because of some motives but the main was the direct observations of the Abrikosov state. First of all because of it the concept of vortex lattice melting appeared and became very popular.

\

\noindent
\textit{2.5. Experimental corroboration of the Abrikosov prediction.}

First experimental corroboration of the periodic magnetic flux structure predicted by Abrikosov was obtained by Cribier et al. in 1964 [22]. They investigated neutron diffraction through the interaction of the magnetic moment of the neutron with the magnetic field gradients in the mixed state and has found a periodic magnetic flux structure with triangular but no square symmetry predicted in the famous Abrikosov work [1]. The latter discrepancy was explained in the same year [2]. According to the mean field approximation a real state should correspond a minimum value of the GL free-energy. Abrikosov searched periodic structure corresponds this minimum and has found that it is a square lattice. Kleiner, Roth and Autler [2] had shown that a triangular periodic lattice structure corresponds a lower value of the GL free-energy. 

\begin{figure}
\includegraphics{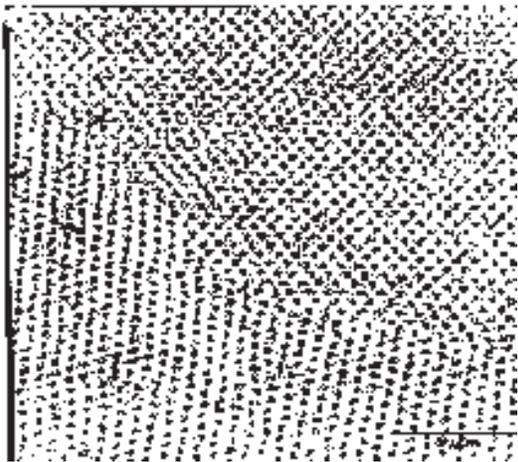}
\caption{\label{fig:epsart} The flux-line lattice observed at the surface of a type II superconductor in an electron microscopy after decoration with $Fe$ micro-particles. A remanent magnetic field of $H = 0.007 \ T$ yields a flux-line spacing $a = (2\Phi_{0}/3^{1/2}H)^{1/2} \approx  580 \ nm$.}
\end{figure}

The most direct proof of the triangular vortex structure of the mixed state has been obtained from experiments by Essmann and Trauble [23] utilizing a Bitter method in conjunction with electron microscopy. In such an experiment the magnetic flux structure is decorated using small magnetic particles and subsequently observed with some optical technique. The microscopic vortex structure in the mixed state and the magnetic field distribution have been investigated from the decoration, neutron diffraction [24-26] and other methods [27-29]. A great number of vortex structure images with and without crystalline order was obtained by many authors (see at Fig.1 a typical picture obtained by the Bitter method). Not only the vortex lattice but even its defects were observed with help of the method by Essmann and Trauble [21]. Such seeming corroboration of the prediction [2] obtained in the mean field approximation caught imagination although not only triangular vortex structure was observed.

The numerous experimental investigations of high quality samples with weak disorders have shown that dominant feature of the mixed state is the existence of the triangular vortex structure corresponding to the minimum of the GL free-energy. But it is important that because of the interaction between the vortex lattice and the crystal lattice of the superconductor, the orientation of the vortex lattice can be correlated with the orientation of the crystal lattice [21]. Departure from the triangular lattice structure can occur sometimes in the form of deviation from triangular symmetry [21,30]. The deviation from the triangular lattice structure are associated with the interaction between the vortex structure and the crystal lattice of the superconductor. This interaction can induce a transformation from triangular to square symmetry [31-33]. 

\bigskip

\noindent
\textit{2.6. Absence of the crystalline long-range order of vortices in any real sample. The Larkin's result.} 

An important result obtained by Larkin in 1970 [34] shows that the crystalline long-rang order of the vortex lattice is unstable against the introduction of random pinning disorder. This result means that crystalline long rang order of vortex lattice is absent in any real superconductor having pinning disorders and the numerous direct observations demonstrate the crystalline order of vortex lattice in sample with enough weak pinning can not prove its existence. The Larkin's result raises a question: ``Can the crystalline order observed in the vortex structure be spontaneous or it is determined by asymmetry (pinning disorders for example) of space?''. The numerous observations of the deviation from the triangular lattice structure because of the interaction between the vortex structure and the crystal lattice and because of pinning disorders show that the vortex structure is rather like to a fishing net stretched with help of stakes than to a spontaneous vortex lattice. There is an important question: ``What can be in the ideal case isotropic infinite space without disorder, i.e. when stakes are taken away?'' An answer on this question is possible only in limits of fluctuation theory since the mean field approximation is not valid just in the ideal case considered by Abrikosov. 

\bigskip

\noindent
\textit{2.7. Invalidity of the Abrikosov solution for the ideal case. The result by Maki and Takayama.} 

Most physicists have got accustomed to assume that thermal fluctuations give only slight corrections to a result obtained in the mean field approximation. But it is not so in the case of the Abrikosov state. Taking into account of thermal fluctuations changes in essence the habitual notion about the mixed state of type II superconductor built because of the Abrikosov result. One of the qualitative changes in this notion connects with the invalidity of the Abrikosov solution just in the ideal case for which it was obtained. It is evidence from the calculation of fluctuation correction in the mixed state. 

First an effective method of the calculation of the fluctuation correction to the Abrikosov solution was proposed by Eilenberger [35]. He took the function $\varphi _{0}(r)$ describing the triangular vortex lattice found in [2], shifted this function by a vector $\lambda = (\lambda _{x}, \lambda_{y})$ in the x-y plane perpendicular to the magnetic field and multiplied it by $\exp(i2\pi \lambda_{y}xH/\Phi_{0})$ to take care of the vector potential. The new function $\varphi_{\lambda} (r)$ describes also the triangular vortex lattice shifted on the vector $\lambda $, $\vert \varphi_{\lambda}(r)\vert^{2} = \vert \varphi_{0} (r+\lambda )\vert^{2}$, Fig.2. An set of functions $\varphi_{\lambda} (r)$, where $\lambda_{x} = (l_{x}^{2}/L_{y})k_{x}$; $\lambda_{y} = (l_{y}^{2}/L_{x})k_{y}$ is orthogonal and spans the complete function subspace corresponded the lowest Landau level of a superconductor with sizes $L_{x} \times  L_{y}$ across magnetic field. Here $l_{x}$ and $l_{y}$ are periods of the vortex lattice; $k_{x}$, $k_{y}$ are nteger numbers. 

\begin{figure}
\includegraphics{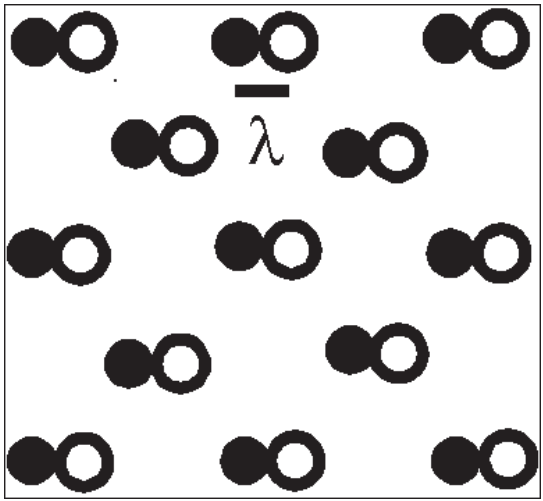}
\caption{\label{fig:epsart} Two triangular vortex lattice shifted on a vector $\lambda $.}
\end{figure}

It is important that all triangular vortex structure described by $\varphi_{\lambda}(r)$ have the same value of the GL free-energy. This means that the vortex lattice is absolutely unstable just in the ideal case considered by Abrikosov. All functions $\varphi_{\lambda} (r)$ should make the same contribution to the thermodynamic average $<\vert \Psi \vert ^{2}>$ since they are equal in rights in the homogeneous, symmetric, infinite space. But the Abrikosov state can be stabilized by pinning disorders, asymmetry of space (because of asymmetry of the crystal lattice of the superconductor) and finite sizes $L_{x} \times  L_{y}$ of superconductor. Therefore it is important to calculate the fluctuation corrections to the solution [2] obtained in the mean field approximation for superconductor with sizes $L_{x} \times  L_{y}$. 

The dependence of the fluctuation corrections $\Delta n_{s,fl}$ on superconductor size $L_{x} = L_{y} =  L$ across magnetic field direction was calculated in the linear approximation by Maki and Takayama in 1971 [36]. Using the method proposed by Eilenberger [35] Maki and Takayama [36] have shown that the fluctuation correction $\Delta n_{s,fl}$ calculated in the linear approximation for a homogeneous, symmetric superconductor with sizes $L$ is proportional $\ln (L/\xi )$ in three-dimensional case and $(L/\xi )^{2}$ in two-dimensional one.$\xi $ is the coherence length of the superconductor. This result confirms the obvious conclusion made above that the Abrikosov state should be absolutely unstable in a homogeneous, symmetric, infinite superconductor.

Nevertheless the Maki-Takayama result [36] seemed very queer for most scientists, because they think that it contradicts to the direct observations of the Abrikosov state. Almost nobody has believed in a reality of the Maki-Takayama result, even author. Maki (with Thompson) attempted to correct this result in [37]. But this work [37] by Maki and Thompson has an important mistake and therefore it can not be considered as a correction of the Maki-Takayama result. Only because of this mistake the fluctuation correction $\Delta n_{s,fl}$ calculated in [37] is finite for infinite superconductor. The Maki-Takayama result [36] is confirmed in the theoretical works [38-44].

It is important to emphasize that there is no real contradiction between the Maki-Takayama result [36] and the direct observation of the Abrikosov state. Moreover experimental results [45,46] corroborate the inapplicability of the Abrikosov theory in the regions where the Maki-Takayama theory predicts this inapplicability. One should take into account that in superconductor with real sizes $L$ this inapplicability can become apparent in a wide region below $H_{c2}$ only in two-dimensional superconductor (for example in thin film) where $\Delta n_{s,fl}$ is proportional $(L/\xi )^{2}$. In three-dimensional, i.e. bulk, superconductor, where $\Delta n_{s,fl}$ is proportional $\ln(L/\xi )$, the fluctuation correction to $n_{s}$ exceeds the one calculated in the mean field approximation only in a narrow region near $H_{c2}$ for all real $L$ values. One should also take into account that pinning disorders can stabilize the Abrikosov state. Therefore only experimental results obtained at investigations of samples without or with enough weak pinning disorders can be used for verification of the 
Maki-Takayama result [36]. 

Such results corroborate qualitative difference between three- and two-dimensional superconductors [45,46]. In conventional bulk superconductor with weak pinning the transition into the Abrikosov state is only slightly lower the second critical field [45] whereas in thin film of like superconductor this transition is not observed down to very low magnetic field, much lower than $H_{c2}$ [46]. This difference between three- and two-dimensional superconductors, observed also at investigation 
of high-Tc superconductors, corresponds to the Maki-Takayama result [36]. 

We can conclude from the comparison of the experimental result and the Maki-Takayama theory that the thermal fluctuation destroy the Abrikosov state almost in all region below $H_{c2}$ in thin film with weak pinning and only a very narrow region near $H_{c2}$ in bulk superconductor with any real and even unreal sizes. This can be explained by the great difference between the $(L/\xi )^{2}$ and $\ln(L/\xi )$ values. For example, for a real samples with $L  = 1\ mm$ and $\xi  = 10^{5}\ mm$, $(L/\xi )^{2} = 10^{10}$ and therefore the Abrikosov solution is not valid almost in all mixed state region in thin film whereas $\ln(L/\xi ) = 11.5$ and therefore it is not valid only in a narrow region near $H_{c2}$. It could be noted that width of this region increases only in 2 time in sample with unreal size $L = 10^{5} \ mm = 100 \ m$. But it is important that $\ln(L/\xi )$ is infinite in the thermodynamic limit. This means that the numerous corroboration of the Abrikosov solution is only outward appearances since the experiment and theory deal with different matters. Although it is obvious that the Abrikosov state exists, this does not prove that the Abrikosov solution is valid in the ideal case of a boundless superconductor for which it was obtained. Together with the Larkin's result [34] this inapplicability means that neither the direct observation [21-33] nor the Abrikosov solution [1] are not evidence of spontaneous crystalline long-range order of the vortex lattice. 

According to the mean field approximation the triangular vortex lattice [2] 
should be in the ideal case isotropic infinite space and two long-range 
orders exist in the Abrikosov state. But according to the fluctuation theory 
this approximation is not valid in the thermodynamic limit, i.e. just in 
this ideal case. The mean field approximation is valid if the fluctuation 
correction is small. But the theory shows that the value of this correction 
calculated in the linear approximation depends on superconductor sizes and 
is infinite for the infinite sizes. The later does not mean that the 
fluctuations are indeed infinite but it means that the mean field 
approximation, i.e. the Abrikosov solution, is not valid in the ideal case 
of a boundless superconductor without pinning disorder.

\

\noindent
\textit{2.8. Absence of the second order phase transition at the second critical field.} 

Other qualitative change because of consideration of thermal fluctuations connects with the phase transition from the normal to the Abrikosov state. It was assumed during a long time [47-49] that a second-order phase transition between these states occurs at $H_{c2}$. This assumption was based on the observation of the specific-heat jump at $H_{c2}$ which should be at the second-order phase transition in accordance with the theoretical result obtained in the mean field approximation. But it is not correct to conclude only on base the mean field approximation and the specific-heat jump about the availability of the phase transition. The mean field approximation disregards the sample dimensionality and the specific-heat jump can be observed in sample with any dimensionality whereas the second order phase transition is possible only in three-dimensional case [50]. Therefore it is important that the effective dimensionality of superconductor decreases on two near $H_{c2}$ because of the Landau quantization. The thermodynamic peculiarity characteristic of second order phase transitions is in dependencies calculated for one-dimensional superconductor only in linear approximation and it is remove already in the Hartree approximation [50].

\begin{figure}
\includegraphics{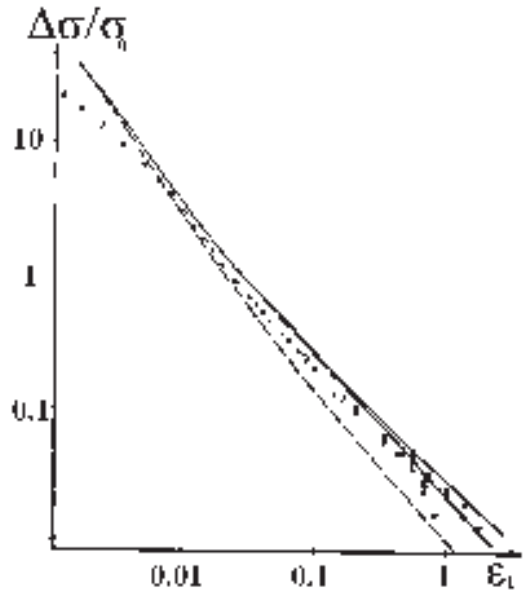}
\caption{\label{fig:epsart} The temperature dependence of excess conductivity $\Delta \sigma = \sigma - \sigma_{n}$ of single-crystal  $V_{3}Ge$ ($H_{c2}(0) = 14.7 \ T$) in parallel magnetic field $h = H/H_{c2}(0) = 0.075$ obtained in [51]. The lines show theoretical dependencies for the paraconductivity calculated in [52]: the contribution of the Aslamasov-Larkin type $\sigma_{AL}$ and the total with the one of the Maki -Thompson type, $\sigma_{AL} + \sigma_{1}$ and$\sigma_{AL} + \sigma_{1} + \sigma_{2}$.  $\epsilon_{t}  = T/T_{c2} - 1$; $\sigma_{0} = \pi e^{2}/2^{3/2}\hbar \xi (0) = 560 (\Omega  cm)^{-1}$; $\sigma_{0}/\sigma_{n}  = 0.034$.}
\end{figure}

The reduction of the effective dimensionality is obvious in the linear approximation of the fluctuation theory. It was noticed first by Lee and Shenoy in the paper ``Effective Dimensionality Change of Fluctuations in Superconductors in a Magnetic Field'' [3]. The dimensionality of 
fluctuations above $T_{c}$ is determined be value of three superconductor dimensions, $L_{x}, L_{y}, L_{z}$, relatively the coherence length $\xi (T) = \hbar /(2m\alpha )^{1 / 2}$. Any gradient, for example $\partial \Psi /\partial x$, gives large contribution to the GL free-energy 
$$f_{GL} = \alpha (T)[\vert \Psi \vert ^{2} + {\frac{{\beta} }{{2\alpha (T)}}}\vert \Psi \vert ^{4} + \xi ^{2}(T)\vert ({\frac{{\partial }}{{\partial x}}} + {\frac{{\partial} }{{\partial y}}} + {\frac{{\partial }}{{\partial z}}})\Psi \vert ^{2}] \eqno{(7)}$$

\noindent
at $L_{x} < \xi (T)$, $\xi^{2}(T)\vert \partial \Psi /\partial x\vert ^{2} > (\xi (T)/L_{x})^{2}\vert \Psi \vert ^{2} > \vert \Psi \vert ^{2}$. Therefore the approximation $\vert \Psi \vert ^{2} = constant$ along x is admissible and $(\partial /\partial x + \partial / \partial y + \partial /\partial z)$ can be replaced by $(\partial / \partial y + \partial /\partial z)$ in the two-dimensional case. In a one-dimensional superconductor, with $L_{x},L_{y} < \xi (T)$, $(\partial /\partial x + \partial / \partial y + \partial /\partial z)$ can be replaced by $\partial /\partial z$. In a bulk superconductor the gradient $(\partial /\partial x + \partial /\partial y)$ vanish near $H_{c2}$ because of the Landau quantization. Therefore values of the fluctuation specific heat $C$ [3], paraconductivity (i.e. excess conductivity because of superconducting fluctuations) along magnetic field $\Delta \sigma_{\vert \vert}$ and others of bulk superconductor are proportional $(H/H_{c2}-1)^{ - 3 / 2}$ in the linear approximation region above $H_{c2}$ as in one-dimensional, where $C, \Delta \sigma_{\vert \vert} \propto (T/T_{c}-1)^{- 3 / 2}$ but not as in three-dimensional superconductor, where $C, \Delta \sigma_{\vert \vert} \propto  (T/T_{c} - 1)^{ - 1 / 2}$. The $(H/H_{c2}-1)^{ - 3 / 2}$ or $(T/T_{c2}-1)^{ - 3 / 2}$ dependence in the linear approximation region near  $H_{c2}$ is corroborate by experimental data [51] and exact theoretical calculation [52]. For example, one can see from the data presented on Fig.3 that both the theoretical and experimental (in the linear approximation region $T/T_{c2} - 1 > 0.005$) of paraconductivity are close to $(T/T_{c2}-1)^{ - 3 / 2}$.

\begin{figure}
\includegraphics{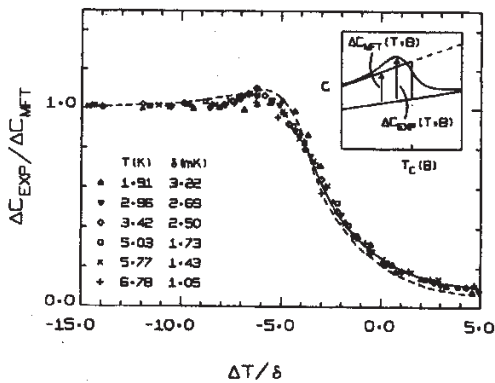}
\caption{\label{fig:epsart} Dependencies of the normalized  heat capacity $\Delta C_{exp}(T,H)/\Delta C_{MFT}$ on the reduced temperature $\Delta T = T-T_{c2}$ of a very pure niobium crystal at different values of magnetic field measured in [7]. $\Delta C_{exp}$ and $\Delta C_{MFT}$ (the specific-heat discontinuity expected from mean-field theory) are defined in the inset. A reduced temperature scale in units of $\delta  = T(2e/\hbar)(k_{B}H/8\pi \Delta C)^{2/3}|TdH_{c2}/dT|^{1/3}$ is used in accordance of the scaling law proposed by Thouless [6]. The dashed curve is the temperature dependence expected for an ideal one-dimensional superconductor.}
\end{figure}

The fluctuation values $C, \Delta \sigma_{\vert \vert} $ and other calculated in the linear approximation diverge at $T = T_{ c}$ both in three- and one-dimensional superconductors. But taking into account fluctuation interaction removes the divergence in one-dimensional case but it remains in three-dimensional superconductor. The divergence is removed already in the Hartree approximation, first approximation taking into account fluctuation interaction, in one-dimensional superconductor whereas it remains in three-dimensional case even in the exact solution. Because of the latter the $\lambda $ anomaly should be observed on the of specific heat dependence for three-dimensional superconductor at $T = T_{c}$. 

This $\lambda $ anomaly of the specific heat dependence is main experimental evidence of the second order phase transition. But all experimental investigations show that the $\lambda $ anomaly is absent at $H_{c2}$ and the specific-heat ``jump'' at $H_{c2}$ is described very well by theoretical dependence for one-dimensional model. Comparison of the experimental dependence of the specific-heat near $H_{c2}$ with the theoretical one obtained in the Hartree approximation was made first in 1973 [4]. Bray has shown that the theoretical dependence calculated in the Hartre-Fock approximation with screened potential [5] describe better experimental data. It is important that thermodynamic values can be calculated exactly [53] for one-dimensional superconductor. The temperature dependencies of the fluctuation specific-heat of a pure niobium crystal measured at different value of magnetic field in [7] are obeyed the scaling law valid in the lowest Landau level approximation and are described very well by the exact theoretical dependence [6] for a one-dimensional superconductor, Fig.4. All measurements 
of the specific heat show that a $\lambda $ anomaly characteristic of second order phase transitions is not observed at $H_{c2}$. The $\lambda $ anomaly is absence since second order phase transitions can not be in any one-dimensional system. 

\

\noindent
\textit{2.9. The discovery of the lost transition into the Abrikosov state.}

Thus, in contrast to the belief in the second order phase transition at $H_{c2}$ the reduction of the effective dimensionality and the investigations made in the middle of 70 years prove the absence of this transition. On the other hand the direct observations of the Abrikosov vortices, i.e. singularities in the mixed state with long-rang phase coherence, and many other results prove the presence the long-rang phase coherence and consequently a phase transition should be somewhere below $H_{c2}$. This phase transition into the Abrikosov state lost in 70 years was discovered first at investigation of paraconductivity in bulk conventional superconductors [8]. 

A sharp qualitative change of the resistive properties was observed in [8] against a background of a smooth dependence of paraconductivity in perpendicular magnetic field Fig.5. The current-voltage curves become non-Ohmic, the resistance increases sharply and its value measured at enough low current becomes equal zero in a narrow region, much narrower than the ``jump'' of specific-heat at $H_{c2}$ Fig.4. Above this sharp change the current-voltage curves are Ohmic and the smooth dependence of paraconductivity was described by a theoretical one obtained for a one-dimensional model both above and below $H_{c2}$ Fig.5. The second critical field was determined in this work from investigation of paraconductivity in the region of applicability of the linear approximation above $H_{c2}$ [45]. The resistive feature observed in [8] was interpreted as a phase transition from fluctuation one-dimensional state to the Abrikosov state. This interpretation is obvious since the resistance can be zero only in a state with long-rang phase coherence and should be non-zero in any state without phase coherence. 

\begin{figure}
\includegraphics{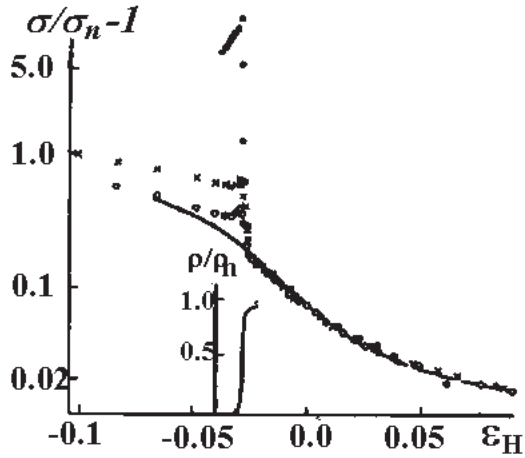}
\caption{\label{fig:epsart} Magnetic field dependence of the excess conductivity for different measuring currents: $\bullet  - j = 0.5 \ A/cm^{2}$, $\times - j = 2 \ A/cm^{2}$, $o - j = 10 \ A/cm^{2}$. The line  is the theoretical dependence for paraconductivity obtained in the Hartree approximation. The resistive transition shown at the bottom was measured at $j = 0.05 \ A/cm^{2}$ [8].}
\end{figure}

It is important that the transition into the Abrikosov state is observed below $H_{c2}$. The difference of the position of this phase transition from $H_{c2}$ is enough small in bulk conventional superconductors. It was discovered accidentally at a comparison of 
experimental and theoretical dependencies of paraconductivity in the linear approximation region [45]. The comparison of excess conductivity $\Delta \sigma (T,H) = \sigma (T,H) - \sigma_{n}$ of single-crystal $V_{3}Ge$ measured above $H_{c2}$ with dependencies calculated numerically on base of results of a theory by Ami-Maki [52] showed that perfect accordance of the experimental data with the theoretical dependence takes place only if the transition into the Abrikosov state occurs at $H  \approx  0.98H_{c2}$. Then it was assumed first that the transition into the Abrikosov state takes place below the second critical field. 

The second critical field was determined as a rule by the position of the resistive transition. But this method has not any theoretical ground in contrast to the comparison of paraconductivity dependencies used in [45]. Because of the absence of any transition at $H_{c2}$ only method of the $H_{c2}$ determination, which can be grounded theoretically in the framework of the mean field approximation, is magnetization measurement.

\begin{figure}
\includegraphics{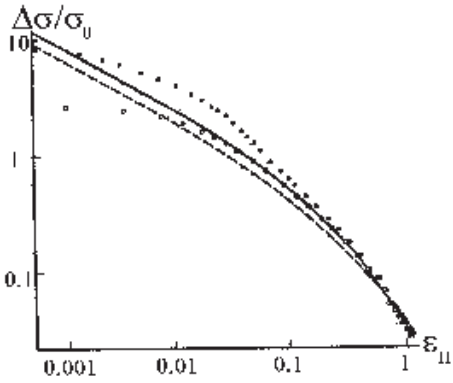}
\caption{\label{fig:epsart} Comparison of experimental and theoretical dependencies for paraconductivity in the linear approximation region at different values of the second critical field made in [45]: $\bullet  - H_{c2}$ corresponds to the position of the resistive transition;$\circ $ - with the value of shifted $0.02H_{c2}$ upwards. The excess conductivity $\Delta \sigma = \sigma - \sigma_{n}$ was measured on single-crystal  $V_{3}Ge$ ($T_{c} = 6.1 \ K$) in perpendicular magnetic field at T = 4.2 K. $\epsilon_{H} = H/H_{c2} - 1$. The lines show theoretical dependencies calculated on the base of the Ami-Maki theory [52]: solid line -  $\sigma_{AL} + \sigma_{MT}$; dashed -  $\sigma_{AL}$.}
\end{figure}

\begin{figure}
\includegraphics{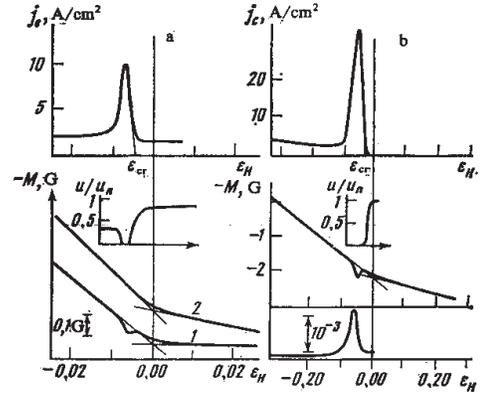}
\caption{\label{fig:epsart} Dependencies of the critical current density jc, of the voltage u/un and of the magnetization M on the relative magnetic $\epsilon_{H} = H/H_{c2} - 1$ at $T = 4.2 \ K$. a - for $Nb_{94.3}Mo_{5.7}$, $T_{c} =7.5 \ K$, $H_{c2}(4.2) = 4.03 \ kOe$, the parameter of the GL theory $\kappa  = 3$, $u/u_{n}$ was measured at $j = 7 \ A/cm^{2}$; b - for $V_{3}Ge$, $T_{c} =6.1 \ K$, $H_{c2}(4.2) = 26.4 \ kOe$, $\kappa  = 13$, $u/u_{n}$ was measured at $j$ from 0.1 to 1 $A/cm^{2}$. Curve 1 - $M(H)$ measured without a screening container, 2 - with the container. To the right, below a real part of the dynamic susceptibility is shown}
\end{figure}

\noindent
According to the mean field approximation the magnetization of type II superconductor $--M  \propto  (H_{c2} - H)$ at $H_{c2}/3 < H < H_{c2}$ and $-M = 0$ at $H > H_{c2}$. The fracture of the $-M(H)$ dependence at $H_{c2}$ becomes smooth because of fluctuation. But the linear dependence $-M  \propto  (H_{c2} - H)$ observed in a wide region allows to determine $H_{c2}$ from the crossing of the linear dependencies $-M(H)$ above and below $H_{c2}$. The simultaneous measurements of the magnetization and resistive properties in [54] have confirmed the difference of the position of the transition into the Abrikosov state in $V_{3}Ge$ found in [45] and this difference was determined in a superconductor $Nb_{94.3}Mo_{5.7}$ with other parameters Fig.7. It was found [45] that the appearance of a critical current, resistive transition and other features connected with the phase coherence are observed below $H_{c2}$, Fig.7. It is difficult to determine exactly the field $H_{cr}$ of phase coherence appearance but it is important that $H_{cr} < H_{c2}$ both for $V_{3}Ge$ and for $Nb_{94.3}Mo_{5.7}$ and that the $(H_{c2} - H_{cr})$ values differ for this superconductors: at $T = 4.2 \ K$, $H_{c2} - H_{cr} = 300 \div 600 \ Oe$ for $V_{3}Ge$ (at $H_{c2} \approx  26 kOe$) and $H_{c2} - H_{cr} = 10 \div 30 \ Oe$ for $Nb_{94.3}Mo_{5.7}$ (at $H_{c2} \approx  4 \ kOe$) [45]. 

\begin{figure}
\includegraphics{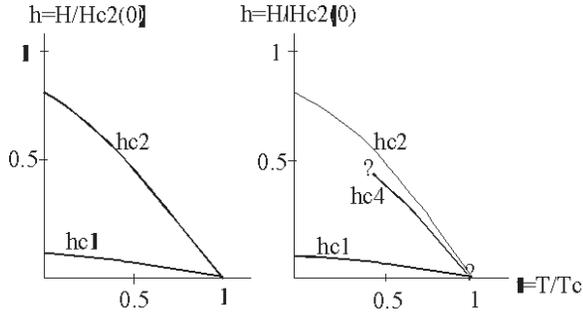}
\caption{\label{fig:epsart} The change of the phase diagram of type II superconductors in a magnetic field caused by the thermal fluctuation investigation. On the left side the mean field phase diagram supposed before the fluctuation investigation is shown. On the right side the phase diagram to the results of fluctuation investigation of bulk conventional superconductor [51,54,55] is shown. }
\end{figure}

First attempt to explain the difference of the position of the transition into the Abrikosov state was made in 1990 [9]. The homogeneous, symmetric, infinite space was considered as well as in the Abrikosov work. It was shown that if the transition into the Abrikosov state exists in this ideal case it should be below $H_{c2}$ because of thermal fluctuations. The position of this transition was denoted $H_{c4}$. The difference of the $(H_{c2}- H_{c4})/H_{c2}$ values observed in $V_{3}Ge$ and $Nb_{94.3}Mo_{5.7}$ was explained in [9] by a scaling law obtained in the lowest Landau level approximation. According to this scaling law the transition into the Abrikosov state should be at the 
same value $\epsilon_{c4} = Gi^{-1/3}(T/T_{c})^{-2/3} (H_{c2}-H_{c4})/H_{c2}^{2/3}(T)H_{c2}^{1/3}(0)$. Where $Gi$ is the Ginzburg number; $H_{c2}(T) = H_{c2}(0)(1 - T/T_{c})$; $H_{c2}(0) = -T_{c}(dH_{c2}/dT)_{T = Tc}$ is the GL second critical field at $T = 0$. The decreasing of the $(H_{c2}-H_{c4})$ difference near $T_{c}$ following from the scaling law $(H_{c2}- H_{c4})  \propto  H_{c2}^{2/3}(T) \propto (1 - T/T_{c})^{2 / 3}$ corresponds to the experimental results [51,55]. The experimental investigations of fluctuation influence on the transition into the Abrikosov state in conventional bulk superconductors have allowed to alter the phase diagram of type II superconductor in a magnetic field, Fig.8. 

\bigskip

\noindent
\textit{2.10. The expansion of attention to fluctuation phenomena in the mixed state after the discovery of the high-Tc superconductors. The concept of vortex lattice melting.}

The problem of fluctuation phenomena in the mixed state of type II superconductors became very popular soon after the discovery of high-Tc superconductors (HTSC) in 1986 since these phenomena are very appreciable in these superconductors with very small coherence length. Many new scientists enlisted the investigation of superconductivity because of the HTSC discovery. Most of they did not know about the results of the specific-heat investigations [4-7] refuting the assumption on the second order phase transition at $H_{c2}$ and their knowledge was limit by the old notions from the [21,47-49] and other books. Therefore it was strange for they that the resistance of HTSC does not fall down to zero just below $H_{c2}$ but decreases smoothly in a wide region. The result [8] shown on Fig.5 was little known. Therefore when this sharp change, called in some paper [10-12,56] resistive kink, was observed in $YBa_{2}Cu_{3}O_{7 - x}$, Fig.9, it was misapprehended as a new result observed only in HTSC. But this result could be foreseen from the observations [8] and the scaling law. The comparison shows that there is only quantitative but not qualitative difference between $YBa_{2}Cu_{3}O_{7 - x}$ and bulk conventional superconductors. The difference of the ``kink'' position from $H_{c2}$ observed in $YBa_{2}Cu_{3}O_{7 - x}$ corresponds the same value $\epsilon_{c4}$ which was observed in $V_{3}Ge$ and $Nb_{94.3}Mo_{5.7}$ [57]. Therefore I will designate the position of the ``kink'' $H_{c4}$. 

The resistive kink was observed first only in four years after the HTSC discovery, in the beginning of the nineties, since this sharp change of resistive properties can be observed only in high quality samples, enough homogeneous and with weak pinning disorders. It will be explained below why the transition into the Abrikosov state can be narrow only at weak pinning disorders and why it is wide in most samples. The resistive kink observed in $YBa_{2}Cu_{3}O_{7 - x}$ was interpreted as vortex lattice melting [10-12]. This interpretation became very popular [58-62] because most scientists did not understand enough clear that the transition into the Abrikosov state is not observed at $H_{c2}$. If anyone thinks that the transition into the Abrikosov state occurs at $H_{c2}$ and this state is vortex lattice then he can interpret other transition observed below $H_{c2}$ as vortex lattice melting. Now some experts understand that no transition is at $H_{c2}$. Moreover some of they understand that two long rang order were predicted by Abrikosov. Then, vortex lattice melting is a transition from the Abrikosov state in the normal state (exactly in a fluctuation mixed state without phase coherence) and vortex liquid is a state without vortices. This interpretation could coincide with the one proposed in 1981 [8] if the Abrikosov state is vortex lattice with long-rang phase coherence. Then two long-rang orders disappear at $H_{c4}$. 

\begin{figure}
\includegraphics{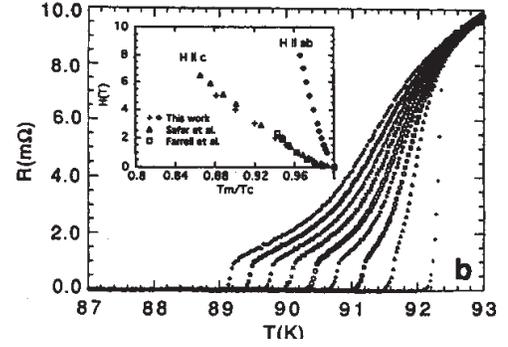}
\caption{\label{fig:epsart} Resistive transition in perpendicular magnetic fields  of 0, 1, 2, 3, 4, 5, 6, 7 and 8 T for H || (a,b) of an untwinned $YBa_{2}Cu_{3}O_{7 - x}$, crystal obtained in [12]. Inset: Phase diagram of the transition into the Abricosov state.  }
\end{figure}

This interpretation follows directly from habitual determination of phase coherence used by all theorists. When a theorist determines phase coherence as a coherence between two points $<\varphi (r)\varphi (r+r')>$ he should conclude that long-rang phase coherence should disappear in the Abrikosov state at any disappearance of crystalline long-rang order of vortex lattice. According to this habitual determination long-rang phase coherence is absent in the Abrikosov state observed in any real sample because pinning disorders destroy crystalline long-rang order of vortex lattice [34]. But why could zero resistance be observed in the real Abrikosov state if phase coherence is absent? This contradiction shows clear that the habitual determination of phase coherence is not valid for multi-connected superconducting state, such as the Abrikosov state. 

The Abrikosov vortices appear since a magnetic flux can not penetrate inside superconductor with long-phase coherence and without singularity (the Meissner effect). Therefore a existence of the vortices is already evidence of long-range phase coherence. The existence of long-rang phase coherence is obvious in any vortex state, solid or liquid, when $\oint_{l} dl\nabla \varphi = 2\pi n $, where $n$ is a number of the Abrikosov vortices inside the closed path l. This long-rang phase coherence exists at any distribution of vortices inside $l$ and its existence does not connect with crystalline long-rang order. Therefore vortex lattice melting is not only poor name. It is error. Numerous theories of vortex lattice melting [58,59,62-64] describe indeed vortex lattice melting at which only crystalline long-rang order disappears whereas long-rang phase coherence remains. Therefore they describe a transition which is not observed whereas the transition observed both in conventional superconductors and HTSC is not described for the present. In order to describe it a new determination of phase coherence should be used. 

Some experts understand that vortices can not exist in the ``vortex liquid''. But vortex liquid without vortices it seems very strange for many scientists. Indeed it sounds strangely. Therefore it is better to call the state above $H_{c4}$ as mixed state without phase coherence by contrast with the Abrikosov state which is the mixed state with long-rang phase coherence. Although phase coherence is absent the density of superconducting electrons $n_{s}$ in this state does not differ strongly from the one in the Abrikosov state and it can be enough high as it takes place for example in two-dimensional superconductor at magnetic field much lower $H_{c2}$ [46]. It should be emphasized that the appearance of the long-rang phase coherence, i.e. the transition into the Abrikosov state, is not connected directly with the $n_{s}$ appearance and its value. The specific-heat jump [4,7] and the fracture of magnetization dependence [54] connected with the $n_{s}$ appearance are observed at $H_{c2}$ whereas the zero resistance connected with the long-rang phase coherence appears at $H_{c4}< H_{c2}$ [8,10-12].

\bigskip

\noindent
\textit{2.11. Experimental evidence of first order nature of the transition in the Abrikosov state.} 

Numerous investigations of the fluctuation phenomena in the mixed state of HTSC repeat in the main the results obtained at investigations of these phenomena in LTSC. Only qualitatively new, trustworthy result obtained at 
investigations of HTSC is experimental evidence of first order phase 
transition. It was assumed in [8] that a narrow thermodynamic feature should observed at the transition into the Abrikosov state. Measurements of magnetization [13] and specific heat [14,65] have shown that a jump of 
thermodynamic quantities is observed at this transition in samples with 
enough weak pinning disorders. As it will be shown below the appearance of the long-rang phase coherence should be first order phase transition in the ideal or enough quality samples. 

\bigskip

\noindent
\textbf{3. Two long rang order predicted by Abrikosov's solution.} 

The famous Abrikosov solutions [1], as well as all following solutions [2] obtained in the mean field approximation for an ideal superconductor, predict two long-rang orders in the Abrikosov state: crystalline long-rang order of vortex lattice and long-rang phase coherence. First order is more visual but the second one is more evident. 

\bigskip

\noindent
\textit{3.1. Long rang phase coherence.} 

In order to understand why the existence of long-rang phase coherence is more evident than the existence of vortex lattice one should recollect the Meisner effect. This effect has very simple, mathematical explanation when the wave function $\Psi (r) = \vert \Psi \vert \exp(i\varphi )$ introduced in the GL theory is used. It is well known that an integral along any closed path $l$ of gradient of a function equals zero $\oint_{l} dl\nabla \varphi = 0 $ if this function does not have singularities inside the closed path $l$. It is important that the gradient of phase of the wave function is proportional to 
the momentum of superconducting pairs $\hbar \nabla \varphi  = p = mv + 2eA$. Where $A$ is the vector potential; $v$ is the velocity of the superconducting pairs. Therefore when $\oint_{l} dl\nabla \varphi = 0 $ then the magnetic flux $\Phi = \oint_{l} dlA $ contained within the closed path $l$ should be equal zero $\Phi = \oint_{l} dlA  = (\hbar / 2e)\oint_{l}dl\nabla \varphi - (m / 2e)\oint_{l} dlv = 0 $ if the velocity circulation equals zero $\oint_{l} dlv = 0 $.

\begin{figure}
\includegraphics{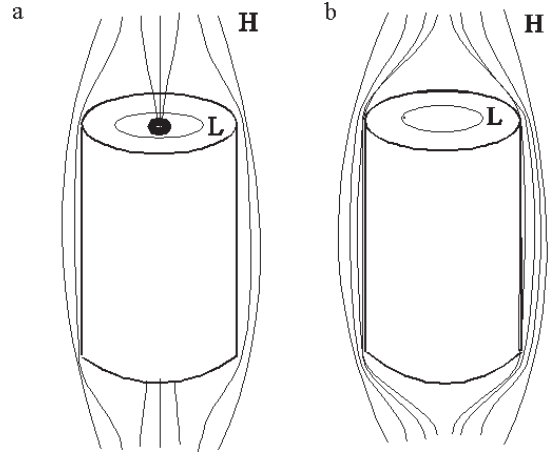}
\caption{\label{fig:epsart} The magnetic flux $\Phi = n\Phi_{0} \neq 0$ can be inside a superconductor if only a singularity exists inside it (a). If a singularity is absent $n = 0$ and Meissner effect takes place (b). }
\end{figure}

Thus, magnetic flux can not penetrate in a superconducting state with long rang phase coherence and without singularities since superconductivity is a 
macroscopic quantum phenomenon. This effect may be considered as a 
particular case of the flux quantization. The phase $\varphi $ of the wave function $\Psi (r) = \vert \Psi \vert \exp(i\varphi )$ of superconducting electrons can cohere all over the volume of a superconductor. The relation for the superconducting current $j_{s} = n_{s}2ev$
$$\oint\limits_{l} dl\lambda _{L}^{2} j_{s} = \frac{\Phi _{0}}{2\pi }\oint\limits_{l} dl\nabla \varphi - \Phi \eqno{(8)}$$

\noindent
is valid in the region where the phase coherence exists. $\Phi_{0} = \pi \hbar/e$ is the flux quantum; $n_{s}$ is the superconducting electron density; $\lambda_{L} = (mc/e^{2}n_{s})^{0.5}$ is the London penetration depth; $l$ is a closed path of integration. If a singularity is absent, then $\oint_{l} dl\nabla \varphi = 0 $ since the closed path of integration $l$ can be tightened down to point if the wave function $\Psi (r) = \vert \Psi \vert \exp(i\varphi )$ does not have any singularities. In this case the relation (8) is the equation postulated by F. and H.London for the explanation of the Meissner effect (see [21,66]). The magnetic flux can not penetrate in this 
case inside a superconductor with long-rang phase coherence, Fig.10. 

\bigskip

\noindent
\textit{3.2. The Abrikosov vortex is singularity in the mixed state with long-range phase coherence.}

But the magnetic flux can be inside a superconductor if the wave function $\Psi (r) = \vert \Psi \vert \exp(i\varphi )$ has singularities Fig.10. In this case the closed path of integration $l$ can not be tightened down to point. Therefore in the relation $\oint_{l} dl\nabla \varphi = n2\pi $, $n$ can be any integer number and consequently $\Phi  = n\Phi_{0} \neq 0$ even at $j_{s} = 0$ according to (8). 

\begin{figure}
\includegraphics{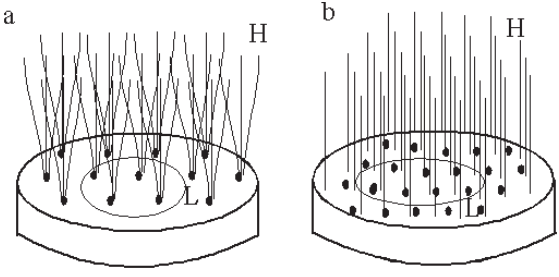}
\caption{\label{fig:epsart} The singularities in type-II superconductor are the Abrikosov vortices. The $n = \Phi/\Phi_{0}$ vortices exist both at inhomogeneous (a), when the distance $a \approx  (\Phi_{0}/H)^{1/2}$ between the vortices exceeds the penetration depth, $a > \lambda_{L}$, and homogeneous (b), when $a \ll  \lambda_{L}$, magnetic field. }
\end{figure}

The penetration of the magnetic field in the mixed state of type II superconductor can be explain by both Mendelssohn's and Abrikosov's model. In both case the magnetic flux penetrates because of singularities. The two main difference between these models is that in the first case singularities exist always and in the second case they, i.e. the Abrikosov vortices, appear because of magnetic field. Thus, the Abrikosov vortices are singularity in the mixed state with long rang phase coherence and therefore vortex existence is evidence of phase coherence existence in any case. 

The existence of the phase coherence does not depend on any order of the 
vortex distribution since the relation $\oint_{l} dl\nabla \varphi = n2\pi  $, where n is number of the Abrikosov vortices inside $l$, is valid at any distribution of vortices inside $l$. Therefore the crystalline long-rang order of vortex lattice is not necessary order for the Abrikosov state. 

\bigskip

\noindent
\textit{3.3. Crystalline long-rang order of vortex lattice.} 

This order was predicted in the Abrikosov work [1] since an ideal case was considered in this work. It is impossible to obtain any other result besides a periodical structure for the case of homogeneous, symmetric, infinite space. The search of the periodical structure corresponded to minimum of the GL free-energy (1) can be reduced to the search of a function $\Psi $ corresponded to the lowest Landau level and smallest value of the Abrikosov's parameter $\beta_{A} = <\vert \Psi \vert ^{4}>/<\vert \Psi \vert^{2}>^{2}$. The GL free-energy (1) can be written in the form
$$f_{GL} = \sum\limits_{k} [\alpha + (n + 0.5)\frac{2e\hbar H}{mc} + \frac{\hbar ^{2}q_{z}^{2}}{2m}]\vert \Psi _{k} \vert ^{2} +$$
$$ \frac{\beta}{2}\sum\limits_{k_{i}} V_{k_{1} k_{2} k_{3} k_{4}}  \Psi^{*} _{k_{1}} \Psi^{*} _{k_{2}} \Psi _{k_{3}} \Psi _{k_{4}}  \eqno{(9)}$$

\noindent
when the wave function $\Psi (r)$ is expanded by the eigenfunctions $\varphi_{k}(r)$: $\Psi (r) = \sum_{k}\Psi _{k}\varphi _{k}(r)$ corresponded different Landau level. Here $k = (n,l,q)$; $n$ is the number of the Landau level; $l$ is the index of the Landau level functions; $q$ is the longitudinal (along magnetic field) wavevector; $V_{k_{1} k_{2} k_{3} k_{4}}  = \int_{V}d^{3}r \varphi^{*}_{k_{1}} \varphi^{*}_{k_{2}} \varphi_{k_{3}} \varphi_{k_{4}} $. At $H < H_{c2} = -\alpha mc/e\hbar $ the eigenvalues $\Psi_{k} = \Psi_{0,l,0}$ corresponded to the lowest Landau level $n = 0$ and $q = 0$ can give negative value of the GL free-energy. In order to find a solution corresponded to the minimum free-energy at $H_{c2}/3 < H < H_{c2}$ only the contribution from the lowest Landau level and $q = 0$ should be considered. There is not difference between tree and two-dimensional superconductors in the mean field approximation. The GL free-energy in the both case is 
$$f_{GL} = \frac{e\hbar }{mc}(H - H_{c2} ) < \vert \Psi \vert ^{2} > + \frac{\beta \beta _{A}}{2} < \vert \Psi \vert ^{2} > ^{2} \eqno{(10)}$$

\noindent
suitable for this case the superconducting state with the average density of superconducting pairs $< n_{s} > = < \vert \Psi \vert ^{2} > = V^{-1}\int_{V} d^{3}r\vert \Psi (r)\vert ^{2} = \sum\limits_{l} \vert \Psi_{l} \vert ^{2}= (e\hbar /m\beta \beta_{A}) (H_{c2} - H)$ and an smallest value of the Abrikosov's parameter $\beta_{A} = <\vert \Psi \vert ^{4}>/<\vert \Psi \vert^{2}>^{2}$ is thermodynamically most stable. The smallest $\beta _{A}$ value could be equal 1 if $\Psi (r)$ is any function. But for the functions $\Psi (r)$ corresponded to the lowest Landau level the $\beta_{A}$ value should exceed 1. Abrikosov found [1] that a periodical square structure gives the value $\beta_{A} \approx  1.18$. Later [2] it was shown that the triangular lattice yield the smaller value $\beta_{A} \approx  1.16$. 

\

\noindent
\textbf{4. Delusions provoked by the direct observation of the Abrikosov state. }

Magnetic flux structures at a low field was observed at first experimental corroboration of the Abrikosov prediction. Therefore the visualization of the Abrikosov state as periodic flux line structure is predominated up to now. This visualization used already during forty years has provoked some delusions and even no quite correct names. These delusions provoked in one's turn erroneous notion about the transition from the Abrikosov state to the normal state. Therefore they should be indicated here. 

\bigskip

\noindent
\textit{4.1. Abrikosov vortices are not magnetic flux lines.} 

It is important to emphasize first of all that the Abrikosov vortices are no magnetic flux lines but are singularities in the mixed state with phase coherence. Because of the Bohr quantization the momentum circulation of superconducting pair along any closed path around single vortex should be equal $2\pi \hbar $
$$\oint_{l} dlp = \oint_{l} dl(mv + 2eA) = m\oint_{l} dlv + 2e\Phi  = 2\pi \hbar \eqno{(11)}$$

The value $\oint_{l} dlp / 2e $ was called fluxoid by F.London [66] and $2\pi \hbar /2e$ is the fluxoid quantum. The relation (11) describes completely the velocity $v$ distribution around the vortex. Most direct observations of the Abrikosov state were made in a low magnetic field, when a distance between vortices exceeds the penetration length $a \gg \lambda $, contours $l$ exist for which $\oint_{l} dlv = 0 $ and $\Phi  = \Phi_{0}$ and the gradient of magnetic field induced by the current $j_{s} = n_{s}2ev$ is enough large. In this case the Abrikosov vortex seems a solenoid in which the current $j_{s} = n_{s}2ev$ indices magnetic flux, Fig.11a. Therefore it is misapprehended as a flux line [21,47-49,59]. But the nature of the vortex can not change in the opposite limit $a \ll  \lambda $ when the current $j_{s} = n_{s}2ev$ induces only negligible flux and it is obvious that the magnetic flux is induced by an external magnet Fig.11b. This flux changes very slightly when the vortices disappear at the transition from the Abrikosov state. If the vortices are flux lines why can magnetic flux remains when the vortices disappear? 

Therefore it is obvious that the Abrikosov vortices are singularities in the mixed state with phase coherence. These singularities allow the magnetic flux to penetrate into a superconductor with long-range phase coherence but do not create the magnetic flux. A singularity can not exist without a medium. Consequently the existence of the vortices is an evidence of the existence of the phase coherence. 

It is important to understand that the Abrikosov vortex is not magnetic flux line since this incorrect conception provoked some other no quite correct conception: about causes of non-zero and zero resistance in the Abrikosov state and first of all about the vortex lattice melting. 

\bigskip

\noindent
\textit{4.2. Flux flow resistance is not induced by flux flow.} 

It is written in all books devoted the mixed state of type II superconductor that non-zero voltage in the Abrikosov state is induced by a motion of magnetic flux structure [21,47-49,59]. The magnetic flux lines move (with a velocity $v_{vor}$) under the influence of an electric current (with a density $j$) when the Lorentz force $f_{L} = j\times H$ exceeds a pinning force $f_{p}$ [21,49,67], i.e. when the applied current exceeds a critical current, $j > j_{c} = f_{p}/B$. The steady-state motion of the magnetic flux structure induced by the applied electric current causes the time-averaged macroscopic electric field following Faraday's law $E = -grad V = - (v_{vor}\times H)$ [21,49,67]. The forces regulating the flux motion satisfy the following equation
$$f_{L} - f_{p} =\eta v_{vor} \eqno{(12)}$$

\noindent
where $\eta v_{vor}$ is the damping force, $\eta $ is the viscosity coefficient. According to this equation a potential difference along x, $dV/dx = (H^{2}/\eta)(j - j_{c})$ when the magnetic field $H$ is directed along $z$, the current $j$ - along x and vortices move along y. The slope of the current-voltage characteristic at $j > j_{c}$ is called flux-flow resistivity $\rho_{f} = H^{2}/\eta $. 

\begin{figure}
\includegraphics{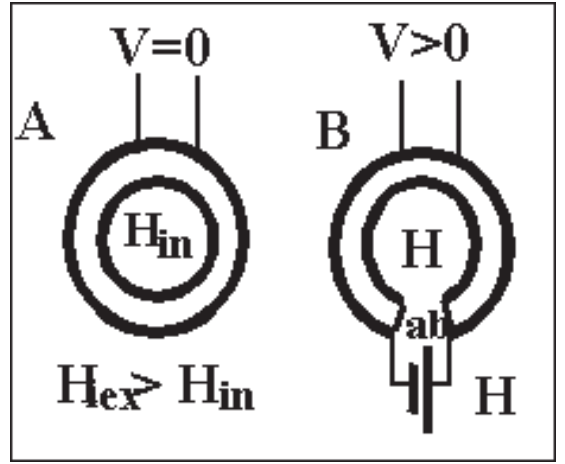}
\caption{\label{fig:epsart} The vortex flow in closed (A) and unclosed (B) type II superconductors placed in  a magnetic field under the influence of an applied electric current. The current with the same density j is induced by a difference of magnetic fields inside $H_{in}$ and outside $H_{ex}$ the closed loop  the case A and by a dc power source in the case B. The value of the magnetic flux changes with a speed $n_{vor}\Phi_{0}2\pi Rv_{vor}$ inside the closed loop (A) and does not change inside the unclosed loop (B). }
\end{figure}

But it is enough easy to show that this wide-spread conception of the flux-flow resistivity is not correct. Two cases of the vortex flow under the influence of an applied electric current are shown on Fig.12. The same current density $j$ is induced by a difference of the magnetic field values inside $H_{in}$ and outside $H_{ex}$ of a closed superconducting cylinder in the case A and by an external dc power source in a unclosed cylinder in the case B. The vortices move with the same velocity $v_{vor} = (f_{L} - f_{p})/\eta $ from the outside to the inside of the both cylinders. It is obvious that the magnetic flux flows in the case A and does not flow in the case B since its value changes in time inside the closed cylinder and does not change inside the unclosed one. But it is obvious also that a potential difference $V  \neq  0$ can be observed only in the case B when the flux flow is absent and can not be observed in the case A when it takes place.

Thus, in the contrast to conception of the flux-flow resistivity the potential difference is observed when the flux flow is absent and is not observed when the magnetic flux flows. This seeming paradox can be solved if we take into account that the momentum (phase gradient of wave function) is not the gauge-invariant value because of the vector potential $A$. A gauge-invariant value, such as velocity of superconducting pairs should be considered in order to find the real cause of the potential difference observed in the Abrikosov state. It is obvious that the integral $\int_{l} dlv $ along a path between points a-b (see Fig.12, B) does not change when vortices shift on a period of the vortex lattice if the current density $j$ is constant in time. The same is right also for a closed path $l$ in the case A. The current density $j = (H_{ex} - H_{in})/w$ changes in time in this case. But this change at vortex shifting on one period $\Delta j = -\Delta H_{in}/w = 2\Phi _{0}/Rw$ is very small at a large radius $R$ of the cylinder and therefore can be disregarded.

The constancy of the velocity in the case A is following from the relation for the momentum circulation (11) $m\oint_{l} dlv =  2\pi \hbar n - 2e\Phi $. It is important here that the integral of the vector potential along a closed path is the gauge-invariant value $\oint_{l} dlA = \Phi $. The momentum circulation changes on $2\pi \hbar $ and the integral of the phase gradient $\oint_{l} dl\nabla \varphi $ changes on $2\pi $ when an Abrikosov vortex crosses the closed path $l$. But in the same time the magnetic flux $\Phi $ inside $l$ changes on $\Phi_{0}$. Therefore when $n'$ vortices have crossed the closed path $l$ the value $2\pi \hbar n - 2e\Phi $ has not changed since $2\pi \hbar n' - 2en'\Phi_{0} = 0$. Consequently the velocity does not change in the case A since the magnetic flux flow compensates for the change of the phase difference $\oint_{l} dl\nabla \varphi  $ because of the vortex motion. 

In the case B the flux flow is absent. Therefore a potential difference $V$ should compensate for the change of the phase difference. The voltage $V$ and the velocity of the phase difference change $d\varphi /dt$ is connected by the Josephson relation $d\varphi /dt = 2eV/\hbar $ [68]. Since the phase difference $\Delta \varphi $ between two points changes by $2\pi $ when a vortex crosses a line connecting these points then $d\varphi /dt = v_{vor}n_{vor}2\pi $ along a length unit. Consequently, the macroscopic voltage along a length unit is equal to $E = -grad V = (\hbar /2e)d\varphi /dt = v_{vor}n_{vor}(\pi \hbar/e) = v_{vor}n_{vor}\Phi_{0}$. Where $v_{vor}$ is the velocity of the vortex flow, $n_{vor}$ is the density of the vortices.

This result coincides nominally with the one obtained by the Faraday's law $E = -v_{vor}\times  H$, because $H = n_{vor}\Phi_{0}$ in the Abrikosov state. Therefore, it became possible that the resistivity in the Abrikosov state is considered as a consequence of flux flow in all textbooks [21,47-49] and majority of papers, although it is obvious that the magnetic flux does not flow in a superconductors in this case. It is obvious also that the Lorentz force can not be the driving force on an Abrikosov vortex [69]. One should use the more correct notation "vortex flow resistivity" instead of "flux flow resistivity" since the latter is not quite correct notation. The vortex should consider as singularity moving of which induces a change of phase difference.

\bigskip

\noindent
\textit{4.3. Long range phase coherence but no vortex pinning is main reason of zero resistance in the Abrikosov state.} 

Although the conception about the Abrikosov state as magnetic flux structure is not correct it allowed to describe many properties and features of the mixed state with long-rang phase coherence of type II superconductors. It was explained in the Section 4.2 that the incorrect conception of the flux flow resistivity described enough well resistive properties of the Abrikosov 
state since the result given by it does not differ in the main from the one given by the correct conception. The conception of the magnetic flux structure has made greatest progress in the description of vortex pinning [70]. First of all because of this progress it is very popular. There is not a difference between this conception and the correct conception until only the mixed state with long-rang phase coherence is considered. But there is a difference based on principle when the transition from the Abrikosov state is considered. 

According to the conception of vortex pinning the resistance can be equal zero in the Abrikosov state since the pinning force prevents from the vortex flow [70]. The pinning force exists since superconductor disorders eliminate the homogeneity of the space in which the vortex structure exists. The pinning disorders are very important for the Abrikosov state. As it was shown in the Section 2.6 this state is absolutely unstable in homogeneous infinite space. Without pinning disorders (including superconductor boundaries) vortex structure should move under the influence of anyhow weak electric current. Therefore the vortex pinning is considered as the reason of non-dissipation current in the Abrikosov state. It is correct. 

But one should remember at a consideration of the transition from the Abrikosov state that any non-dissipation current is observed in superconductors first of all because of long-range phase coherence. One should remember that the Abrikosov vortices are singularities in the mixed state with the long-rang phase coherence. Therefore the vortex pinning is also a consequence of the phase coherence. The vortex pinning can not exist in a state without the phase coherence. Therefore the transition between the mixed states with and without long-range phase coherence can be fixed first of all by the use of an observation of disappearance of the non-dissipation current and the vortex pinning. Changes of the resistive properties should be observed first of all at the transition into the Abrikosov state, because a transition from the paraconductivity regime to the vortex flow regime must occur and the vortex pinning can appear at the appearance of the long-range phase coherence

\bigskip

\noindent
\textit{4.4. Fundamental differences between vortex ``lattice'' and crystal lattice.} 

Many scientists consider the Abrikosov state as flux line lattice like an 
atom lattice, or a lattice of long molecules. Therefore it is important to 
emphasize that the Abrikosov state differs in the main on any crystalline 
lattice. One should once again say that the Abrikosov vortex is singularity 
but is not a flux line. The flux lines can be likened interacted particles 
which can form a lattice. Whereas singularities witness an other long-rang 
order. The energy of vortices is inseparable from the energy of their 
interaction whereas the energy of atoms ore molecules is much larger than 
the energy of their interaction. Therefore atoms and molecules remain almost 
invariable when their lattice melts. But it is impossible to say about the 
vortices. Therefore the popular image of flux line lattice misleads only and 
does not correspond to the facts. 

All atom and molecule lattices are formed in homogeneous symmetric space 
only because of their mutual interaction. Just therefore the melting of 
these lattice is phase transition. The distinguishing feature of phase 
transition is spontaneous change of a symmetry [71]. Any change of a 
symmetry can be only spontaneous in any homogeneous symmetric space if any 
external influence does not break the symmetry. An external influence can 
make a change of a symmetry non-spontaneous and then the phase transition 
can disappear. Such disappearance takes place with the ferromagnetic 
transition in a strong magnetic field. 

The vortex lattice predicted by Abrikosov breaks displacement symmetry. The 
same symmetry is broken by pinning disorders. Therefore the existence of 
pinning disorders is the external influence because of which the breach of 
displacement symmetry at the transition into the Abrikosov state can be 
non-spontaneous. It is important to remember that the real Abrikosov state 
is observed in an inhomogeneous space with broken displacement symmetry. 
Since the experimental results show strong influence of pinning disorders 
and symmetry of the crystal lattice of the superconductor on the vortex 
structure [21,31-33,72-76] it is important to answer on the question: ``What state could be in homogeneous symmetric infinite space, i.e. in the space in 
which atom and molecule lattices exist?'' 

\bigskip

\noindent
\textbf{5. Qualitative change because of taking into account of thermal fluctuation.} 

Fluctuation theory does not limit oneself to the search of a state corresponded to minimum of the GL free-energy (1) but considers all possible 
states and calculates thermodynamic average of different quantities. The 
thermodynamic average of any quantity, for example order parameter, is 
$$ < \vert \Psi \vert ^{2} > = \frac{\sum \vert \Psi \vert ^{2}\exp (-F_{GL}/k_{B}T)}{\sum \exp (-F_{GL}/k_{B} T)} \eqno{(13)}$$
In this case the total relation (1) of the GL free-energy should be used. But near the second critical field $H_{c2}$ this relation can be simplified since at $h = H/H_{c2}(T=0) \gg  (k_{B}T_{c}/H_{c}^{2}(0)\xi ^{3}(0))^{2}$, $H - H_{c2}(T) \ll  2H$ above $H_{c2}$ and $H > H_{c2}/3$ below $H_{c2}$ the main contribution to any thermodynamic average gives functions corresponded to the lowest Landau level with $n = 0$. Main qualitative changes because of taking into account of thermal fluctuation are obvious in this lowest Landau level approximation. Therefore it will be considered first of all in this paper. 

\bigskip

\noindent
\textit{5.1. The lowest Landau level approximation.}

The density of the GL free-energy in the lowest Landau level (LLL) approximation is
$$f_{GL} = \sum\limits_{k} [\frac{e\hbar}{m}(H - H_{c2} ) + \frac{\hbar ^{2}q_{z}^{2}}{2m}]\vert \Psi _{k} \vert ^{2} + $$ 
$$\frac{\beta }{2}\sum\limits_{k_{i}}  V_{k_{1} k_{2} k_{3} k_{4}}  \Psi _{k_{1}} ^{*}  \Psi _{k_{2}} ^{*}  \Psi _{k_{3}}  \Psi _{k_{4}}  \eqno{(14)}$$

\noindent
where $\Psi_{k} = \Psi_{0,l,q}$. The summation by the longitudinal wave-vector $q = 2\pi l/L_{z}$ depends on a superconductor size $L_{z}$ along magnetic field, where $l$ is a integer number. In bulk sample the difference between values $\hbar ^{2}q^{2}/2m$ corresponded adjacent values of wave-vector is much lesser than the first addendum in (14) $\hbar ^{2}q^{2}(l+1)/2m - \hbar^{2}q^{2}(l)/2m \approx  (2\pi \hbar)^{2}/2mL_{z}^{2}\ll  (e\hbar/m)\vert H-H_{c2}\vert  < (e\hbar /m)H_{c2} = (e\hbar/m)\Phi_{0}/2\pi \xi^{2}(T) = \hbar^{2}/2m\xi ^{2}(T)$. Therefore the summation can be replaced by integration in tree-dimensional superconductor with $L_{z} \gg  2\pi\xi (T)$. In the opposed limit $L_{z} \ll  2\pi\xi (T)$ of two-dimensional superconductor the contribution to any thermodynamic average of the functions with $q = 0$ is much larger than the one of all other functions with $q > 0$. Therefore one can limit oneself to $q = 0$ in (14) in case of thin film and layered superconductor 
$$f_{GL} = \frac{e\hbar}{mc}(H - H_{c2} ) \sum\limits_{k} \vert \Psi_{k} \vert ^{2} + $$ 
$$\frac{\beta}{2}\sum\limits_{k_{i}} V_{k_{1} k_{2} k_{3} k_{4}}  \Psi _{k_{1}} ^{*}  \Psi _{k_{2}} ^{*}  \Psi_{k_{3}}  \Psi _{k_{4}} \eqno{(15)}$$

\noindent
where $\Psi_{k} = \Psi_{0,l,0} = \Psi_{l}$. This relation repeats the relation (10): $\sum_{l}\vert \Psi_{l}\vert^{2} = <\vert \Psi _{l}\vert^{2}>$ and $\sum_{l}V_{l1,l2,l3,l4}\Psi_{l1}\Psi_{l2}\Psi_{l3}\Psi_{l4} = <\vert \Psi \vert^{4}> = \beta_{a}<\vert \Psi \vert^{2}>^{2}$. But it is very difficult to calculate thermodynamic average of the Abrikosov's parameter $\beta_{A}$ and to find the transition into the Abrikosov state taking into account of thermal fluctuation even for this simplest case. According to the mean field approximation the Abrikosov's parameter $\beta_{A}$ can make sense only below $H_{c2}$ since the density of superconducting pairs $<\vert \Psi \vert^{2}> = 0$ at $H > H_{c2}$. But because of fluctuations $<\vert \Psi \vert^{2}> \ > 0$ both below and above $H_{c2}$. According to fluctuation theory thermodynamic average of the Abrikosov's parameter $<\beta_{a}>$ changes from $<\beta_{a}> = 2$ in fluctuation region enough above $H_{c2}$ to $\beta _{A} \approx  1.159595$, calculated for triangular structure [2], in the mixed state enough below $H_{c2}$ (see for example [77]). The dependence $<\beta_{a}>(H/H_{c2})$ was calculated for the present only in some approximations and only for two-dimensional superconductor [77, 44]. These results does not give an answer having a single meaning on existence of the Abrikosov state. Although a total solution about the transition into the Abrikosov state is not obtained for the present we can make some conclusions about its possible position considering the GL free-energy in the LLL approximation (14). 

\bigskip

\noindent
\textit{5.2. Reduction of the effective dimensionality near $H_{c2}$.}

It is important that the GL free-energy of bulk superconductor in the LLL approximation (14) is like the one of one-dimensional superconductor. There is no difference between superconductors with different dimensionality according to the mean field approximation. Therefore it is assumed during a long time that the second law phase transition takes place at $H_{c2}$ since the ``jump'' of the specific heat is observed at $H_{c2}$ as well as at $T_{c}$. This ``jump'' $C_{s} - C_{n} = \alpha_{0}^{2} /\beta $ should be observed according to the mean field approximation. Therefore the observation only this ``jump'' is not evidence of the second order phase transition. This ``jump'' should be observed in superconductor with any dimensionality but the second order phase transition can be only in three-dimensional superconductor and $\lambda $ anomaly of the specific heat dependence should be observed at this transition. 

\begin{figure}
\includegraphics{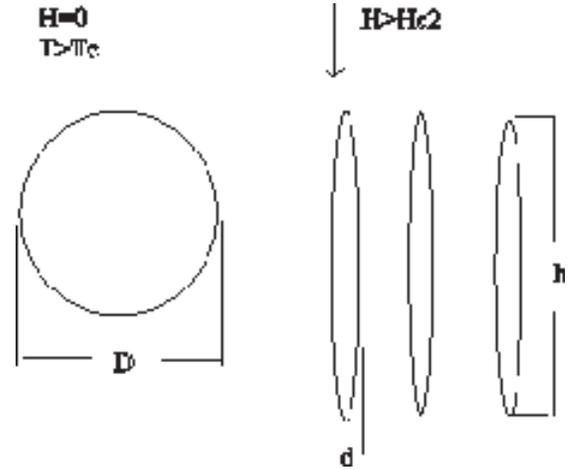}
\caption{\label{fig:epsart} Superconducting drops above $T_{c}$ and above $H_{c2}$. $D = \xi (T) = \xi (0)(T/T_{c} - 1)^{-1/2}$; $h = \xi _{z} = (\Phi_{0}/2\pi (H-H_{c2}))^{1/2}$;  $d = \xi _{\rho }\approx  (\Phi_{0}/H){1/2}$}
\end{figure}

But the $\lambda $ anomaly was not observed in superconductors for the present. It is not observed at $T_{c}$ because the coherence length of superconductors is very large. Therefore the width of phase transition is very small and the $\lambda $ anomaly can not be observed in a real experiment. It is assumed during a long time that the same takes place also at $H_{c2}$. Therefore the experimental results obtained in the seventieth years [4-7] are very important. They have shown that the dependence (``jump'') of the specific heat of bulk superconductor near $H_{c2}$ can be described by theoretical dependence for a one-dimensional superconductor and therefore the second order phase transition into the Abrikosov state can not be at $H_{c2}$.

\begin{figure}
\includegraphics{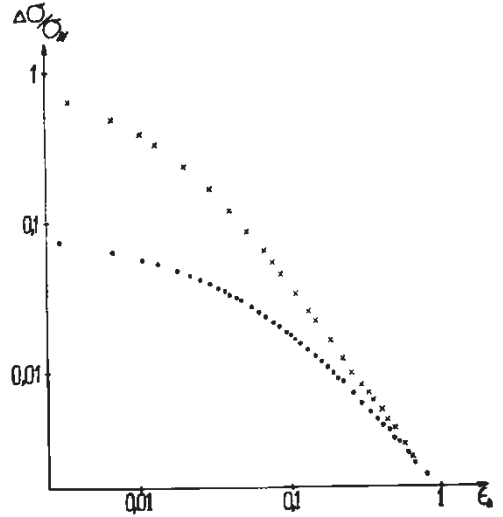}
\caption{\label{fig:epsart} Dependencies of excess conductivity $\Delta \sigma /\sigma _{n} = \sigma (T,H)/\sigma_{n} - 1$ of single-crystal  $V_{3}Ge$ on the value of parallel and perpendicular magnetic field at $T = 4.2 \ K$ measured in [51]. $\epsilon_{H} = H/H_{c2} - 1$, $H_{c2} = 4.32 \ T$.}
\end{figure}

The $\lambda $ anomaly is observed because of a divergence of the coherence length $\xi (T)$ at the transition. The coherence function for bulk superconductor in a linear approximation region above $T_{c}$ is $g(R) = <\Psi^{*}(r)\Psi(r+R)> = (mk_{B}T/2\pi \hbar^{2})exp[-R/\xi (T)]/R$ at $R \gg  \xi (T)$ [49] and typical size of superconducting fluctuation drops above $T_{c}$ equals $\xi (T)$. 

\begin{figure}
\includegraphics{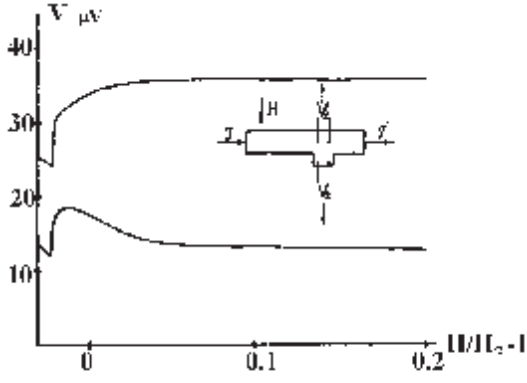}
\caption{\label{fig:epsart} The magnetic field dependencies of potential difference measured on a branch $V_{1}$ (lower curve) and on a main part $V_{2}$ (upper curve) of a single crystal $V_{3}Ge$ sample at constant current through the main part of the sample and $T = 4.2 \ K$ [51,79]. The non-local resistance observed near $H_{c2}$ may be interpreted as a consequence of the anisotropy of the conductivity, Fig.14, or superconducting drops, Fig.13.}
\end{figure}

\noindent
The fluctuation drop is spherical when magnetic field is absent: $R= (x^{2}+y^{2}+z^{2})^{1/2}$, Fig.13 and its radius increases together with $\xi (T)$ at $T \to T_{c}$. The second low phase transition takes place when $\xi(T)$ (radius fluctuation drop) mounts to infinity (to sample size in real case) and phase coherence spreads out the whole of superconductor. According to the linear approximation of the fluctuation theory $\xi (T) = \xi (0)(T/T_{c} - 1)^{1 / 2}$ both in three- and one-dimensional superconductor. But already in the Hartree approximation, first approximation taking into account fluctuation interaction, qualitative difference is between three- and one-dimensional superconductor. Taking into account fluctuation interaction removes the divergence of the coherence length in one-dimensional case but this divergence remains in three-dimensional superconductor. Therefore the second order phase transition and the $\lambda $ anomaly of the specific heat take place at $T_{c}$ in bulk superconductor and are absent in one-dimensional case.

The coherence function for bulk superconductor near $H_{c2}$ is like to one for one-dimensional superconductor. The coherence length increases only along magnetic field at $H \to H_{c2}$. In the linear approximation $g(\rho ,z)  \propto  exp(-\vert z\vert /\xi_{z})exp(-\rho /\xi_{\rho}^{2})$ [49], where $\xi_{z} = [\Phi_{0}/2\pi (H-H_{c2})]^{1 / 2}$ and $\xi_{\rho}  = (2\Phi_{0}/\pi H)^{1/2}$ are the coherence lengths along and across magnetic field; $\rho  = (x^{2}+y^{2})^{1/2}$. Size of superconducting drop along magnetic field $\xi_{z}$ exceeds considerably the one across the $H$ direction, $\xi _{z}\gg  \xi _{\rho} $, near $H_{c2}$, at $\vert H-H_{c2}\vert \ll  H$, Fig.13. It is manifested in the anisotropy of paraconductivity [78,79] along and across magnetic field observed near $H_{c2}$, Fig.14, and the non-local resistance, Fig.15. The anisotropy increases near $H_{c2}$, Fig.14, as well as the $\xi_{z}/\xi_{\rho}$ relation. The $\xi_{z}$ value increases at $H  \to H_{c2}$ but remains finite not only at $H = H_{c2}$ but also $H < H_{c2}$, as well as the conductivity along magnetic field, Fig.15, since taking into account fluctuation interaction removes the divergence in one-dimensional case.

\bigskip

\noindent
\textit{5.3. Scaling law.} 

Since the GL free-energy of whole superconductor $F_{GL} = Vf_{GL}$ is used in the relations for thermodynamic average (13) it is more handy to replace of $\Psi _{k}$ in the relation (14) for $Vf_{GL}$ by $\Psi'_{k} = V^{1 /2}\Psi_{k}$. Then 
$$\frac{F_{GL}}{k_{B} T} = \sum\limits_{k} [\frac{e\hbar}{mk_{B}T}(H - H_{c2} ) + \frac{\hbar ^{2}q_{z}^{2}}{2m}]\vert \Psi '_{k}\vert ^{2} + $$
$$\frac{\beta}{2Vk_{B} T}\sum\limits_{k_{i}} V_{k_{1} k_{2} k_{3} k_{4}}  \Psi_{k_{1}}^{'*}\Psi_{k_{2}}^{'*} \Psi _{k_{3}}^{'} \Psi_{k_{4}}^{'} \eqno{(16)}$$
It is handy to use a dimensionless unit system in which $(e\hbar H_{c2}(0)/mk_{B}T)^{1/2}Gi_{H}^{1/2}\Psi'_{k} = \Psi''_{k}$, a length unit across magnetic field is $(\Phi_{0}/H)^{1/2}$ and a length unit along magnetic field is $\xi (0)Gi_{H}$ [9,57], where $Gi_{H} = Gi^{1/3}(th)^{2/3}$ is the Ginzburg number in magnetic field; $Gi = (k_{B}T_{c}/H_{c}^{2}(0)\xi ^{3}(0))^{2}$ is the Ginzburg number; $t = T/T_{c}$; $h = H/H_{c2}(0)$; $H_{c2}(0)$ is the GL second critical field at $T = 0$; $H_{c}(0)$ is the thermodynamic critical field at $T = 0$. It is followed from (13) and the relation for the GL free-energy in the dimensionless unit system 
$$\frac{F_{GL}}{k_{B}T} = \sum\limits_{k}(\epsilon_{n} + q^{2})\vert \Psi \vert ^{2} + \frac{1}{2V}\sum\limits_{k_{i}}V_{k_{1}k_{2}k_{3} k_{4}}\Psi _{k_{1}}^{*}\Psi _{k_{2}} ^{*}\Psi _{k_{3}}\Psi _{k_{4}}  \eqno{(17)}$$

\noindent
that thermodynamic average of all values is universal function of $\epsilon = (h-h_{c2})/Gi_{H} = (h-h_{c2})(ht)^{-2/3}Gi^{ -1/3}$ in the LLL approximation. 

In order to find the transition into the Abrikosov state the veritable free-energy 
$$F = -k_{B}T \ln \sum_{\Psi_{k}} \exp(-\frac{F_{GL}}{k_{B} T} \eqno{(18)}$$
can be considered. Since the free-energy depends on only parameter $\epsilon $ all theories should give an identical dependence of a critical field $H_{c4}$ position on temperature and on the Ginzburg number defined by the relation $\epsilon_{c4} = (h-h_{c2})(ht)^{-2/3}Gi^{-1/3}$. This scaling law should be observed in the region $h \gg  Gi$, $h-h_{c2} \ll  2h$ and $h > h_{c2}/3$ where the LLL approximation is valid. This means that a coincidence of a theoretical dependence $H_{c4}(T,Gi)$ with the experimental one does not prove correctness of the theory. This proves only that the LLL approximation is valid. 

\begin{figure}
\includegraphics{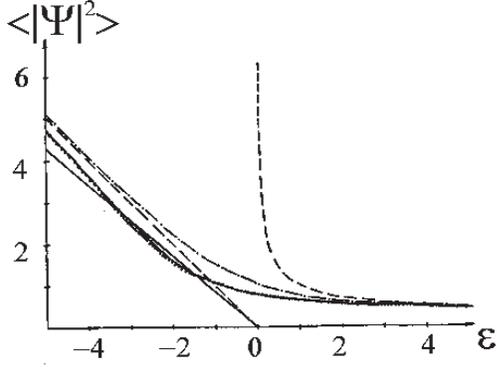}
\caption{\label{fig:epsart} Temperature dependencies of density of superconducting pairs in the dimensionless unit system calculated on the base of the GL free energy (19) in different approximations [51]: the solid straight line is the mean field approximation for bulk type II superconductor [1] near $H_{c2}$; the dashed straight line is the same for a 1D superconductor; the solid curve is exact fluctuation theory for 1D superconductor [80]; the dot curve is Hartre-Fock approximation with  screened potential [5]; the dash-dot curve is Hartre approximation; the dash curve is the linear approximation of the fluctuation theory. }
\end{figure}

The relation (17) in the LLL approximation is close to the one for a one-dimensional superconductor. We may write this relation in following form
$$\frac{F_{GL}}{k_{B} T} = V(\epsilon <\vert \Psi \vert^{2}> + \frac{1}{V}\int\limits_{V} dr^{3}\Psi ^{*}\frac{d^{2}}{dz^{2}}\Psi  + \frac{\beta _{a}}{2}<\vert \Psi \vert ^{2}> ^{2}) \eqno{(19)}$$
The main difference the relation (19) from the one for a one-dimensional superconductor consists in the generalized Abrikosov parameter value $\beta_{a}$ which is determined by a distribution of the order parameter in the space. Much below the transition $\Psi (r) = const$ in one-dimensional superconductor and consequently $\beta_{a} = 1$. But function $\Psi (r) = const$ is not belong to the LLL functions. This has a simple interpretation: a magnetic field must penetrate through superconductor. Therefore $<\beta_{a}>$ changes from 2 at $H \gg  H_{c2}$ [77] to $\beta_{A} \approx  1.159595$ at $H \ll  H_{c2}$ [2].

\bigskip

\noindent
\textit{5.4. Density of superconducting pairs and phase coherence.}

Although the thermodynamic average of the Abrikosov parameter $<\beta_{a}>$ is not calculated exactly for the present for the whole LLL approximation region the dependence of density of superconducting pairs $<n_{s}> = <\vert \Psi \vert^{2}>$ can be calculated enough exactly from (19) since it is known that $2 > \ <\beta_{a}>  \ge  1.159595$. It is important that the appearance of phase coherence is connected rather with the $<\beta_{a}>$ value than with $<n_{s}>$. 

\begin{figure}
\includegraphics{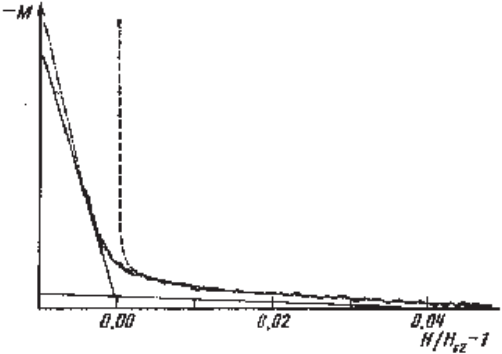}
\caption{\label{fig:epsart} Dependence of the magnetization  on the magnetic field in the fluctuation region near $H_{c2}$ of a single-crystal $Nb_{94.3}Mo_{5.7}$ sample at $T = 4.2 \ K$ [54]: wave curve is experimental dependence; dashed curve is linear approximation; dash-dot - Hartre-Fock approximation with  screened potential; solid - mean field approximation.}
\end{figure}

According to the mean field approximation a non-zero density of superconducting pairs $n_{s}$ and long-rang phase coherence appear simultaneously at $H = H_{c2}$. According to the fluctuation 
theory $<n_{s}>$ appears already above $H_{c2}$. Appearance of phase coherence loses a connection with the second critical field but the $H_{c2}$ remains a particular point for the $n_{s}(H)$ dependence. 

The theoretical dependencies of the thermodynamic average density of superconducting pairs $<\vert \Psi \vert^{2}>$ for one-dimensional superconductor and a bulk superconductor in the LLL approximation are compared on Fig.16. The dependence calculated in the Hartre-Fock approximation with screened potential [5] coincides almost completely with the one obtained by the exact solution for one-dimensional model [80]. The dependence obtained in the Hartree approximation differs from the exact one near the critical point. All three dependencies have the same asymptotic $<\vert \Psi \vert^{2}> = -\epsilon $ at $-\epsilon \gg  1$. It differs from the asymptotic for the LLL approximation $<\vert \Psi \vert^{2}> = -\epsilon /\beta_{A}$ at $-\epsilon \gg  1$ since for both one-dimensional superconductor and Hartry (Hartree-Fock) approximations $\beta _{a} = 1$ at $-\epsilon \gg  1$. 

\begin{figure}
\includegraphics{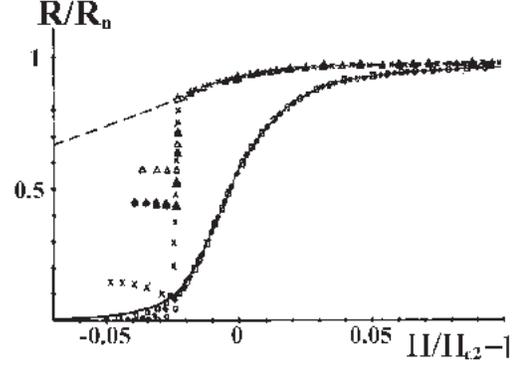}
\caption{\label{fig:epsart} The resistive transition of a single-crystal $V_{3}Ge$ sample for different current densities for perpendicular and parallel magnetic field. The lines denote the theoretical dependencies obtained in the Hartree approximation for perpendicular (dashed) and parallel (solid) magnetic fields [51].}
\end{figure}

According to (18) and (19) the magnetization $M = -dF/dH$ is proportional to $<n_{s}>$ and the specific heat $C = -T(d^{2}F/dT^{2})$ is proportional to $T(d<n_{s}>/dT)$ in the lowest Landau level approximation region. These and also some other values are determined first of all by average density of superconducting pairs and depend weakly from existence of phase coherence. Therefore the one-dimensional model is described enough well the specific heat Fig.4 and magnetization dependencies Fig.17 in the mixed states both without and with long-rang phase coherence. Dynamical properties, such as resistance, are described enough well by the one-dimensional model only in 
the mixed state without phase coherence, Fig.18. Sharp deviation of the 
experimental resistance dependencies from the theoretical one obtained in 
the one-dimensional model is observed at the transition into the Abrikosov 
state, Fig.18. This deviation is more appreciable for resistance dependence 
in perpendicular magnetic field because of high anisotropy of 
paraconductivity at $H_{c4}$, Fig.18. The length of phase coherence changes 
first of all across magnetic field at the sharp transition into the 
Abrikosov state. 

\bigskip

\noindent
\textbf{6. Transition into the Abrikosov state.} 

Thus, dependence near $H_{c2}$ of the quantities connected with 
the average density of superconducting pairs, such as magnetization, 
specific-heat, paraconductivity, can be enough easy described in the LLL 
approximation. The same can be made for the resistance dependencies in the 
mixed state without phase coherence. The sharp deviation from the 
theoretical dependence should be interpreted as a consequence of the 
transition into the Abrikosov state. This interpretation is having one 
meaning since just such qualitative change of the resistive properties 
should be observed at the long-rang phase coherence appearance. The 
resistance can be equal zero because of the vortex pinning in the Abrikosov 
state, i.e. in the mixed state with long rang phase coherence, whereas in 
the fluctuation mixed state without long rang phase coherence it has a 
non-zero value.

The sharp qualitative change of the resistive properties is observed at $H_{c4} < H_{c2}$ in all high qualitative bulk superconductors with enough weak 
pinning disorders, both conventional [8] and HTSC [11,12,56]. The great difference of the $(H_{c2}-H_{c4})$ value for superconductors $Nb_{94,3}Mo_{5,7}$, $V_{3}Ge$ and $YBa_{2}Cu_{3}O_{7-x}$ corresponds to the great difference of the Ginzburg number and $(H_{c2}(0)$ value [57]: $Nb_{94,3}Mo_{5,7}$ $Gi = 10^{-9}$, $H_{c2}(0) = 0.8 \ T$; $V_{3}Ge$, $Gi = 10^{-6}$, $H_{c2}(0) = 12 \ T$; $YBa_{2}Cu_{3}O_{7-x}$ $Gi = 10^{-2}$, $H_{c2}(0) = 200 \ T$. The same experimental value $\epsilon_{c4} \approx  1$ for these superconductors conforms the scaling law of the LLL approximation. Therefore there is a base to think 
that the experimental determination of the transition into the Abrikosov 
state in bulk superconductors with weak pinning disorders is enough 
reliable. But theoretical description of the phase coherence appearance is absent for the present. This problem is not solved completely up to now. Some theories, [9,81] and others including the vortex lattice melting theories, calculate the value $(1 - H_{c4}/H_{c2})$ enough close to the one observed experimentally in bulk superconductors, both conventional and $YBa_{2}Cu_{3}O_{7-x}$. But this 
coincidence does not mean that these theories describe indeed the transition into the Abrikosov state, since it may be only a consequence of the scaling law.

\bigskip

\noindent
\textit{6.1. The phase transition into the Abrikosov state of an ideal superconductor must be first order.}

Although the transition into the Abrikosov state is not described theoretically for the present we can conclude anything about it from 
consideration of difference between states with and without long rang phase coherence. We can show that it should be first order phase transition and 
should occurs below the second critical field. 

The transition into the Abrikosov state can not be second order since size of superconducting drops increases only along magnetic field near $H_{c2}$, Fig.13. They have approximately the same size across magnetic field $\xi_{\rho} = (2\Phi _{0}/\pi H)^{1/2}$ in all region near the second critical field, at $H \approx H_{c2}$. This size is much smaller than a size $L$ of a real sample, $(2\Phi_{0}/\pi H)^{1/2} \ll  L$. For example in work [8], the sharp transition was observed in a sample with $L  \approx 1 \ mm$ at $H_{c4} \approx  40 \ kOe$, what corresponds $\xi_{\rho} = (2\Phi_{0}/\pi H_{c4})^{1/2} \approx  2 \ 10^{-5} \ mm$. The observation of the non-dissipation current below $H_{c4}$ means that phase coherence spreads over whole superconductor with macroscopic sizes $L \gg  (2\Phi_{0}/\pi H)^{1/2}$. If the sample is absolutely homogeneous only these two characteristic lengths $L$ and $(2\Phi_{0}/\pi H)^{1/2}$ exist across magnetic field. Consequently phase coherence 
should appear by jump from $(2\Phi_{0}/\pi H)^{1/2}$ to L and the transition into the Abrikosov state should be first order in an ideal superconductor without disorders. Thus, the result obtained in the fluctuation theory differs qualitatively from the one obtained in the mean field approximation. The transition into the Abrikosov state can not be second order phase transition. It must be first order phase transition in ideal superconductors 
without disorder.

The size of superconducting drops across magnetic field $\xi_{\rho} = (2\Phi_{0}/\pi H)^{1/2}$ in the mixed state without phase coherence is approximately equal the 
distance between the vortices $(2\Phi_{0}/\surd 3 H)^{1/2}$ in the Abrikosov state. But these two mixed states have different topology. The Abrikosov state is multi-connected superconducting state. Therefore the definition of the phase 
coherence by the correlation function is unsuited for it. The phase 
difference 
$$\varphi _{1} - \varphi _{2} = \int_{l} dl\nabla \varphi = \int_{l} dlp/\hbar = (m/\hbar )\int_{l} dlv + 2e\int_{l} dlA \eqno{(20)}$$

\noindent
is not a gauge-invariant value. But it is not main obstacle for the 
definition of phase coherence. Different gauge-invariant values, $(\varphi_{1} - \varphi _{2} ) - 2e\int_{l} {dlA} $ and others, were used in some papers for the definition of the phase coherence through the correlation function [39,47,82,83]. Different definitions of the phase of the gauge-invariant order parameter in the mixed state of type II 
superconductors are compared in [84,85]. These definitions give different 
result [84,85]. But it is main that all definitions of the phase through the 
correlation function are not valid for the Abrikosov state. Indeed, it is 
obvious that if the phase coherence is defined as any correlation between 
two points then long-rang phase coherence is absent in the Abrikosov state 
with a chaotic distribution of vortices, since any shift of a vortex alter 
in a point not only the phase, which is not a gauge-invariant value, but 
also gauge-invariant values, for example density of superconducting current. 
The chaotic distribution is observed in samples with strong pinning 
disorders in which the non-dissipation current (critical current) has 
highest value. Thus, there is obvious contradiction since the 
non-dissipation current is a consequence of long-range phase coherence. We 
should emphasize once again that existence of singularities (i.e. Abrikosov 
vortices) of the mixed state with long-rang phase coherence is evidence of 
long-rang phase coherence. Therefore it is natural to define phase coherence 
through the relation 
$$\oint_{l} dl\nabla \varphi = n2\pi \eqno{(21)}$$

\noindent
where n is the number of the fluxoids inside a closed path $l$. The long-rang 
phase coherence exists if this relation makes sense and is correct for any 
closed path $l$ which does not cross any non-superconducting region. This 
definition is valid for both the Meissner and Abrikosov states. It is valid 
for the Abrikosov state without depending on vortex distribution, for 
samples both without and with strong pinning disorders. 

It is important that phase coherence appearance in these two opposite limits 
may be qualitatively different. It was shown above that length of phase coherence should change by jump and the transition into the Abrikosov state should be first order in an ideal superconductor in which only two 
characteristic lengths $L$ and $(2\Phi_{0}/\pi H)^{1/2}$ exist across magnetic field. Pinning disorders give an additional characteristic length - distance between pinning centers $d_{p}$. In this case the length of phase coherence can jump from $(2\Phi_{0}/\pi H)^{1/2}$ to $d_{p}$. If $d_{p} \approx (2\Phi_{0}/\pi H)^{1/2}$ then the jump is absent. Strong pinning changes type of this transition. It has numerous experimental corroboration. The sharp transition into the Abrikosov state is observed only in samples with enough weak pinning disorders. In the overwhelming majority of samples this transition is smooth. 

Because of the importance of pinning disorders for type of the transition in the Abrikosov state we will consider separately superconductor with weak and strong pinning disorders. There is also an important qualitative difference between three- and two-dimensional superconductors. 

\bigskip

\noindent
\textit{6.2. Transition into the Abrikosov state in bulk superconductors with weak pinning disorders.}

The narrow transition into the Abrikosov state was found first only in 1981 
since most samples of type II superconductors is not enough homogeneous and 
have strong pinning disorders. The samples have different width of the 
resistive transition because of heterogeneousness and pinning disorders [51]. For example the width of the resistive transition in perpendicular magnetic field of different single-crystal and polycrystalline $V_{3}Ge$ samples used in [45] was from 0.002 to 0.15 $H_{c2}$. The intrinsic 
(only because of thermal fluctuation) width of the specific heat ``jump'' of 
similar superconductor at similar conditions equals some percent $H_{c2}$ 
[4]. Therefore the discrepancy of the width of the transition into the 
Abrikosov state with the width of critical region at $H_{c2}$ could be found 
at measurement of the resistive properties only high quality samples. 

\begin{figure}
\includegraphics{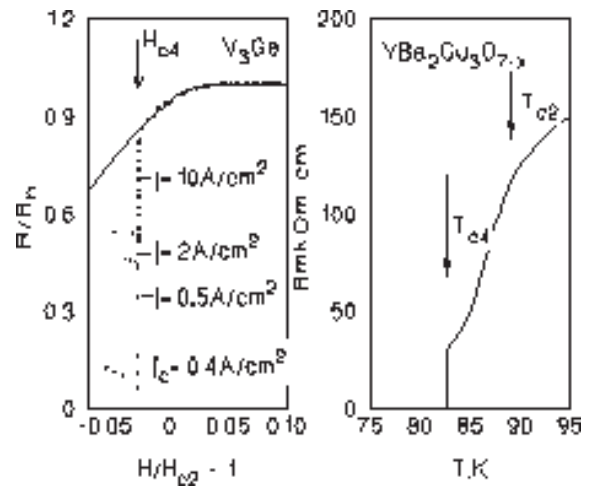}
\caption{\label{fig:epsart} Comparison of the resistive transition of bulk conventional superconductor $V_{3}Ge$ and bulk HTSC $YBa_{2}Cu_{3}O_{7-x}$. The paraconductivity regime above $H_{c4}$ ($T_{c4}$) and the sharp change of the resistive properties at the transition into the Abrikosov state are clear observed in both superconductors. The line on the left figure is the theoretical dependence for paraconductivity in the Hartree approximation.}
\end{figure}

It was difficult to find this discrepancy in conventional superconductors first of all because of relatively narrow fluctuation region. The width of 
the critical region, where fluctuations are strong and its interaction is 
essential, can be estimated by the Ginzburg number. Its value in 
conventional superconductors is enough small. Without magnetic field, for 
example, for $V_{3}Ge$ it equals approximately $Gi \approx  10^{-6}$. This means that width of the critical region at $T_{c}$ is very narrow. It is important that in a magnetic field, at $H_{c2}$, the critical region widens essentially because of the reduction of the 
effective dimensionality. Because of it the Ginzburg number for $H_{c2}$ is $Gi_{H} = Gi^{1/3}(th)^{2/3}$. For example $Gi^{1/3} \approx  10^{-2}$ at $Gi \approx  10^{-6}$. Therefore the dependencies of the specific heat [4,7] and magnetization [54] can be measured in the critical region near $H_{c2}$ in contrast to the one near $T_{c}$. 

One of the characteristic features of HTSC discovered after 1986 is more value of thermal fluctuations. One should note that it is not connected with 
the higher critical temperature. The higher $T_{c}$ value is compensated in the Ginzburg number $Gi = (k_{B}T_{c}/H_{c}^{2}(0)\xi ^{3}(0))^{2}$ by higher value of the thermodynamic critical field $H_{c}(0)$. The relation $T_{c}/H_{c}^{2}(0)$ can be even smaller in HTSC than in conventional 
superconductors. Thermal fluctuations are more appreciable in HTSC first of 
all because of small value of the coherence length $\xi (0)$. Fluctuation effects 
become stronger in layered HTSC, such as BiSrCaCuO, because of lesser 
dimensionality (in a high magnetic field perpendicular to layer it is 
zero-dimensional). In this Section only bulk HTSC, $YBa_{2}Cu_{3}O_{7-x}$ is considered. 

The region of the mixed state without phase coherence below $H_{c2}$ in $YBa_{2}Cu_{3}O_{7-x}$ is much wider than in conventional bulk superconductors. Therefore the difference of the ``kink'' position from $H_{c2}$ arrests one's attention. The resistance decrease because of the paraconductivity before the transition is more appreciable in HTCS than in conventional superconductor: $R(H=H_{c4})  \approx  0.83R_{n}$ in $V_{3}Ge$ and $R(H=H_{c4})  \approx  0.2R_{n}$ in $YBa_{2}Cu_{3}O_{7-x}$, Fig.19. But it is important that there is not qualitative difference between $YBa_{2}Cu_{3}O_{7-x}$ and conventional bulk superconductors. It is obvious that the same transition is observed in both cases since the same qualitative changes of the resistive properties are observed and the difference of the $(H_{c2}-H_{c4})$ values conforms the scaling law of the LLL approximation. 

Unfortunately the shape change of the resistive properties (the ``kink'') observed in $YBa_{2}Cu_{3}O_{7-x}$ is interpreted as vortex lattice melting [10-14]. This interpretation is very popular [58-64,86-99] but it can not be correct [100,101]. In ten years 
before the first observation of the ``kink'' in $YBa_{2}Cu_{3}O_{7-x}$ [10-12] the correct interpretation of the shape feature of the resistive properties observed below $H_{c2}$ was proposed [8]. This interpretation of the resistive phenomenon connected with vortex pinning appearance as the transition into the Abrikosov state is correct in any case. It could coincide with the vortex lattice melting interpretation if the Abrikosov state is vortex lattice with spontaneous crystalline long-rang order. But the Abrikosov state is first of all the mixed state with long-rang phase coherence. 

\begin{figure}
\includegraphics{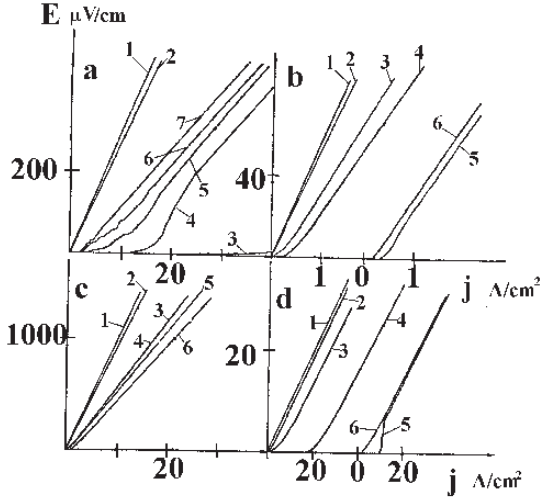}
\caption{\label{fig:epsart} The current voltage curves $E(j)$ of some bulk type II superconductors with weak pinning in different values of perpendicular magnetic field close to $H_{c2}$: a) $V_{3}Ge$, $T = 4.2 \ K$, $H_{c2} = 2.6 \ T$, $\kappa  = 20$, at $H/H_{c2} \gg  1$ (1), $H/H_{c2}$ = 0.987 (2), 0.940 (3), 0.925 (4), 0.918 (5), 0.910 (6), 0.894 (7); b) $V_{3}Ge$, $T = 4.2 \ K$, $H_{c2} = 4.4 \ T$, $\kappa  = 40$, at $H/H_{c2} \gg  1$ (1), $H/H_{c2}$ = 1.000 (2), 0.980 (3), 0.977 (4), 0.970 (5), 0.922 (6); c) $Ti_{84}Mo_{16}$, $T = 2.35 \ K$, $H_{c2} = 4.3 \ T$, $\kappa  = 67$, at $H/H_{c2} \gg  1$ (1), $H/H_{c2}$ = 1.000 (2), 0.977 (3), 0.954 (4), 0.907 (5), 0.861 (6); d) $Nb_{94.3}Ge_{5.7}$, $T = 1.96 \ K$, $H_{c2} = 0.66 \ T$, $\kappa  = 3$, at $H/H_{c2} \gg  1$ (1), $H/H_{c2}$ = 1.000 (2), 0.997 (3), 0.994 (4), 0.988 (5), 0.975 (6) [51]. The curves 5 and 6 for b) and d) are shifted along the j axis. }
\end{figure}

It is important that according to the habitual definition of phase coherence 
used by all theorists the long-rang phase coherence can not exist in the 
Abrikosov state without the crystalline long-rang order of vortex lattice. 
It is main reason preserved delusion about vortex lattice melting among 
theorists. If the long-rang phase coherence can not exist without the 
crystalline long-rang order then there is not a difference between vortex 
lattice melting and the phase transition from the Abrikosov state at which 
the long-rang phase coherence disappears. Therefore it is important to 
emphasize once again that the definition of phase coherence by the 
correlation function is not valid for the multi-connected mixed state. 
According to the correct definition of phase coherence (21), natural for the 
Abrikosov state, the existence of long-rang coherence does not depend on any 
order of vortex distribution. Therefore two phase transition could be 
observed on the way from the Abrikosov state into the normal state if the 
Abrikosov state is vortex lattice with spontaneous crystalline long-rang 
order.

Experimental evidence of the first order phase transition observed at $H_{c4}$ is considered as main beyond argument that vortex 
lattice melting exists since this melting should be first order phase 
transition according to most theories. But as it is shown above phase 
coherence disappearance should be also first order phase transition. 
Therefore if vortices could form spontaneous crystalline order than two 
first order phase transitions should be observed on the way from the 
Abrikosov state into the normal state. But only phase transition is observed 
on this way. It is obvious that it is phase coherence disappearance since 
just at this transition the non-dissipation current can disappear. 

\bigskip

\noindent
\textit{6.3. Vortex creep induced by thermally activated depinning of vortices just below $H_{c4}$.}

The energy dissipation is absent in the Abrikosov state when an applied electric current does not exceed a critical value, $j < j_{c}$. The value of the critical current is determined first of all by a force of vortex pinning $j_{p}$. According to the relation (12) the vortex 
velocity $v_{vor} > 0$ at $f_{L} = jH > f_{p}$, i.e. $j_{c} = f_{p}/H$. The $j_{c}$ value can change with the applied current and real current voltage curves may have intricate form [51]. A typical current-voltage curve has three different 
regime: $j < j_{cs}$ where the voltage equals zero, an intermediate regime where the current-voltage curve is non-linear and the linear regime $E = \rho_{f}(j - j_{cd})$ observed at an enough high current. The static $j_{cs}$ and dynamic $j_{cd}$ critical currents have different values. The static critical current 
is measured in practice by a voltage level $V_{l}$. The value $j_{cs}$ can depend strongly or weakly on the $V_{l}$ value at different conditions Fig.21. 

The dependence can be strong because of vortex creep induced by thermally 
activated depinning of vortices. The HTSC stand out against a background of conventional superconductors by strong vortex pinning. But it is not 
qualitative difference. A noticeable vortex creep was observed near $H_{c4}$ in conventional bulk superconductors Fig.22. It is interesting that in this case the vortex creep is combined with the ``peak effect'' observed in the critical current dependence, Fig.23. Both the creep increase and the ``peak effect'' can be explained by the decrease of the correlated volume of pinning V$_{{\rm p}}$ because of the softening of elastic modulis 
of the vortex lattice near $H_{c2}$. According to the theory of weak collective pinning [102] the average collective pinning force can be estimated by a relation $F_{p} = j_{c}H = (W/V_{p})^{1 / 2}$. Where $W = n_{p}<f_{p,i}^{2}>_{pins}$ is the average pinning force squared, with $n_{p}$ the volume density of pins and $f_{p,i}$ the actual force exerted by the ith pin on the vortex lattice. The sizes of the correlated volume across $R_{p} = 8\pi r_{p}2c_{44}^{1/2}c_{66}^{3/2}/W$ and along $L_{p} = (c_{44}/c_{66})^{1/2}R_{p}$ magnetic field depend on the elastic modulis $c_{44}$ and $c_{66}$. Here $r_{p}$ is the range of the pinning forces. The narrow peak of the $j_{c}(H)$ dependence is observed near the second critical field since $c_{66}  \to 0$ at $H \to H_{c2}$ [103] and the collective pinning result goes over into the direct summation $F_{p} \approx  n_{p}f_{p}$. 

\begin{figure}
\includegraphics{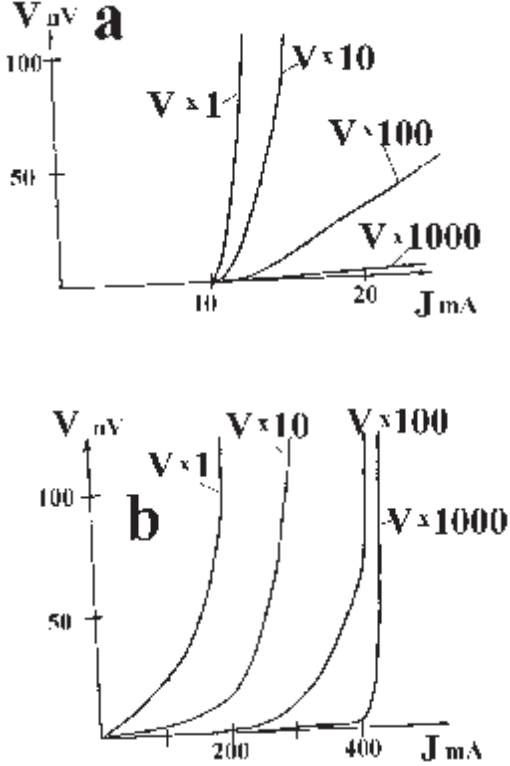}
\caption{\label{fig:epsart} The current voltage curves of $V_{3}Ge$,single-crystal ($T = 4.2 \ K$, $H_{c2} = 2.6 \ T$, $\kappa  = 20$) in perpendicular magnetic field $H/H_{c2} = 0.82$, corresponded the region below the peak of the $j_{c}(H)$  dependence (a) and $H/H_{c2} = 0.95$, corresponded the  top of the peak (b) [51].}
\end{figure}

The peak effect of the true critical current measured by magnetization 
irreversibility, Fig.23, may transform with temperature increase into the one of untrue critical current Fig.21 because of the vortex creep Fig.22. According to the creep theory by Kim and Anderson [104] the electric field $E(T,H,j)$ caused by thermally activated vortex jumps out of pinning centers equals 
$$E(j) = E_{0} \sinh (\frac{j}{j_{0}}) \eqno{(22)}$$
Where $E_{0}= \rho_{p}j_{p} \exp(-U/k_{B}T)$; $j_{0} = j_{p}k_{B}T/U$; $\rho_{p}$ and $j_{p}$ are phenomenological parameters; $U(T,U)$ is activation energy for vortex jumps [59]. The value $U$ of the activation energy for vortex jumps is proportional to an activated (jumping) volume of the vortex lattice which can jump independently. The activated volume, as well as the correlated volume, decreases at $H \to H_{c2}$. Therefore the peak effect observed at a high temperature may be no true, Fig.21. It can be observed on the $j_{cs}$ dependencies measured by a high voltage level $V_{l}$ and can be absent at a lower V$_{l}$. For example, according to the current-voltage curves for a single-crystal $V_{3}Ge$ sample ($T_{c} = 6.1 \ K$) measured at $T = 4.2 \ K$ (see Fig.22) the peak effect is observed at $V_{l} > 10 \ nV/cm$ and is absent at $V_{1} < 10 \ nV/cm$. The peak effect of the true critical current appears in 
this sample at lower temperature. For example, at $T = 2.37 \ K$ the critical current value measured by magnetization irreversibility, Fig.23, increases from $j_{cs} = 20 \ A/cm^{2}$ at $H \approx 0.8 H_{c2}$ up to $j_{cs} = 900 \ A/cm^{2}$ at $H \approx  0.9 H_{c2}$. The low, but finite resistance because of the vortex creep gives an feature of the dynamic magnetic susceptibility in the region of the peak effect (see Fig.7 and [54]). 

\begin{figure}
\includegraphics{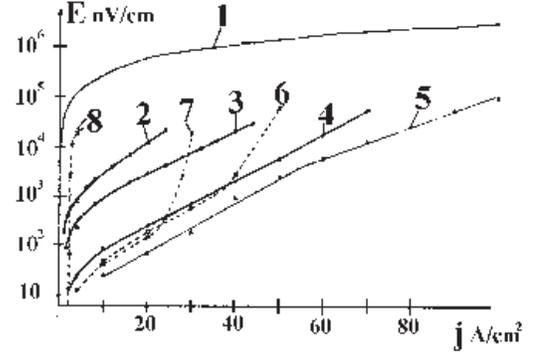}
\caption{\label{fig:epsart} The current voltage curves, $lg(E)$ on $j$, of $V_{3}Ge$ single-crystal at $T = 4.2 \ K$ ($H_{c2} = 2.6 \ T$) in different values of perpendicular magnetic field: $H/H_{c2} \gg 1$ (1) (normal resistance), $H/H_{c2}$ = 0.960 (2), 0.952 (3), 0.946 (4), 0.940 (5), 0.933 (6), 0.925 (7), 0.821 (7). The solid lines (2), (3), (4) are the theoretical dependencies obtained from the Kim-Anderson relation (22) with $E_{0} = 657 \ nV/cm$, $j_{0} = 7 \ A/cm^{2}$ (2), $E_{0} = 350 \ nV/cm$, $j_{0} = 10 \ A/cm^{2}$ (3), $E_{0} = 29 \ nV/cm$, $j_{0} = 9 \ A/cm^{2}$ (4) [51].}
\end{figure}

There is a question: ``Could the true critical current be observed just below $H_{c4}$?'' According to the data presented on Fig.22 and Fig.23 it is absent both at $T = 4.2 \ K$ and $T = 2.37 \ K$. But it does not mean that the true critical current can be observed just below $H_{c4}$ in any case. Since sizes of the correlated volume $V_{p}$ are inversely proportional to the average pinning force squared $W$ we can expect that it is possible in a sample with very weak pinning disorders. For example if the vortex pinning takes place only on sample boundaries than the correlated volume should equal the volume of the sample. It should be emphasized that the absence of the true critical current because of the vortex pinning does not mean the absence of long-rang phase coherence. Therefore rather the transformation of the current-voltage curves from Ohmic $E = \rho j$ to non-Ohmic $E = \rho _{f}(j - j_{cd})$ than the appearance the true non-dissipation current can be experimental evidence of the transition into the Abrikosov state. 

\begin{figure}
\includegraphics{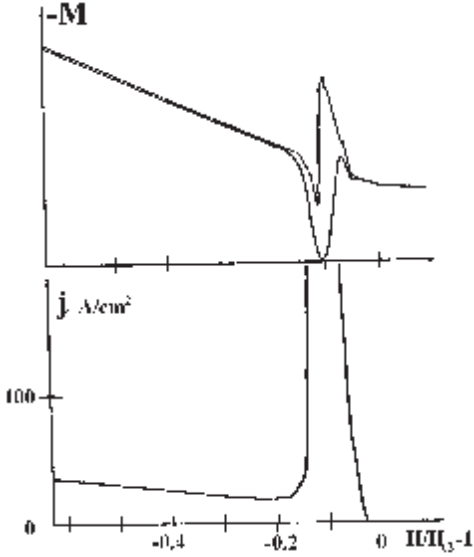}
\caption{\label{fig:epsart} Dependencies of the magnetization $M$ (a) and the critical current density jc measured at $E = 100 \ nV/cm$ (b) on the magnetic field of $V_{3}Ge$ single-crystal at $T = 2.37 \ K$ ($H_{c2} = 4.8 \ T$). The value of the true critical current evaluated from the magnetization irreversibility is $j_{c} = 20 \ A/cm^{2}$ below of the peak and the maximum value at the peak is $j_{c} = 900 \ A/cm^{2}$. The first is close to the one on (b). The high values $j_{c} > 200 A/cm^{2}$ were not measured by the resistive method because of overheating [51]. }
\end{figure}

\bigskip

\noindent
\textit{6.4. Sharp change of the vortex flow resistance at the transition into the Abrikosov state.}

It is important that the slope of the linear part of the current-voltage 
curves, i.e. the vortex flow resistance $\rho _{f}$, decreases sharply at $H_{c4}$ simultaneously with the $j_{cd} > 0$ appearance, Fig.20. According to the mean field approximation the 
transition to the vortex flow regime occurs at $H_{c2}$. and the vortex flow resistance, $\rho _{f}$, near $H_{c2}$ can be described by the relation 
$$\frac{\rho_{f}}{\rho _{n}} = \gamma (1 - \frac{H}{H_{c2}}) \eqno{(23)}$$
 $\rho_{n}$ is the resistance in the normal state. The coefficient $\gamma $ was calculated in many works (see [67]). According to (23) the resistivity of ideal superconductor does not change at the long-rang phase coherence appearance in $H_{c2}$.

The fluctuations change qualitatively the vortex flow resistance $\rho_{f}(H)$ dependence. First of all they displace the transition into the Abrikosov state and consequently the vortex flow regime from $H_{c2}$ to $H_{c4}$. Between $H_{c2}$ and $H_{c4}$ the paraconductivity regime is observed as well as above $H_{c2}$. In additional, according to the fluctuation theory the value the vortex flow resistance near the transition should be lower than the one predicted by the mean field approximation (23) as well as the resistance above the transition is lower than the normal one $\rho_{n}$. Thus, the thermal fluctuations decrease the resistance value both above and below the transition even without taking into account the vortex pinning. This effect should increase near the transition. Therefore, a feature ought be expected at $H_{c4}$, i.e. at the transition from the paraconducting regime to the vortex flow regime. According to the Maki-Takayama theory [36] the decrease of the resistance in the vortex flow regime because of the fluctuations should exceed the one above the transition. Therefore the feature of the vortex flow resistance should be enough noticeable. 

\begin{figure}
\includegraphics{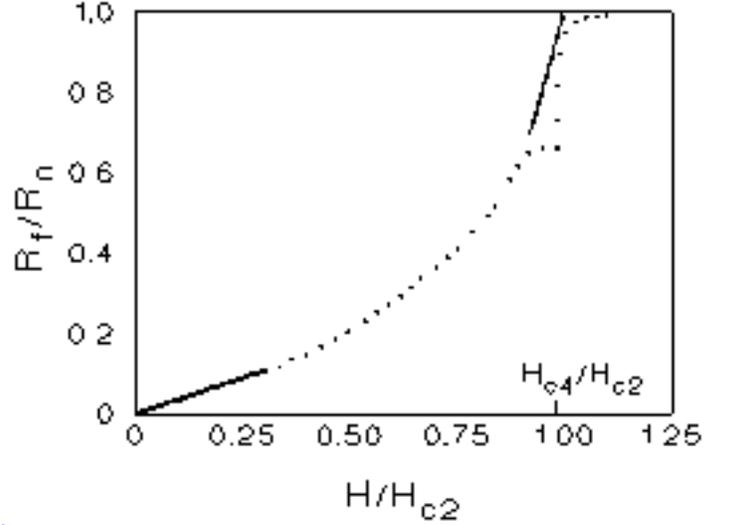}
\caption{\label{fig:epsart} Dependence of the vortex flow resistance on the magnetic field of $Ti_{84}Mo_{16}$ sample at $T = 2.35 \ K$ [51]. The peak effect in the critical current is absent in this sample (see the current voltage curves on Fig.20c). Lines are mean field approximation theoretical dependencies.}
\end{figure}

The detailed investigations made in [51] have shown that the feature is indeed observed in all enough homogeneous bulk samples of conventional superconductors used in this work (see Fig. 5 in [45] and Fig.24, Fig.25. The feature is absent and the $\rho _{f}(H)$ dependence is like to the one predicted by the mean field approximation (23) only in no enough homogeneous sample, Fig.26, or sample with no enough weak pinning disorders. In all high quality samples a sharp change of vortex flow resistance was observed at $H_{c4}$ and the step, Fig.24,25, or even a minimum, Fig. 5 in [45], were observed on the $\rho_{f}(H)$ dependencies below the transition into the Abrikosov state. Such dependence differs qualitatively from the one (23) predicted by the mean-field approximation [67].

Because the step or minimum of the $\rho_{f}(H)$ dependence is observed near $H_{c4}$ where the peak effect of the critical current may be observed the position of these two features coincides in some samples [54,105-108]. Therefore some authors [109] connect the feature of the $\rho_{f}(H)$ dependencies with the $j_{cs}(H)$ peak effect in the critical current. It should be noted that the peak effect 
can not be a cause of the $\rho_{f}(H)$ feature, as anybody 
assumes, since the latter is observed in samples both with [54,105-108] and without Fig.24 the peak effect. The width of these two features can considerably differ in some samples, Fig.25. 

\begin{figure}
\includegraphics{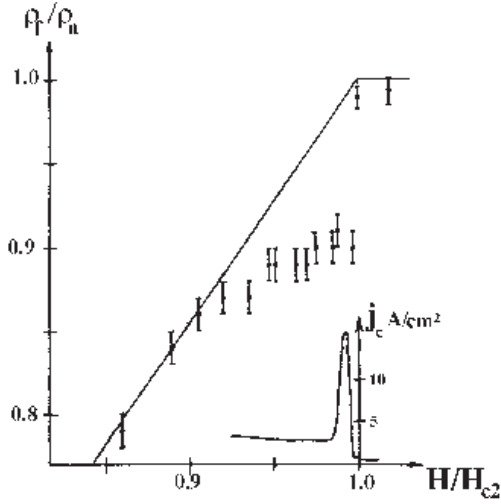}
\caption{\label{fig:epsart}Dependence of the vortex flow resistance on the magnetic field of single-crystal $Nb_{94.3}Ge_{5.7}$ sample at $T = 1.96 \ K$. The line is the mean field approximation theoretical dependence. The peak of the critical current dependence (shown in the bottom) is considerably narrower than the step of the $\rho_{f}(H)$ dependence for this sample [51]. }
\end{figure}

It is important to note at the consideration of HTSC and two-dimensional 
superconductors that the step or minimum of the $\rho_{f}(H)$ dependence is observed below $H_{c4}$ but not below $H_{c2}$. The difference $H_{c2} - H_{c4}$ is no very visible in conventional bulk superconductors whereas in HTSC and in thin films it can be very visible. Therefore one should expect to observe the $\rho_{f}(H)$ feature no at $H_{c2}$ but below the field at which the current-voltage curves become non-Ohmic because of phase coherence appearance. In order to measure correctly the $\rho_{f}(H)$ dependence one should have homogeneous samples with enough weak pinning disorders. The small value of the critical current in most HTSC samples does not mean that pinning disorders are enough weak in these samples since the $j_{cs}$ value is small in HTSC because of strong vortex creep. The quality of HTSC samples does not reach for the present the one of conventional superconductors. Nevertheless the step just below the transition into the Abrikosov state was observed recently [110] on the $\rho_{f}(H)$ dependence obtained from complex impedance 
measurements in an untwinned $YBa_{2}Cu_{3}O_{7-x}$ crystal, Fig.27. This step called unexpected in this paper [110] 
although it can be expected from the results of investigations of 
conventional superconductors [45,51]. The step is manifested in [110] no so distinctly as in high quality samples conventional superconductors [45,51] 
but this result confirms one again that there is not qualitative difference between fluctuation phenomena in $YBa_{2}Cu_{3}O_{7-x}$ and conventional bulk superconductors. One should remember that the $\rho_{f}(H)$ feature below $H_{c4}$ can be observed only in high quality samples and disorders can smear it. 

\begin{figure}
\includegraphics{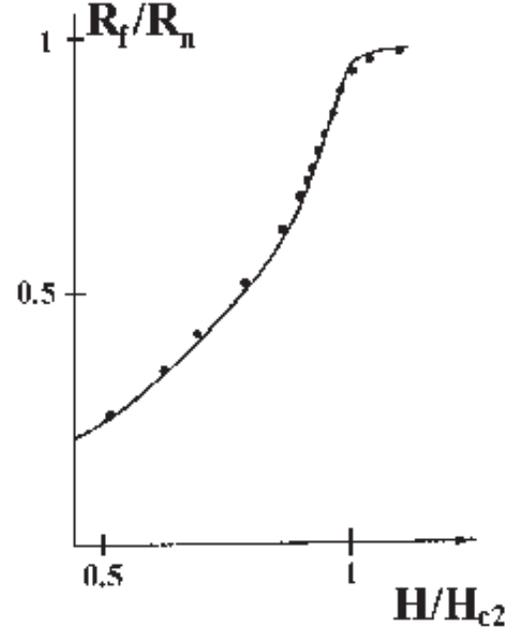}
\caption{\label{fig:epsart} Dependence of the vortex flow resistance on the magnetic field of an inhomogeneous $Ti_{84}Mo_{16}$ sample at $T = 2.35 \ K$. The width of the resistive transition in perpendicular magnetic field $\Delta H \approx  0.08H_{c2}$ of this sample exceeds considerably the one $\Delta H \approx  0.01H_{c2}$ of the homogeneous sample, Fig.24. The line is the $\rho_{f}(H)$ dependence for homogeneous sample, Fig.24, averaged by different $H_{c2}$ values in the region $0.08H_{c2}$ [51].}
\end{figure}

The observation of the $\rho_{f}(H)$ feature just below $H_{c4}$ of bulk superconductors corroborates the prediction of the Maki-Takayama theory [36]. It is very important for the interpretation of the nature of the Abrikosov state since according to the results [36] the mean field approximation is not valid for infinite homogeneous 
superconductor, i.e. just for that ideal case for which the Abrikosov solution [1] was obtained. According to [36] the fluctuation correction $\Delta \sigma _{H<H_{c2}}$ to the vortex flow conductivity is $1.4 \ln(L/\xi )$ times the paraconductivity $\Delta \sigma _{H>H_{c2}}$ above $H_{c2}$ 
$$\Delta \sigma_{H<H_{c2}} = 1.4 \ln(L/\xi )\Delta \sigma_{H>H_{c2}} = $$
$$1.4 \ln(L/\xi )C_{\bot}(t)\sigma_{0}\vert H/H_{c2}- 1\vert^{-1/2} \eqno{(24)}$$

\noindent
where $\sigma _{0} = [\pi e^{2}/2^{3/2}\hbar \xi (0)]$, $1/\sigma _{0} \approx  3.7 \ 10^{4} \xi (0) \ \Omega  cm$; $C_{\bot} (t) $ is a dependence (calculated in [52]) increasing with $t = T/T_{c}$, from $C_{\bot} = 0$ at $T = 0$ to $C_{\bot} \approx  0.1$ at t = 0.5, to $C_{\bot} \approx  1$ at t = 0.9 and so on at $T  \to T_{c}$. It is important that the fluctuation correction (24) calculated in the linear approximation depends on superconductor size L across magnetic field. The dependence (24) does not mean that the resistance in the Abrikosov state $\rho _{H<H_{c2}} = \rho _{f}/(1 + \Delta \sigma_{H<H_{c2}}\rho_{f})$ equals zero without vortex pinning in infinite superconductor $1/L = 0$ or at $H = H_{c2}$. Here $\rho_{f}$ is the vortex flow resistance calculated in the mean field approximation. It means only that the linear approximation, as well as the Abrikosov solution, are not valid in the region where fluctuations are strong. The $L/\xi $ value is very large in a real case but $\ln(L/\xi )$ is not so large. For example, at size of the $V_{3}Ge$ samples used in [45] $L \approx  1 \ mm$ and the coherence length $\xi  \approx  10^{-5}\ mm$, $L/\xi \approx  100000$ and $\ln(L/\xi ) \approx  11.5$. At parameters of the $V_{3}Ge$ samples: $\sigma _{n} = 1/\rho _{f} \approx 36000 \ (\Omega cm)^{-1}$; $\sigma _{0} \approx  400 \ (\Omega cm)^{-1}$; the relation (24) gives for $t = T/T_{s}$ = 4.2/6.1 = 0.69 when $C_{\bot} (t)  \approx  0.27$ 
$$\Delta \sigma_{H<H_{c2}}/\sigma_{n}  \approx  0.05(1 - H/H_{c2})^{-1/2} \eqno{(25)}$$
According to this relation and (23) the $\rho_{H < H_{c2}}(H) = \rho_{f}/(1 + \Delta \sigma_{H < H_{c2}}\rho_{f})$ dependence should have a maximum at $1 -- H/H_{c2} \approx  0.04$. The position of the maximum is closed in order of value to the width of the step of the $\rho _{f}(H)$ dependence observed in the $V_{3}Ge$ samples at $T = 4.2 \ K$ [45]. It is impossible to describe more exactly the $\rho _{f}(H)$ feature since the result [36] obtained in the linear approximation is not 
valid in this region where fluctuations are strong and fluctuation interaction should take into account. 

The thermal fluctuations decreases the vortex flow resistance in whole range below $H_{c4}$. But the $\Delta \sigma_{H<H_{c2}}/\sigma_{n}$ value can be visible 
only when fluctuations become enough strong and the linear approximation, i.e. the Maki-Takayama result [36] becomes not valid. It takes place in the 
critical region before the transition from the Abrikosov state. Therefore the $\rho_{f}(H)$ feature (the step) should be observed (and is observed [51,110]) below $H_{c4}$ but not below $H_{c2}$. Above $H_{c4}$ the Maki-Takayama theory [36], as well as 
the mean field approximation of the vortex flow resistance theory [67], can 
not be valid since any vortex theory can not by valid in the mixed state 
without vortices. The both theories is not valid not only just below $H_{c2}$ but even just below $H_{c4}$. It was not 
take into account by Maki and Thompson which supposed that the Maki-Takayama 
result [36] contradicts to the smooth monotonic decrease of resistance 
observed in HTSC below $H_{c2}$ and tried to correct it making 
an important mistake [37]. The authors of [111] write that the infrared 
divergences in certain quantities in [36] the perturbation theory was 
abandoned. They have shown that these divergences cancel in physical quantities [112,113]. Indeed, the result [37] does not mean that the 
fluctuation corrections to any physical quantities are infinite in an 
infinite superconductor or at $H= H_{c2}$. This result means only that the linear approximation of the fluctuation theory as well as the 
mean field approximation are not valid for the infinite superconductor and 
near and above $H_{c4}$. 

\begin{figure}
\includegraphics{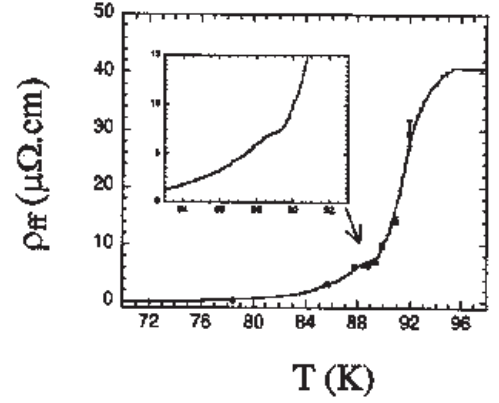}
\caption{\label{fig:epsart} The vortex flow resistance of an untwinned $YBa_{2}Cu_{3}O_{7-x}$ crystal extracted in [110] from the complex impedance measurements. }
\end{figure}

Because of absence of the vortex flow theory valid near $H_{c4}$ it is not quite clear how the vortex flow resistance should change at $H_{c4}$ in an ideal case. The experiment shows that the slop of the current voltage curves can change by jump in high quality (homogeneous 
and weak pinning) bulk samples, [45,51], Fig.20,24,25, and more smooth [51] 
in samples with not enough high quality. Only smooth the $\rho_{f}(H)$ dependence at $H_{c4}$ was observed for the present in 
$YBa_{2}Cu_{3}O_{7-x}$ [110], Fig.27 and it is not quite clear one could observe more sharp change of the $\rho_{f}(H)$ value at $H_{c4}$ of $YBa_{2}Cu_{3}O_{7-x}$ in which fluctuation value is much higher than in conventional bulk superconductor. The step of the $\rho_{f}(H)$ dependence of both $YBa_{2}Cu_{3}O_{7-x}$ and conventional superconductor is observed just below $H_{c4}$ at which the 
current-voltage curves become non-Ohmic. Although the values are very different in these superconductors. The $\rho_{f}(H)$ feature is observed at an interval of fluctuation value in which thermal fluctuations 
are enough strong in order to decrease appreciably the vortex flow 
resistance but are not enough strong in order to destroy long-rang phase 
coherence and induce the transition from the Abrikosov state. This interval 
corresponds to different $H/H_{c2}$ value in superconductors with different value of fluctuations. One may say that the $\rho_{f}(H)$ feature should be observed in a region where the linear approximation of the fluctuation theory becomes not valid but the Abrikosov state still remains, i.e. just below $H_{c4}$.

According to the Maki-Takayama theory [36] the thermal fluctuations in the two-dimensional Abrikosov state are much stronger than in the 
three-dimensional one. The value of fluctuation correction in the Abrikosov state calculated in the linear approximation is $1.4 \ln(L/\xi )$ times the one in the 
normal state for bulk superconductor with $L_{z} \gg \xi $ and is $(L/\xi )^{2}$ times the one for thin film with $L_{z} < \xi $. Here $L_{z}$ is superconductor size along magnetic field. The difference between the values is enormous in a real case. For example at a superconductor size across magnetic field $L  \approx  1 \ mm$ and the coherence length $\xi \approx 10^{-5} \ mm$, $\ln(L/\xi ) \approx 11.5$ and $\ln(L/\xi ) \approx  10^{10}$. The difference remains enormous even for a thin film micro-structure with $L approx 1 \ \mu m$. This means that the linear approximation of the fluctuation theory [36] as 
well as the Abrikosov solution are not valid for the mixed state of thin films with $L_{z} < \xi $. 

Thus, the boundary of the linear approximation [36] applicability shifts 
from magnetic field value correspond some per cent below $H_{c2}$ for bulk ($L_{z} \gg \xi $) conventional superconductors to zero magnetic field for thin film with $L_{z} < \xi $. One may expect that in an intermediate case between $L_{z} \gg \xi $ and $L_{z} < \xi $ this boundary and the $\rho _{f}(H)$ feature, connected with it, can shift with a change of film thickness $L_{z}$. The position of the $\rho _{f}(H)$ feature in films of conventional 
superconductors depends indeed from film thickness. For example, the investigations [106] of amorphous $Nb_{1-x}Ge_{x}$ films, a superconductor with a small coherence length $\xi (0) \approx 7 \ nm$, have shown that the $\rho _{f}(H)$ feature is observed at $H \approx 0.9H_{c2}$ in a film with thickness $L_{z} = 2350 \ nm$, at $H \approx 0.82H_{c2}$ with $L_{z} = 565 \ nm$, at $H \approx 0.75H_{c2}$  with $L_{z} = 205 \ nm$ and at $H \approx 0.62H_{c2}$ with $L_{z} = 92.5 \ nm$. It is important that the $\rho _{f}(H)$ feature is observed below $H_{c4}$ at which the current voltage curves become non-Ohmic. 

\bigskip

\noindent
\textit{6.5. Absence of the transition into the Abrikosov state in two-dimensional superconductors with weak disorder down to very low magnetic field.} 

The shift of the position of the $\rho _{f}(H)$ feature [106] and 
the resistive transition [114] to lower $H/H_{c2}$values at the decrease of film thickness means that the difference between $H_{c2}$ and the position of the transition into the Abrikosov state $H_{c4}$ increases in thin films. This experimental result collaborate 
the difference of fluctuation effects in the Abrikosov state of three- and 
two-dimensional superconductors predicted by the Maki-Takayama theory [36]. 
According to this theory the Abrikosov state can be absent or observed only 
at very low magnetic field $H \ll H_{c4}$ in two-dimensional superconductors i.e. in film with $L_{z} < \xi$. 

It is important to note that the $\rho_{f}(H)$ feature and the dependence of its position from film thickness can be observed only in samples with enough weak pinning. The transition into the Abrikosov state in samples with strong pinning disorders differs qualitative from the one with weak pinning (see [115] and below). The importance of pinning disorders for stabilization of the Abrikosov state is obvious already from the consideration of fluctuation influence in the Eilenberger's conception (see Section 2.6). It is obvious that disorders breaking displacement symmetry can stabilize the Abrikosov state as well as a finite size of superconductor $L$.

\begin{figure}
\includegraphics{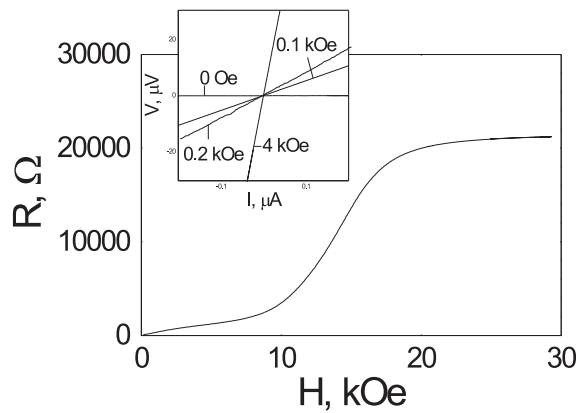}
\caption{\label{fig:epsart} The magnetic dependence of the resistance of the amorphous $NbO_{x}$ film structure with thickness $d = 20 \ nm$, width $w = 10 \ mm$ and length $L = 2.25 \ mm$. In the inset the current-voltage curves in perpendicular magnetic fields, H = 0, 100 Oe, 200 Oe, 4 kOe, are shown. $T = 1.6 \ K$, $H_{c2} = 16.8 \ kOe$ [46].}
\end{figure}

There is possibility to compare experimental results obtained for thin films 
of amorphous $a-Nb_{3}Ge$ [109] and of amorphous $NbO_{x}$ [46] with closed parameters, $L_{z} = 18 \ nm$, $\xi (0) = 7.4 \ nm$, $T_{c} = 2.91 \ K$ K in the first case and $L_{z} = 20 \ nm$, $\xi (0) = 7.8 \ nm$, $T_{c} = 2.37 \ K$ in the second case, but with slightly different 
amount of pinning disorders. In the $a-Nb_{3}Ge$ film at T = 1.55 K the current voltage curves become non-Ohmic at $H  \approx  0.45H_{c2}$ ($H  \approx  1.2 \ T$) and below this field the $\rho _{f}(H)$ feature is observed, Fig.7 in [109]. In the $NbO_{x}$ film at 
T = 1.6 K the current voltage curves remain Ohmic down to $H = 0.006H_{c2}$ ($H  \approx  0.01 \ T$) and the $\rho _{f}(H)$ feature is not observed, Fig.28. On base of the measurements made in [46] the static critical current is absent at $H  \approx  0.01 \ T$ to within $10000 \ A/m^{2}$ and at $H \approx  0.4H_{c2}$ to within $200 \ A/m^{2}$. The critical current the $NbO_{x}$ film appears at $H < 0.003H_{c2}$ and increases in an narrow interval of the magnetic field from $j_{c} < 10^{4}\ A/m^{2}$ at H = 0.005 T to $j_{c} > 10^{10}\ A/m^{2}$ at H = 0 (at $T = 1.6 \ K$, $H_{c2} = 2.2 \ T$). Such immense change of the critical current in a low magnetic field $H \ll H_{c2}$ is evidence of extremely weak pinning. The weak $j_{c}(H)$ dependence of low critical current of some HTSC samples observed in low magnetic field is 
evidence of weak links but not weak pinning.

The appearance of the critical current in the $NbO_{x}$ films at very 
low magnetic field $H < 0.003H_{c2}$ may be interpreted as the 
transition into the Abrikosov state. But an incomprehensible anomaly is 
observed in this region [116]. It is impossible to understand for the 
present why the peaks of the resistance dependence $R(H)$ of the $NbO_{x}$ strips can be observed in low magnetic field $H < 0.003H_{c2}$ and why the resistance can considerably exceed the vortex flow 
resistance. But it is important that the width of the anomaly $H_{w}$ 
and the peak position $H_{p}$ are inversely proportional to width of 
the strip $w_{st}$: $H_{w} = 0.0006 \ T$ and $H_{p} = 0.0002 \ T$ for the $NbO_{x}$ strip with $w_{st} = 50 \ \mu m$ and $H_{w} = 0.004 \ T$ and the position of main maximum $H_{p} = 0.001 \ T$ for the $NbO_{x}$ strip with $w_{st} = 10 \ \mu m$ [116]. It gives a possibility to assume that the anomaly observed in thin films with extremely weak pinning disorders may be connected with a 
vortex pinning on strip boundaries. This may mean that vortex pinning is 
only on superconductor boundaries in the amorphous $NbO_{x}$ films used in [46]. 

Comparison of the critical current values shows that the amorphous $a-Nb_{3}Ge$ film used in [109] has no so weak pinning disorders as the amorphous $NbO_{x}$ films used in [46]. For example, the dynamic critical current $j_{cd}$ is absent to within $200 \ A/m^{2}$ at $H \approx  0.4H_{c2}$ in the $NbO_{x}$ films whereas the $a-Nb_{3}Ge$ film (see Fig.7 [109]) in the current voltage curves are appreciably non-Ohmic at this magnetic field in the used interval of the measuring current density $1500 \div 4.5 \ 10^{5}\ A/m^{2}$. The observation of the $\rho _{f}(H)$ feature at $H \approx  0.4H_{c2}$ in the amorphous $a-Nb_{3}Ge$ film with weak pinning (Fig.7 [109]) and the absence of any $\rho_{f}(H)$ feature down to $H = 0.006H_{c2}$ in the amorphous $NbO_{x}$ films with weaker pinning are evidence of strong influence of pinning disorders on 
the position of the transition into the Abrikosov state of thin films which can be considered as two-dimensional superconductor. 

Authors of [109] state that the transition into the Abrikosov state is not observed in [46] since the critical current in the amorphous $NbO_{x}$ films is extremely small in comparison to the measuring current and therefore the appearance of vortex pinning can not be visible. Indeed, the transition from the Ohmic, $E = \rho j$, to non-Ohmic, $E = \rho _{f}(j -- j_{cd})$, regime can be not visible if $j_{cd} \ll  j$. But the $\rho_{f}(H)$ feature observed near the transition into the Abrikosov state in all enough high quality sample should be observed at any relation $j_{cd}/j$. The $\rho_{f}(H)$ feature is observed at the transition into the Abrikosov state in the $a-Nb_{3}Ge$ film (see Fig.7 [109]) with an intermediate amount of pinning disorders. Therefore it should be manifested more distinctly at this transition in the $NbO_{x}$ films with weaker pinning since all investigations show that the $\rho_{f}(H)$ feature is manifested more distinctly in high quality samples with weaker pinning. The absence of this feature the amorphous $NbO_{x}$ films, Fig.28, is evidence of the absence of the transition into the Abrikosov state down to $H = 0.006H_{c2}$ [46].

\begin{figure}
\includegraphics{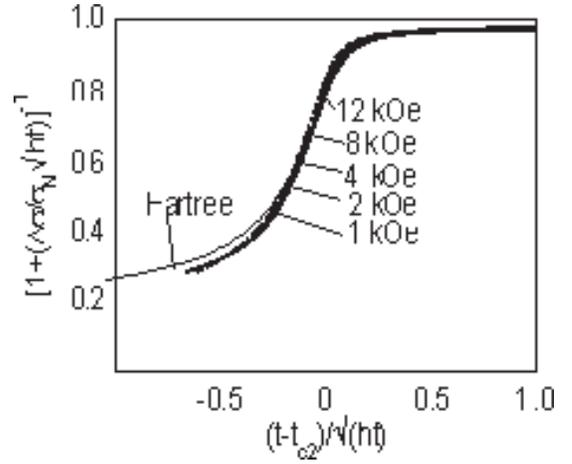}
\caption{\label{fig:epsart} Dependencies of the excess conductivity on temperature $\Delta \sigma (T)$ of the amorphous $NbO_{x}$ film in the co-ordinates following from the scaling law of the LLL approximation in different magnetic fields [46]. The line denotes the theoretical dependence of paraconductivity obtained in the Hartree approximation of fluctuation theory.}
\end{figure}

It is shown in the Sections 5.3, 5.4 that the dependence of the values connected only with average density of superconducting pairs should obey the scaling law in the region where the LLL approximation is valid. The resistive properties depends first of all on existence or absent of phase coherence. But one may expect that the paraconductivity may also obey the scaling law not only above $H_{c2}$ but also below $H_{c2}$, until the mixed state without phase coherence remains. It was shown in [46] that the temperature dependencies of the excess conductivity $\Delta \sigma (T)$ of the amorphous $NbO_{x}$ films measured at different field $H = 0.1 - 1.2 \ T$ obey the scaling law and are close to the theoretical dependence for paraconductivity calculated in the Hartree 
approximation down to temperature considerably lower $T_{c2}$, Fig.29. The existence of phase coherence should cause a deviation of the $\Delta \sigma (T)$ dependencies from the scaling law. This deviation can be evidence of phase coherence appearance. The absence of the deviation on Fig.29 confirms one again that that phase coherence is absent both above and below $H_{c2}$ of the amorphous $NbO_{x}$ films used in [46]. It is important to note that the scaling law is observed below $H_{c2}$ only in superconductors with enough weak pinning. The deviation from the scaling law can be observed already above $H_{c2}$ in films with strong pinning [115], Fig.35. The deviation from the scaling law observed in [109] may be also connected with phase coherence appearance induced by pinning disorders. 

\bigskip

\noindent
\textit{6.6. Experimental confirmation of phase coherence absence below $H_{c2}$ of thin films with weak pinning disorders.}

The absence of phase coherence below $H_{c2}$ in the amorphous $NbO_{x}$ films is confirmed by a investigation of a non-local 
resistance [117]. The non-local resistance can be observed near the critical temperature when size of superconducting drops becomes close to sizes of a microstructure. The potential difference measured on the branch of the strip with a current of the microstructure shown on Fig.30 should depend on the distance from the strip because of the gradient of the current density leaking in the branch. For the amorphous $NbO_{x}$ microstructure with the branch length equals $8 \ \mu m$ and the width of the branch and strip equals $4 \ \mu m$ the potential difference between contacts 3-3, 4-4 and 2-2 equal in the normal state, at temperature much higher $T_{c}$: $V_{33,n} = I \times 185 \ \Omega$, $V_{44,n} = I \times 111 \ \Omega$ and $V_{22,n} = I \times 60 \ \Omega$. The temperature dependencies of the relation $v_{ii}(T) = V_{ii}(T)/V_{ii,n}$ (see Fig.30) show that the gradient of the current density leaking in the branch decreases near $T_{c}$. The reduction of the gradient takes place because no current gradient should be inside superconducting drops according to (8). The integral of superconducting current $j_{s}$ along a closed path inside a drop $\oint_{l} dlj_{s} = 0 $ since $\oint_{l} dl\nabla \varphi = 0$ and $\Phi  = 0$ without external magnetic field and at weak screening, i.e. when the magnetic field induced by superconducting current $j_{s}$ is neglected low. The gradient of the current density decreases since size of superconducting drops increases near $T_{c}$. 

\begin{figure}
\includegraphics{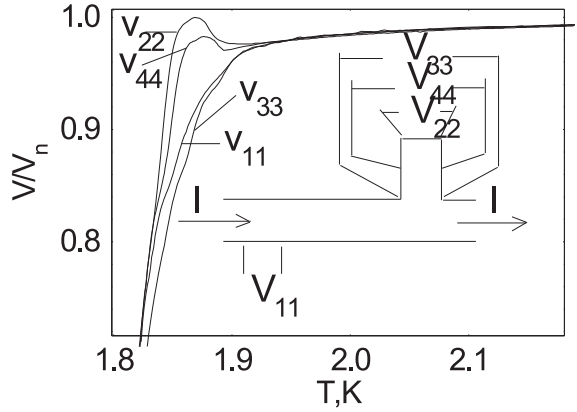}
\caption{\label{fig:epsart} The temperature dependencies of voltage relation, $V/V_{n}$, on different potential contacts in zero magnetic field. $I = 2 \ \mu A$. The $NbO_{x}$ film (d = 20 nm) structure and the location of the potential contacts are shown schematically in the inset. The width of the strip and branch equals  4 mm [117].}
\end{figure}

The screening in a superconducting structure is weak when its size is smaller than the penetration length of the magnetic field. The perpendicular penetration length of a thin film with thickness d equals $\lambda _{2d} = \lambda _{L}^{2}/d$. The value $\lambda_{2d} = 7.6 \ \mu m$ (for the amorphous $NbO_{x}$ films with the London penetration length $\lambda _{L}(T=0) = 390 \ nm$ and $d = 20 \ nm$) is close to the branch sizes even at T = 0. The penetration length increases near $T_{c}$ in many time. Therefore the gradient of the current density should be absent in any region where phase coherence exists, independently with or without the Abrikosov vortices. 

The suppression of the non-local resistance observed in very low magnetic 
field means that magnetic field destroys phase coherence. It decreases size 
of superconducting drops above $T_{c}$ down to ($\Phi _{0}/H)^{1/2}$. The decreasing of the non-local 
resistance, observed near the maximum of the $v_{44}(T)$ 
dependence, becomes to be appreciable at H = 20 Oe = 0.002 T, Fig.31. This 
means that superconducting drops larger than ($\Phi _{0}/20 \ Oe)^{1/2} = 1 \ \mu m$ contribute to the reduction of 
the current density gradient just above the sharp decrease of the resistance. It is enough natural since the sharp decrease of the resistance should be observed when size of superconducting drops mounts a size of the structure. 

\begin{figure}
\includegraphics{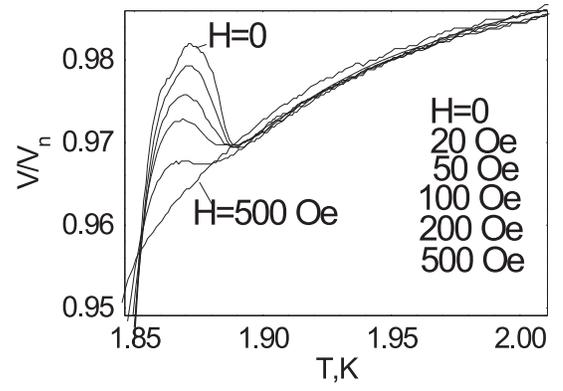}
\caption{\label{fig:epsart} The temperature dependencies of voltage relation on the potential contacts 4-4, $V_{44}/V_{44n}$, (see the inset on Fig.29) at different values of perpendicular magnetic field [117]. }
\end{figure}

It is important to note that the magnetic field first of all destroys phase coherence. The paraconductivity value does not change appreciably up to H = 200 Oe whereas the non-local resistance is suppressed strongly in this field, Fig.31. The small change observed in the paraconductivity is consistent with small change in the mean-field critical temperature, $T_{c2} - T_{c} = H/(dH_{c2}/dT) = -H/(2.2 \ T/K)$: at H = 20 Oe = 0.002 T, $T_{c2} - T_{c} =  -0.0009 \ K$; at H = 200 Oe = 0.02 T, $T_{c2} - T_{c} =  -0.009 \ K$. 

The temperature dependencies of the reduced voltage $v_{ii}(T) = V_{ii}(T)/V_{ii,n}$ measured on the branch $v_{44}(T)$ differs from the one $v_{11}(T)$ measured on the control contacts up to H = 500 Oe = 0.05 T, Fig.32. The difference is visible near $T_{c}$ since the coherence length has maximum value in this region. The difference between $v_{44}(T)$ and $v_{11}(T)$ becomes not visible at H = 2000 Oe = 0.2 T. The non-local resistance is not observed both above and much below $T_{c2}$, Fig.32. 

\begin{figure}
\includegraphics{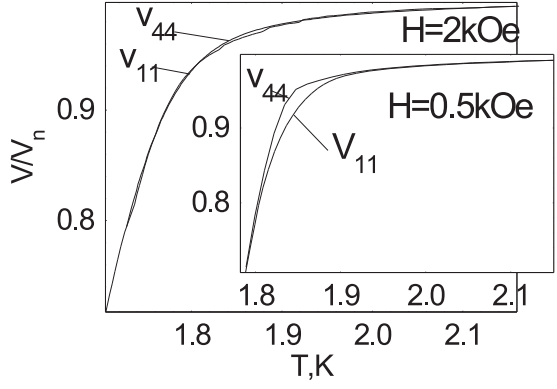}
\caption{\label{fig:epsart} The temperature dependencies of voltage relation on the potential contacts 4-4, $V_{44}/V_{44n}$, and 1-1, $V_{11}/V_{11n}$, (see the inset on Fig.30) in perpendicular magnetic field $H = 0.5 \ kOe$ and $H = 2 \ kOe$. $I = 2 \ \mu A$ [117].}
\end{figure}

\noindent
This means that phase coherence is absent in the amorphous $NbO_{x}$ film placed in a magnetic field both above and below $T_{c2}$, although its resistance is appreciably lower at $T < T_{c}$ than $T > T_{c}$ Fig.32.

\bigskip

\noindent
\textbf{7. Influence of Pinning Disorders on the Nature and Position of the Transition into the Abrikosov State.} 

The sharp, first order phase transition into the Abrikosov state is observed 
only in few high quality bulk samples [8,11-14] as well as the absence of 
this transition down to very low magnetic field is observed only in few high quality thin films [46]. The behaviour of overwhelming majority of real 
samples both bulk and thin film differs qualitatively from these almost ideal cases because of pinning disorders. This difference of the behaviour of most real superconductors from an ideal case is one of the main difficult 
of investigation of the Abrikosov state and the reason of many errors. 
Pinning disorders are not only the cause of the non-dissipation current, as 
one assumed during long time, they change the nature of the transition into 
the Abrikosov state and can stabilize this state. 

\bigskip

\noindent
\textit{7.1. Pinning disorders smooth out the sharp transition observed only in high quality samples.} 

The possibility of an influence of disorders on the nature of the transition 
into the Abrikosov state is obvious from the consideration of the ideal case 
in the Section 6.1. The appearance of long-range phase coherence should 
occur through first order phase transition since only two characteristic 
lengths $L$ and $(2\Phi _{0}/\pi H)^{1/2}$ exist across magnetic field in the ideal superconductor. Therefore the sharp 
change of the resistive properties is observed in few bulk samples with weak 
disorder. But this transition may qualitatively change in a superconductor 
with pinning disorders in which distance $d_{p}$ between pinning centers is an additional characteristic lengths. The sharp transition should remain when $d_{p} \gg (2\Phi _{0}/\pi H)^{1/2}$. Just this takes place in real bulk samples with weak pinning [8,11-14]. But the phase coherence appearance can become a smooth transition 
when the distance $d_{p}$ between pinning centers is comparable in 
order of value with $(2\Phi _{0}/\pi H)^{1/2}$. Therefore no sharp change is observed and the resistive transition is smooth in samples with strong disorder. The transition into the Abrikosov state becomes smooth with increasing of disorder amount and differs 
qualitative from the ideal case considered by Abrikosov. 

\begin{figure}
\includegraphics{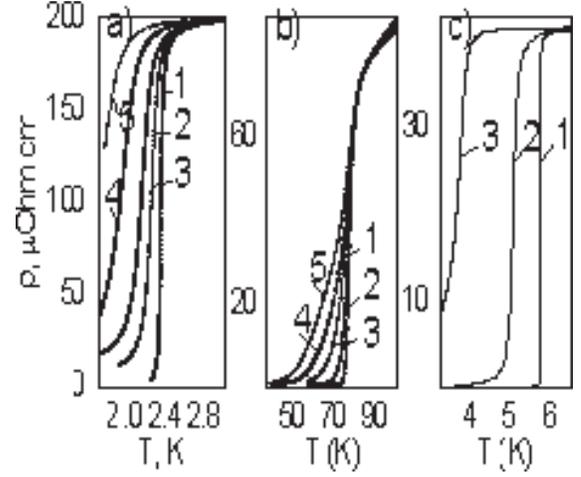}
\caption{\label{fig:epsart} The resistive transition in perpendicular magnetic field of a) the amorphous $NbO_{x}$ film (d = 20 nm) at H = 0 (1), H = 0.2 T (2), H = 0.4 T (3), H = 0.8 T (4), H = 1.2 T (5); b) the $Bi_{2}Sr_{2}CaCu_{2}O_{8+x}$ in-plane (along ab) at H = 0 (1), H = 0.044 T (2), H = 0.2 T (3), H = 0.5 T (4), H = 1 T (5); and c) the $NbO_{x}$ film with fine grain structure (d = 20 nm) at H = 0 (1), H = 0.4 T (2), H = 1.2 T (3) [115].}
\end{figure}

The resistive transition of real type II superconductors may be enough diverse, Fig.33, because of different pinning disorders and other reason and 
can not be described by only dependence. There is a problem with detection of phase coherence appearance. 
The resistive dependencies may not differ qualitatively for the cases with and without phase coherence. For example, the resistive transition for the amorphous $NbO_{x}$ film and the $NbO_{x}$ film with fine grain structure presented on Fig.33 are 
qualitatively similar but in the first case of with weak pinning disorders 
phase coherence is absent down to low resistance value whereas in the second 
case of thin film with strong pinning it appears already near $T_{c2}$. 

The non-Ohmic current voltage curves can be observed only in the mixed state with phase coherence. Therefore the non-Omic regime can be used as the 
evidence of phase coherence. The current voltage curves can become non-Ohmic in different samples at different reduced values $H/H_{c2}$ of magnetic field. For example, it takes place at $H/H_{c2} \approx  0.002$ (at $T/T_{c} = 0.77$) in the amorphous $NbO_{x}$ film, Fig.34. In the $NbO_{x}$ film with the same thickness d = 20 nm but with fine grain structure it takes place at $H/H_{c2} \approx  0.7$ (at $T/T_{c} = 0.74$), Fig.34. It is well known that the current voltage curves of layered HTSC remain Ohmic down to very low magnetic field. But it is important to note that the $H_{c4}/H_{c2}$ value in conventional thin film can be lower than in layered HTSC. 

\begin{figure}
\includegraphics{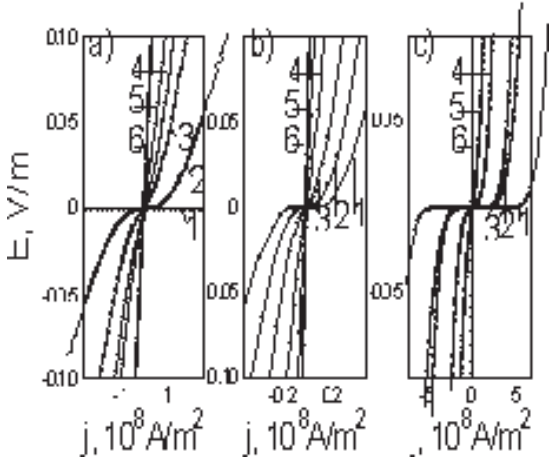}
\caption{\label{fig:epsart} Current-voltage characteristics of a) the amorphous $NbO_{x}$ film (d = 20 nm) at T = 1.75 K ($T/T_{c} = 0.77$, $H_{c2} =1.14 \ T$) and H = 0 (1), H = 0.001 T (2), H = 0.002 T (3), H = 0.004 T (4), H = 0.006T (5), H = 0.01 T (6); b) the $Bi_{2}Sr_{2}CaCu_{2}O_{8+x}$ in-plane (along ab) at T = 60 K ($T/T_{c} = 0.75$, $H_{c2} = 20 \ T$) and  H = 0 (1), H = 0.02 T (2), H = 0.05 T (3), H = 0.1 T (4), H = 0.2 T (5), H = 0.4 T (6); c) the $NbO_{x}$ film with small grain structure (with d = 20 nm) at T = 4.2 K ($T/T_{c} = 0.74$, $H_{c2} = 0.9 \ T$) and H = 0.05 (1), H = 0.1 T (2), H = 0.2 T (3), H = 0.4 T (4), H = 0.5 T (5), H = 0.7 T (6) [115].}
\end{figure}

\noindent
The current voltage curves of a $Bi_{2}Sr_{2}CaCu_{2}O_{8+x}$ sample shown on Fig.34 become non-Ohmic at a higher $H/H_{c2}  \approx  0.004$ value (at $T/T_{c} = 0.75$) than it is observed the amorphous $NbO_{x}$ film. One should emphasize that the observation of the Ohmic current voltage curves does not vouch for the absence of phase coherence. The current voltage curves can be Ohmic because of very strong vortex creep. The Kim-Anderson expression (22) gives a linear dependence $E(j)$ at $j/j_{0} \ll  1$. The thermally activated linear vortex flow resistance is observed first of all in HTSC [59]. 

In superconductors with strong pinning disorders the length of phase coherence $\xi _{p.c}$ changes no by jump from $(2\Phi _{0}/\pi H)^{1/2}$ to L as in the 
ideal case but continuously in any region of temperature or magnetic field 
values. The current voltage curves can be Ohmic in this transitional. In 
order to detect where the length of phase coherence $\xi _{p.c}$ becomes to differ from $(2\Phi _{0}/\pi H)^{1/2}$ the scaling law of the LLL approximation can be used. As it was shown in [46] (see also the Section 6.5) the temperature dependencies of excess conductivity for different magnetic field can be described by an universal law. But it is possible only without phase 
coherence i.e. at $\xi _{p.c} =(2\Phi _{0}/\pi H)^{1/2}$ since a change of the length of phase coherence should strongly influence on the conductivity. According the 
experimental data presented on Fig.35 the temperature dependencies of excess 
conductivity of the $NbO_{x}$ film with fine grain structure deviate 
from the universal dependence describing the $\Delta \sigma (T,H)$ of the amorphous $NbO_{x}$ film already above $H_{c2}$. 

\begin{figure}
\includegraphics{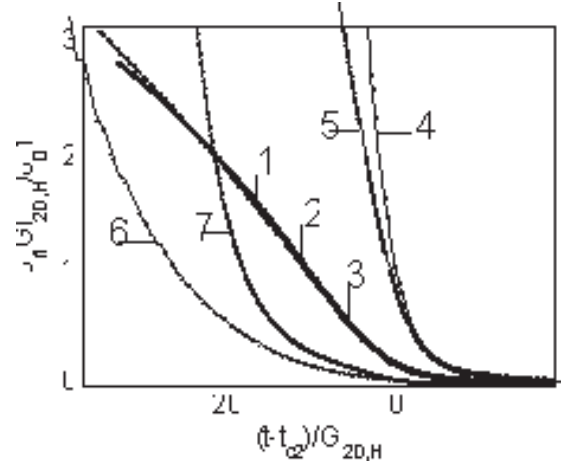}
\caption{\label{fig:epsart}Dependencies of the excess conductivity on temperature $\Delta \sigma (T)$ in the co-ordinates following from the scaling law of the LLL approximation for: the amorphous $NbO_{x}$ film ($T_{c} = 2.37 \ K$; $Gi_{2D} \approx  0.0005$) at H = 0.1 T (1), H = 0.4 T (2), H = 1.2 T (3); the $NbO_{x}$ film with small grain structure ($T_{c} = 5.7 \ K$; $Gi_{2D} \approx   0.0001$) at H = 0.4 T (h = 0.11) (4), H = 1.2 T (h = 0.33) (5) and the $Bi_{2}Sr_{2}CaCu_{2}O_{8+x}$  ($T_{c} = 79 \ K$; $Gi_{2D} \approx   0.02$) at H = 0.1 T (h = 0.0012) (6), H = 1 T (h = 0.012) (7) [115].}
\end{figure}

\noindent
One should remember that the scaling law can be valid only in the LLL approximation region. In particular it is not valid at $h = H/H_{c2}(0) < Gi$. A LLL crossover field $H_{LLL} = H_{c2}(0)Gi$ of conventional superconductors is much lower the values of magnetic field $H = 0.1 \div  10 \ T$ used for investigation of type II superconductors. But the LLL crossover field of layered HTSC is very high, $H_{LLL} = 10 \ T$ for $Bi_{2}Sr_{2}CaCu_{2}O_{8+x}$. Therefore the discrepancy of the $\Delta \sigma (T,H)$ dependencies for $Bi_{2}Sr_{2}CaCu_{2}O_{8+x}$ with the scaling law, Fig.35, means only that the LLL approximation is not valid for H = 0.1 T and H = 1 T for this HTSC. The weak shift in magnetic field of the top of the 
resistive transition observed in layered HTSC, Fig.33, can be explained from the invalidity of the LLL approximation at $H < H_{LLL}$.

\bigskip

\noindent
\textit{7.2. Abrikosov's model and Mendelssohn's model.} 

It is useful now to recall that before as Abrikosov had obtained his famous 
solution [1] many properties of type II superconductors were explained 
enough well by the Mendelssohn's model [18]. These two models have many 
traits in common. First of all the both models consider the mixed state of 
type II superconductors as multi-connected superconducting state. But 
Mendelssohn assumes that the multi-connected superconducting state appears 
because superconducting alloys are inhomogeneous whereas Abrikosov 
considered an ideal homogeneous superconductor. The main difference between 
these models is that the Abrikosov vortex destroy superconductivity near 
itself whereas in the Mendelssohn's model it occupies a nonsuperconducting 
region. Real superconductor is not ideal homogeneous as well as it is not 
the Mendelssohn's "sponge". Therefore the mixed state with long phase 
coherence of most real type-II superconductors may be considered as 
intermediate case between the Mendelssohn's and Abrikosov's models. 

\begin{figure}
\includegraphics{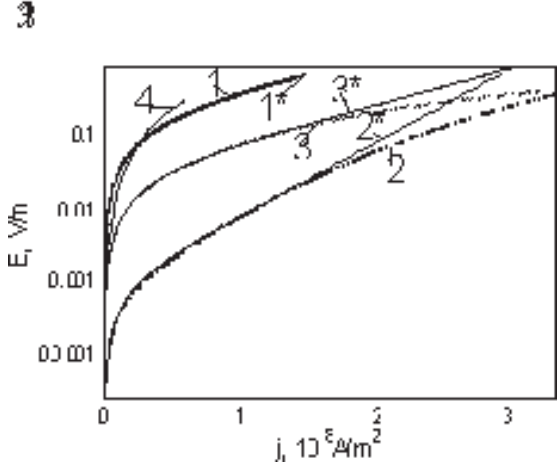}
\caption{\label{fig:epsart} Current-voltage characteristics of the $NbO_{x}$ film with small grain structure (with d = 20 nm) at T = 4.2 K: H = 0.6 T (1) and H = 0.4 T (2), the amorphous $NbO_{x}$ film at T = 1.75 K and H = 0.004 T (3) and the $Bi_{2}Sr_{2}CaCu_{2}O_{8+x}$ at T = 60 K and H = 0.05 T (4). The lines 1*, 2* and 3* denote the theoretical dependencies obtained from the Kim-Anderson relation (22) with $E_{0} = 0.27 \ V/m$, $j_{0} = 9 \ 10^{7} \ A/m^{2}$ (1*), $E_{0} = 0.0015 \ V/m$, $j_{0} = 4.3 \ 10^{7} \ A/m^{2}$ (2*), $E_{0} = 0.05 \ V/m$, $j_{0} = 8.6 \ 10^{7} \ A/m^{2}$ (4) [115].}
\end{figure}

The Mendelssohn's model may be more useful for the description of 
superconductors with strong pinning disorders since superconducting "sponge" 
is the limit case of strong disorder. First of all the Mendelssohn's model 
can explain why the transition into the Abrikosov state is continuous in 
superconductors with strong pinning. The phase coherence appearance in the 
Mendelssohn's "sponge" should be a continuous transition since the length of 
the phase coherence in one-dimensional superconductor does not change by 
jump but increases gradually with the temperature decreasing below $T_{c}$ [53]. 

Thin film with strong pinning disorders in the Abrikosov state may be considered as a sponge of one-dimensional superconductors with a variable width $w(T,H)$ across the magnetic field which equals approximately the distance between normal cores of the vortices: $w(T,H)  \approx a(\Phi_{0}/H)^{1/2} - 2\xi (T)$ [118]. Where a is a number of order 1, in the triangular lattice $a = 2^{1/2}/3^{1/4}$. At a low magnetic field $H \ll  H_{c2}$, $w(T,H)  \approx  2\xi (T)[(H_{c2}/H)^{1/2} - 1]$. This model can be valid when most vortices are disposed on the pinning centers, where superconductivity is suppressed or 
absent. The length of phase coherence changes smoothly in a wide region 
since this model is like to a one-dimensional superconductor. It depends not 
only on temperature but also on the magnetic field value since the width $w(T,H)$ depends both on $T$ and $H$. The deviation of the $\Delta \sigma (T,H)$ dependencies of the $NbO_{x}$ film with fine grain structure from the scaling law already above $H_{c2}$, Fig.35, can be explained qualitatively by the exceeding of the critical field of one-dimensional 
superconductor$H_{c,1D} = 3^{1/2}\Phi _{0}/\pi \xi w$ [49] of the $H_{c2} = \Phi_{0}/2\pi \xi ^{2}$. 

The length of the coherence of the sponge of one-dimensional superconductors 
increases smoothly with the T and H decrease up to $\xi _{p.c.} \gg  L$ but remains finite at non-zero temperature. At low T and H value a crossover to the vortex creep regime takes place. This crossover can be interpreted as a consequence of the increasing of the phase 
coherence length up to sample size. The vortex creep can take place in the 
Mendelssohn's model as well as it takes place in the Abrikosov state. According to the Mendelssohn's model of thin film [118] the parameters $E_{0}$ and $j_{0}$ in the Kim-Anderson relation (22) equal $E_{0} = Hl_{v}\omega _{0}\exp(\xi dwf_{GL}/k_{B}T)$ and $j_{0} = 8\pi ^{2}k_{B}T/\xi d\Phi _{0}$. Here $l_{v}$ is the distance between the vortices; $\omega _{0}$ is an attempt frequency; $f_{GL}$ is the density of the GL free energy, $d$ is the film thickness. Thus, the vortex creep observed in thin film with strong pinning disorders, Fig.36, can be described by both Abrikosov and Mendelssohn's model. The dependence of the $E_{0}$ value on $H/H_{c2}$ observed in the $NbO_{x}$ film with fine grain structure can be explained by the change of the width $w(T,H)$ with $H/H_{c2}$ [115].

\bigskip

\noindent
\textit{7.3. Pinning disorders stabilize the Abrikosov state.}

The comparison of the results presented on Fig.34 and Fig.35 shows that the 
appearance of phase coherence in thin films, which can be considered as 
two-dimensional superconductors, depends strongly on the amount of pinning 
disorders. The current voltage curves become non-Ohmic at  $H/H_{c2}  \approx 0.002$ (at $T/T_{s}  \approx 0.77$) in the amorphous $NbO_{x}$ film, at $H/H_{c2}  \approx  0.45$ 
(at $T/T_{c}\approx  0.53$) in the amorphous $a-Nb_{3}Ge$ film [109] and at $H/H_{c2}  \approx  0.8$ ($T/T_{s}  \approx 0.74$) in the $NbO_{x}$ film with fine grain structure, Fig.34. One may think that the transition from the Ohmic to non-Ohmic regime in the amorphous $a-Nb_{3}Ge$ film [109] is caused by the phase coherence appearance since the $\rho _{f}(H)$ feature is observed 
below $H/H_{c2}  \approx  0.45$. But in the $NbO_{x}$ film with fine grain structure the $\rho _{f}(H)$ feature is not observed and the slop of the Ohmic current voltage curve at $H/H_{c2}  \approx 0.8$ differs strongly from the resistance of the 
amorphous $NbO_{x}$ film at this reduced magnetic field. Just below $H/H_{c2} \approx 0.8$ the $E(j)$ dependencies of the $NbO_{x}$ film with fine grain structure is described by the 
Kim-Anderson relation for the vortex creep (see the curve (1) for $H/H_{c2} \approx  0.7$ on Fig.36). Therefore the transition from the non-Ohmic to Ohmic regime in this $NbO_{x}$ film should be interpreted as a consequence of the vortex creep increase (because the $j/j_{0}$ value becomes much smaller 1) but no phase coherence disappearance. The comparison of the $\Delta \sigma (T,H)$ dependencies for the amorphous $NbO_{x}$ film and the $NbO_{x}$ film with fine grain structure, Fig.35, shows that the phase coherence in the second case appears already above $H_{c2}$. 

\begin{figure}
\includegraphics{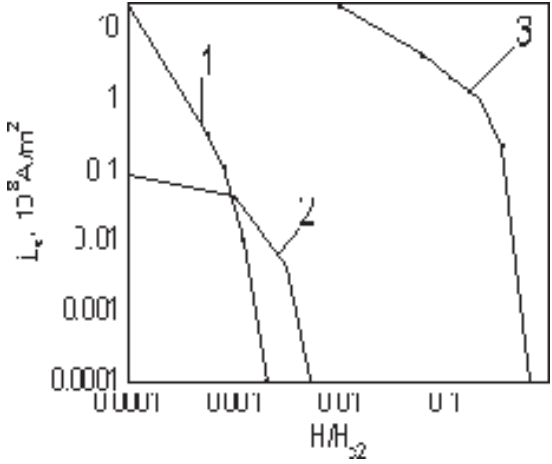}
\caption{\label{fig:epsart} The static critical current jcs dependencies on the reduced magnetic field $H/H{c2}$ of the amorphous $NbO_{x}$ film at T = 1.75 K (1); the $Bi_{2}Sr_{2}CaCu_{2}O_{8+x}$ at T = 60 K (2); the $NbO_{x}$ film with small grain structure at T = 4.2 K (3). The $j_{cs}$ values were determined by the voltage level $0.0001 \ V/m$. $H/H_{c2} = 10^{-4}$ is corresponded to H = 0. $j_{cs} = 10^{4} \ A/m^{2}$ is correspond to $j_{cs} <  10^{4} \ A/m^{2}$ [115].}
\end{figure}

Since the parameters (thickness d, coherence length $\xi (0)$ and others) are closed in the amorphous $NbO_{x}$ film, the amorphous $a-Nb_{3}Ge$ film [109] and the $NbO_{x}$ film with fine grain structure only reason of the qualitative difference observed on Fig.34-36 can be difference amount of pinning disorders. The static critical current $j_{cs} = 10^{6} \ A/m^{2}$ is observed at $H = 0.0012H_{c2}$ (at $T/T_{c} = 0.76$) in the amorphous $NbO_{x}$ film and at $H = 0.45H_{c2}$ (at $T/T_{c} = 0.74$)  in the $NbO_{x}$ film with fine grain structure, Fig.37. The dynamic critical current of the amorphous $a-Nb_{3}Ge$ film estimated from the data presented on Fig.7 of [109], equals in order of value $j_{cd} = 10^{5} \ A/m^{2}$ at $H = 0.8 \ T \approx 0.3H_{c2}$ (at $T/T_{c} = 0.53$). For the comparison $j_{cd} < 10^{5} \ A/m^{2}$ at $H = 0.006 \ T  \approx 0.0052H_{c2}$, $j_{cd} \approx 3 \ 10^{7} \ A/m^{2}$ at $H = 0.004 \ T  \approx 0.0035H_{c2}$, $j_{cd} \approx 5 \ 10^{7} \ A/m^{2}$ at $H = 0.002 \ T  \approx 0.0017H_{c2}$ 
and $j_{cd} \approx 1.5 \ 10^{8} \ A/m^{2}$ at $H = 0.001 \ T  \approx 0.00088H_{c2}$ in the amorphous $NbO_{x}$ film at $T/T_{c} = 0.77$, Fig.34.

This comparison of the critical current values shows that the amorphous $NbO_{x}$ film has very weak strength of pinning disorders, the amorphous $a-Nb_{3}Ge$ film has an intermediate strength and the strength of pinning disorders of the $NbO_{x}$ film with fine grain structure is very strong. We may conclude that the position of the transition into the Abrikosov state is not universal. It depends on the amount of disorder. An intermediate amount of pinning disorders displaces 
the transition into the Abrikosov state of two-dimensional superconductor from the region of very low magnetic field $H \ll  H_{c2}$. This transition in thin film with strong pinning should be described rather by the Mendelssohn's model than the Abrikosov one. The dependence of the position of the transition into the Abrikosov state of two-dimensional superconductors on the amount of disorder is explained by results of the work [44].

\bigskip

\noindent
\textbf{8. What can be in the ideal case considered by Abrikosov?}

Thus, the survey of the experimental and theoretical results shows that the taking into account of thermal fluctuations changes qualitatively interpretation of the Abrikosov state and reveals the determinant of pinning disorders in the mixed state of type II superconductors. Because of fluctuations no the Abrikosov state but the mixed state without phase coherence is observed almost in whole region below $H_{c2}$ of two-dimensional superconductors with enough weak pinning disorders and the transition into the Abrikosov state observed in bulk superconductors with enough weak pinning disorders is first but not second order phase transition. Pinning disorders change the nature of the this transition and displace the appearance of phase coherence in thin films from the region of very low magnetic field. 

The revealed difference between bulk superconductor and thin film and also the feature of the vortex flow resistance observed below $H_{c4}$corroborate the results of the Maki-Takayama theory [36]. According to this theory the mean field approximation is not valid just in the ideal case of homogeneous infinite superconductor considered by Abrikosov. The Abrikosov state is observed in bulk superconductors because the fluctuation correction to the solution [2], proportional to $\ln(L/\xi )$ for three-dimensional superconductor, is small in samples with real size $L$ at the magnetic field enough lower $H_{c2}$. This correction, proportional to $(L/\xi )^{2}$ for two-dimensional superconductor, is much larger for thin films. Therefore the Abrikosov state, i.e. the mixed state with long rang phase coherence, is not observed almost in whole region below $H_{c2}$ of thin films with very weak pinning disorders. Nevertheless the mixed state with phase coherence can be observed in thin films because of pinning disorders. 

There is important question: ``Can the Abrikosov state appear spontaneously in homogeneous symmetric infinite space or it is observed only because of finite size of real superconductors and pinning disorders?'' The absence of phase coherence in thin films with weak pinning down to very low magnetic field $H \ll H_{c2}$ and the observation of the transition into the Abrikosov state of bulk samples at $H_{c4} < H_{c2}$, slightly above the region where the mean filed approximation becomes invalid, allow to assume that in the ideal case could be no the Abrikosov state but the mixed state without phase coherence. In this case the Abrikosov state is not spontaneous but is induced by disorders and finiteness of the space in which it is observed. It has only long range order -- phase coherence since the crystalline long rang order of vortex lattice can not be in the space with disorders. 

It would be useful for the solution of the problem about the state in the ideal infinite superconductor to measure the position of the transition into the Abrikosov state, $H_{c4}$, in sample with different sizes L. It is very difficult to observed the expected dependence of $H_{c4}$ on L in bulk superconductor. The samples should differ in size $L$ in many time in order the expected difference of $H_{c4}$ could be visible. For example ln(L/$\xi $) $ \approx $ 11.5 for a $YBa_{2}Cu_{3}O_{7-x}$ sample with $L = 0.1 \ mm$, $\ln(L/\xi ) \approx  13.8$ at $L = 1 \ mm$, $\ln(L/\xi ) \approx 16.1$ at $L = 1 \ cm$ and $\ln(L/\xi )  \approx 20.7$ at $L = 1 \ m$. The main difficulty is pinning disorders existing in any real sample. In this case the position should be determined by the distance between pinning disorders but not sample size. 

It is even more difficult problem to describe theoretically the transition of bulk type II superconductor from the mixed state without phase coherence into the Abrikosov state. This problem is simplified in two-dimensional case when the last gradient term disappears from the GL free-energy (18) and this relation becomes enough simple 

$$\frac{F_{GL}}{k_{B} T} = V(\epsilon <|\Psi|^{2}> + \frac{\beta_{a}}{2}<|\Psi|^{2}>^{2}) \eqno{(26)}$$
Since the $F_{GL}$ value depends only on the average density of superconductig pairs $n_{s} = <\vert \Psi \vert ^{2}>$ and the Abrikosov parameter $\beta _{a}$ the sum in the relation (18) for the veritable free-energy taken over the function substrate spanned by the LLL can be can be substituted on two integrals [44] 
$$F = -k_{B}T\ln(\int_{0}^{\infty }dn_{s} \int_{\beta_{a}}^{\infty }d\beta_{a}N(n_{s},\beta_{a})\exp(-\frac{F_{GL}}{k_{B}T}) \eqno{(27)} $$ 

\noindent
where $N(n_{s}, \beta _{a}) dn_{s}d\beta _{a}$ is a subspace volume with the given values of $n_{s}$ and $\beta_{a}$. $\beta _{A} \approx  1.159596$, corresponded to the triangular vortex lattice [2], is the minimum value of the Abrikosov parameter $\beta _{a}$ in the subspace spanned by the LLL. 

The number of the eigenfunctions equals to the degeneracy number of the LLL, which equal in the dimensionless unit system to the area $S$ of the two-dimensional superconductor, $V =Sd$. A given $n_{s}$ values lie on a 2S-dimensional sphere with radius $n_{s}^{1/2}$. The area of this sphere is equals to $2\pi ^{S}S!( n_{s}^{1/2})^{2S-1}$. It is important the $\beta _{a}$ value does not depend on the $n_{s}$ value. Therefore $N(n_{s},\beta _{a})$ is proportional to $n_{s}^{S- 0.5}n(\beta_{a})$, where $n(\beta _{a})$ is a fraction of the 2S-dimensional sphere with the given $\beta _{a}$ value. 

The Abrikosov parameter has minimum value $\beta _{a} = \beta_{A}$ in $S$ points of 2S-dimensional sphere. Each point corresponds to the triangular vortex lattice. The transition into the Abrikosov state takes place when the system comes to a stop near a point and can not come, because of thermal fluctuation, in other $S-1$ points with $\beta _{a} = \beta_{A}$. One function from the Eilenberger function basis [35] gives main contribution to the thermodynamic average in the Abrikosov state. Whereas in the mixed state without phase coherence, when the system passes free between all $S$ points with $\beta _{a} = \beta_{A}$ because of thermal fluctuations, all Eilenberger functions equal in rights and give the same contribution to the thermodynamic average. In a region of the 2S-dimensional sphere near $\beta _{a} = \beta_{A}$ the $\beta _{a} - \beta_{A}$ value is small and increases with moving away this point. There is a maximum $(\beta _{a} - \beta_{A})_{c}$ value on the way between two points with $\beta _{a} = \beta_{A}$ on the 2S-dimensional sphere. The problem of the transition into the Abrikosov state can be reduced to the search of this critical value $(\beta _{a} - \beta_{A})_{c}$. The transition takes place at $<\beta _{a}> - \beta_{A} = (\beta _{a} - \beta_{A})_{c}$ [44]. It is enough easy to calculate the thermodynamic average of the Abrikosov parameter below $H_{c2}$ at $-\epsilon \gg 1$: $<\beta _{a}> - \beta_{A}  \approx 2n_{s}^{-2} \approx 2(\beta _{A}/\epsilon )^{2}$ [44]. But it is much more difficult to find $(\beta _{a} - \beta_{A})_{c}$. It was only shown for the present that a value $\beta _{a} - \beta_{A} > (\beta _{a} - \beta_{A})_{c}$, at which the contribution from other Eilenberger functions is not small, is inversely proportional to superconductor size $S^{1/2}$ [44]. 

\bigskip

\noindent
\textbf{9. Why can the erroneous concept of vortex lattice melting appear and become popular?} 

The transition into the Abrikosov state is not described theoretically for the present first of all because of the wide popularity of the conception of vortex lattice melting. It may be that this conception is a greatest and most long-term delusion of modern physics. Hundreds of papers, including many papers with titles "Evidence of the vortex lattice melting ....", have been published in the last years in which the vortex lattice melting is considered. Many papers concerned to the vortex lattice melting are publishing up to now [119-174]. Therefore it is important to explain why this delusion appeared and remains very popular during a long time. 

\bigskip

\noindent
\textit{9.1. The conception of vortex lattice melting could appear because an incorrect interpretation of direct observation of the Abrikosov state.}

According to most naive theories of vortex lattice melting the Abrikosov state is a flux line lattice (FLL) [60,61] like an atom lattice, or a lattice of long molecules [60]. The wide spread opinion that the Abrikosov vortex is a magnetic flux line appeared first of all because of direct observations [22,23] of the Abrikosov state. The magnetic flux structure can be observed only at enough low magnetic field where the persistent current around vortices induces an enough visible gradient of magnetic induction. In this case the Abrikosov vortex seems as a separate magnetic flux induced by the persistent current. Therefore most scientists interpreted the Abrikosov vortex first of all as magnetic flux line. But it was enough to bring up a question: ``If the vortices are magnetic flux, why does not the magnetic flux disappear when the vortices disappear at the transition from the Abrikosov to normal state?'' in order to understand that this interpretation is not quite correct. Also the question: ``If the resistance in the Abrikosov state is induced by flux flow why does the flux stop to flow in the normal state?'' casts doubt on the flux flow resistance. 

Nevertheless the conception about magnetic flux line and flux flow resistance exercise complete sway during a long time. It could be possible 
since there are not manifest contradictions between the Abrikosov vortex as flux line and as singularity in the mixed state with long phase coherence if we consider only this state. It may be that the only difference is that the singularity and the magnetic flux should flow in opposite directions in order to induce a voltage with the same direction. But this question about 
the nature of the Abrikosov vortex becomes a matter of principle when the 
transition from the Abrikosov state to the normal state is considered. When the vortex is considered as a flux line then most people neglect that the 
Abrikosov state is first of all the mixed state with long range phase 
coherence. The flux line is indeed like a long molecule and if it is seemed 
that these long molecules form a lattice then why this lattice could not melt.

\bigskip

\noindent
\textit{9.2. Could vortices be in the vortex liquid?}

If the vortex lattice melts than a state above this transition is vortex liquid. There is an important question: ``Could vortices be in the vortex liquid?'' The attitude to this question is different. Most experimentalist think that vortices should be in the vortex liquid because it is impossible to imagine a vortex liquid without vortices. Some theorists, first of all makers of theories of vortex lattice melting, do not love to answer on this question. But many theorists understand that vortices can not be in the vortex liquid. They say: ``It is only name. We do not should debate about a name. It is not important''. But I can not agree that a name is not important. Just the history of the appearance of the conception of vortex lattice melting manifests that an incorrect name led to unfortunate results. 

Some experts understood that the magnetic flux can not flow in the mixed state of type II superconductors. Nevertheless people read in all textbook [21,47-49] that the resistance in the Abrikosov state is induced by flux flow. Therefore the resistance dependencies observed in the ``vortex liquid'' regime are compared in many experimental work [106] with the theoretical dependence for the flux flow resistance. Such comparison has some grounds if the resistance is induced by flux flow. Magnetic flux is both in the Abrikosov state and in the ``vortex liquid'' regime. Therefore if the flux flows in the Abrikosov state then why can it not flow in the ``vortex liquid''?. But in this case it should flow also in the normal state above $H_{c2}$ since no qualitative change takes place at $H_{c2}$. Thus, the incorrect name carries to the point of absurdity.

It is not important for the theories considering the energy dissipation at the vortex flow [67,175] what flows, magnetic flux or singularity. Therefore the authors of [106] and other papers could compare the experimental dependencies above the ``vortex lattice melting'' with the vortex flow resistance if the vortices are in the vortex liquid. Therefore the vortex liquid is not only name for the experimentalists which compare the experimental data with the vortex flow theory. 

\bigskip

\noindent
\textit{9.3. Lack of information about absence of any phase transition at $H_{c2}$.} 

Some experimentalists can compare the experimental results with a vortex theory since they assume as before that a phase transition occurs at $H_{c2}$. The knowledge about the mixed state of most scientists was formed by the influence of results of the mean field approximation. They thought that the second order phase transition into the Abrikosov state occurs at $H_{c2}$ and that this state is first of all vortex lattice. Only few experts understood in the seventies that thermal fluctuations changes qualitatively this notion. The situation became even worse after the discovery of HTSC in 1986 when many new people turned their 
attention to the fluctuation phenomena in the mixed state. Most scientists are no enough well-informed up to now that the results obtained in the seventies demonstrate clearly the absence of any phase transition at $H_{c2}$. The lack of this information is one of the most reason of popularity of the vortex lattice melting conception. Indeed, if a phase transition can be at $H_{c2}$ then one can assume that the vortices disappear at the second critical field and the first order phase transition observed below $H_{c2}$ is the vortex lattice melting. In this case the vortex liquid is indeed liquid of vortices as some experimentalists assume. 

\bigskip

\noindent
\textit{9.4. Definition of phase coherence.}

But many theorists know that no phase transition can be at the second 
critical field and therefore understand that vortices can not be in the 
vortex liquid. They understand that vortices are singularities in the mixed 
state with long rang phase coherence and therefore the vortex disappearance 
should be a phase transition. Nevertheless these theorists do not repudiate 
the conception of vortex lattice melting. The reason of such attitude is the 
definition of phase coherence used by all theorists. If the phase coherence 
is defined through a correlation function, i.e. as coherence between two 
points, then long-rang phase coherence can exist in the Abrikosov state only 
if the crystalline long rang order of vortex lattice exists. In this case 
the vortex lattice melting is simultaneously the disappearance of long rang 
phase coherence. Then the conception of vortex lattice melting should not be rejected. But if the long rang phase coherence can not be without the crystalline long rang order then it can not be in any real Abrikosov state since as A.I. Larkin showed in 1970 [34] pinning disorders destroy the 
crystalline long rang order of vortex lattice. Moreover even short rang order phase coherence can not be in samples with strong pinning disorders in which the Abrikosov vortices are distributed chaotically. But in this case the vortices can not be also in the Abrikosov state since singularity in the mixed state with phase coherence can not exist without phase coherence. 

Thus, the definition of phase coherence used for the present by theorists carries to the point of absurdity. It was shown by some theorists [176,177] that the phase coherence can be short-ranged even in the vortex solid state. This result means that not only the vortex liquid without vortices but even the vortex lattice without vortices are possible if the definition of phase coherence through a correlation function is used. Therefore the definition of phase coherence through the relation (21) should be used in the Abrikosov state. According to this definition the existence of phase coherence does not depend on vortex distribution, but only existence of vortices is evidence of phase coherence. The independence of long rang order on the existence of the crystalline long rang order of vortex lattice means that 
the vortex lattice melting and phase coherence disappearance are different 
transition. The melting can be without the disappearance of vortices. And 
therefore the conception of the vortex lattice melting should be rejected on 
the base of the experimental data. If only first order phase transition is 
observed on the way from the Abrikosov state to the normal state then it is 
phase coherence disappearance since the existence of the long range phase 
coherence in the Abrikosov state can not be called in question in contrast 
to the vortex lattice. 

The vortex lattice melting is not only name. All theories of the vortex lattice melting describe just vortex lattice melting. That is they describe 
the transition which does not exist whereas the first order phase transition observed in bulk high quality samples at $H_{c4} < H_{c2}$ is not described for the present. In order to describe this transition one should find at which parameters, temperature, magnetic field, superconductor size and others, the phase coherence defined by the relation (21) can appear. It is very difficult theoretical problem. According to this 
definition the long phase coherence exists in the Abrikosov state with any 
amount of pinning disorders. But the sharp transition can be expected to 
find theoretically only in superconductor with weak pinning. It is followed 
from the experimental and simple arguments considered above. The definition 
of phase coherence proposed here diminishes the difference between the Abrikosov and Mendellson's model. It is equally valid for the both models. 
The Mendellson's model but not vortex glass should be used for the 
description of the Abrikosov state with strong pinning disorders. The 
conception of the vortex glass [178-186] is only science fiction as well as 
vortex lattice melting. They are based on the same delusion. The vortex 
glass is not spontaneous glass as well as the vortex lattice. It is 
important to remember that the Abrikosov state is observed in an inhomogeneous finite space in contrast to crystalline lattice and glass.

\bigskip

\noindent
\textbf{References}

\

\noindent
1. A.A.Abrikosov, \textit{Zh.Eksp.Teor.Fiz}. \textbf{32}, 1442 (1957) (\textit{Sov.Phys.-JETP} \textbf{5}, 1174 (1957)).

\noindent
2. W.H.Kleiner, L.M.Roth, S.H.Autler, Phys.Rev. \textbf{133}, A1226 (1964).

\noindent
3. P.A.Lee and S.R.Shenoy, \textit{Phys.Rev.Lett}. \textbf{28}, 1025 (1972).

\noindent
4.R.F.Hassing, R.R.Hake and L.J.Barnes, \textit{Phys.Rev.Lett.} \textbf{30}, 6 (1973).

\noindent
5. A.J.Bray, \textit{Phys.Rev.B} \textbf{9}, 4752 (1974).

\noindent
6. D.J.Thouless, \textit{Phys.Rev.Lett.} \textbf{34}, 946 (1975).

\noindent
7.  S.P.Farrant and C.E.Gough, \textit{Phys.Rev.Lett.} \textbf{34}, 943 (1975).

\noindent
8. V.A. Marchenko, and A.V. Nikulov, \textit{Pisma Zh.Eksp.Teor.Fiz.} \textbf{34}, 19-21 (1981) \textit{ (JETP Lett.} \textbf{34}, 
17-19\textit{} (1981)).

\noindent
9. A.V. Nikulov, \textit{Supercond.Sci.Technol.} \textbf{3}, 377 (1990).

\noindent
10.W.K.Kwok, U.Welp, G.W.Crabtree, et al,, \textit{Phys.Rev.Lett.} \textbf{64}, 966 (1990).

\noindent
11. H.Safar, P.L.Gammel, D.A.Huse, D.J.Bishop, J.P.Rice, and D.M.Ginzberg, \textit{Phys.Rev.Lett.} \textbf{69}, 824 (1992). 

\noindent
12. W.K.Kwok, S.Fleshler, U.Welp, V.M.Vinokur, J.Downey, G.W.Crabtree, and 
M.M.Miller, \textit{Phys.Rev.Lett.} \textbf{69}, 3370 (1992).

\noindent
13. E. Zeldov, D.Majer, M. Konczykowski, et al., \textit{Nature} \textbf{375}, 373 (1995).

\noindent
14. A. Schilling, R.A. Fisher, N.E. Phillips, et al., \textit{Nature} \textbf{382}, 791 (1996).

\noindent
15. W.Meissner and R.Ochsenfeld, \textit{Naturwiss.} \textbf{21}, 787 (1933)

\noindent
16. D.Shoenberg, \textit{Superconductivity.} Cambridge. 1952

\noindent
17. A.A.Abrikosov, \textit{Reports of USSR Academy of Sciences} \textbf{86}, 489 (1952).

\noindent
18. K.Mendelssohn, \textit{Proc.Roy.Soc.} \textbf{152A}, 34 (1935)

\noindent
19. V.L.Ginzburg and L.D. Landau, \textit{Zh.Eksp.Teor.Fiz.} \textbf{20}, 1064 (1950)

\noindent
20. L.D. Landau, \textit{Zh.Eksp.Teor.Fiz.} \textbf{7}, 371 (1937). 

\noindent
21. R.P.Huebener, \textit{Magnetic Flux Structures in Superconductors.} Springer-Verlag. Berlin, Heidelberg, New York. 1979. 

\noindent
22. D.Cribier, B.Jacrot, L.M.Rao and B.Farnoux, \textit{Phys.Lett.} \textbf{9}, 106 (1964). 

\noindent
23. U.Essmann and H.Trauble, \textit{Phys.Lett.} \textbf{24A}, 526 (1967). 

\noindent
24. J.Schelten, H.Ullmaier, and W.Schmatz, \textit{Phys.Status Solidi} (b) \textbf{48}, 619 (1971). 

\noindent
25. H.W.Weber, J.Schelten, G.Lippmann, Phys.Status Solidi (b) 57, 515 (1973). 

\noindent
26. J.Schelten, G.Lippmann, and H.Ullmaier, \textit{J.Low Temp.Phys.} \textbf{14}, 213 (1974).

\noindent
27. A.G.Redfield, \textit{Phys.Rev.} \textbf{162}, 367 (1967). 

\noindent
28. A. Kung, \textit{Phys.Rev.Lett.} \textbf{25}, 1006 (1970). 

\noindent
29. J.M.Delrieu, \textit{J. Low Temp. Phys.} \textbf{6}, 197 (1972)

\noindent
30. B. Obst, \textit{Phys.Status Solidi (b)} \textbf{45}, 467 (1971). 

\noindent
31. J. Schelton, \textit{Anisotropy Effects in Superconductors,} ed. by H.W. Weber (Plenum Press, New York 1977) p.139.

\noindent
32. L.Ya.Vinnikov, T.L.Barkov, S.L.Bud'ko, P.C.Canfield and P.C.Kogan, 
\textit{Phys.Rev.B} \textbf{64}, 024504 (2001)

\noindent
33. T.L.Barkov, L.Ya.Vinnikov, M.V.Kartsovnik and N.D.Kushch, \textit{Physica C} \textbf{385}, 568 
(2003)

\noindent
34. A.I. Larkin, \textit{Zh.Eksp.Teor.Fiz.} \textbf{58}, 1466-1470 (1970) (\textit{Sov.Phys.-JETP} \textbf{31}, 784 (1970)).

\noindent
35. G. Eilenberger, \textit{Phys.Rev}. \textbf{164}, 628 (1967). 

\noindent
36. K. Maki and H. Takayama, \textit{Prog.Theor.Phys.} \textbf{46}, 1651 (1971).

\noindent
37. Maki, K. and Thompson, R.S., \textit{Physica} C \textbf{162-164}, 275 (1989).

\noindent
38. M.A.Moore, Phys.Rev. B \textbf{39}, 136 (1989). 

\noindent
39. M.A.Moore, Phys.Rev. B \textbf{45}, 7336 (1992). 

\noindent
40. J.A.O'Neill and M.A.Moore, Phys.Rev.Lett. \textbf{69}, 2582, (1992). 

\noindent
41. J.A.O'Neill and M.A.Moore, Phys.Rev. B \textbf{48}, 374, (1993).

\noindent
42. N.K.Wilkin and M.A.Moore, Phys.Rev. B \textbf{47}, 957, (1993). 

\noindent
43. Zlatko Tesanovic`, Physica C \textbf{220}, 303, (1994).

\noindent
44. A.V. Nikulov, \textit{Phys.Rev. B} \textbf{52}, 10429 (1995). 

\noindent
45. V.A. Marchenko, and A.V. Nikulov, \textit{Zh.Eksp.Teor.Fiz.} \textbf{80}, 745 (1981) (\textit{Sov.Phys.-JETP} \textbf{53}, 377 
(1981)). 

\noindent
46. A.V. Nikulov, D.Yu. Remisov, and V.A. Oboznov, \textit{Phys.Rev.Lett.} \textbf{75}, 2586 
(1995).

\noindent
47. P.G. De Gennes, \textit{Superconductivity of Metals and Alloys}, Pergamon Press. 1966.

\noindent
48. D.Saint-James, G.Sarma, E.J.Thomas, \textit{Type II Superconductivity.} Pergamon Press, 1969.

\noindent
49. Tinkham, M. \textit{Introduction to Superconductivity}, McGraw-Hill Book Company, New-York. 1975.

\noindent
50. R.F.Hassing and J.W.Wilkins, \textit{Phys.Rev.B} \textbf{7}, 1890 (1973).

\noindent
51. A.V. Nikulov, Fluctuation phenomena in bulk type-II superconductors near 
second critical field, \textit{Thesis.} Inst. of Solid State Physics, USSR Academy of 
Sciences, Chernogolovka, 1985.

\noindent
52. A. Ami and K. Maki, \textit{Phys.Rev.} B \textbf{18}, 4714, (1978). 

\noindent
53. L.W.Grunberg and L.Gunther, \textit{Phys.Lett.} \textbf{38A}, 463 (1972)

\noindent
54. V.A. Marchenko, and A.V. Nikulov, \textit{Zh.Eksp.Teor.Fiz.} \textbf{86}, 1395 (1984) (\textit{Sov.Phys.-JETP} \textbf{59}, 815 (1984)).

\noindent
55. V.A. Marchenko and A.V. Nikulov, \textit{Fiz.Nizk.Temp}. \textbf{9}, 816 (1983).

\noindent
56. W.Jiang, N.-C.Yeh, D.S.Reed, U.Kriplani, and F.Holtzberg, Phys.Rev.Lett. 
\textbf{74}, 1438 (1995). 

\noindent
57. A.V. Nikulov, in \textit{Fluctuation Phenomena in High Temperature Superconductors,} M.Aussloos and A.A.Varlamov (eds.), Kluwer Academic 
Publishers, Dordrecht, p. 271 (1997) 

\noindent
58. G. Blatter, M.V. Feigel'man, V.B. Geshkenbein, A.I. Larkin, and V.M. Vinokur, \textit{Rev.Mod.Phys.} \textbf{66}, 1125 (1994).

\noindent
59. E.H. Brandt, \textit{Rep.Progr.Phys.} \textbf{58}, 1465 (1995).

\noindent
60. D.R. Nelson, \textit{Nature} \textbf{375}, 356 (1995).

\noindent
61. D. Bishop, \textit{Nature} \textbf{382}, 760 (1996).

\noindent
62. T. Giamarchi, S. Bhattacharya, in \textit{``High Magnetic Fields: Applications in Condensed Matter Physics and Spectroscopy"}, ed. C. Berthier et al., Springer-Verlag, 
p. 314, 2002.

\noindent
63. G. P. Mikitik, E. H. Brandt, \textit{Phys. Rev.} $B$ \textbf{68}, 054509 (2003); cond-mat/0304380

\noindent
64. C. Dasgupta, O. T. Valls, to appear in Phys. Rev. Lett., cond-mat/0308084 

\noindent
65. A. Schilling, R.A. Fisher, N.E. Phillips, et al., \textit{Phys.Rev.Lett.} \textbf{78}, 4833 (1997)

\noindent
66. F.London, \textit{Superfluids,} Vol.1. John Wiley and Sons, New York, 1950.

\noindent
67. L.P. Gor'kov, and N.B. Kopnin, \textit{Usp.Fiz.Nauk} \textbf{116}, 413 (1975) (\textit{Sov. Phys.Usp.} \textbf{18}, 496 (1975).)

\noindent
68. A. Barone and G. Paterno, \textit{Physics and Applications of the Josephson Effect.} A Wiley-Interscience Publication, New-York, 1982

\noindent
69. D.-X. Chen, et al., \textit{Phys.Rev. B} \textbf{57}, 5059 (1998).

\noindent
70. A.M.Campbell and J.E.Evetts, \textit{Adv. in Phys.} \textbf{21}, 199 (1972)

\noindent
71. Shang-keng Ma, \textit{Modern Theory of Critical Phenomena.} W.A.Benjamin, Inc. Advanced Book Program Reading, Massachusetts, 1977.

\noindent
72. S.P. Brown, D. Charalambous, E.C. Jones, E.M. Forgan, A. Erb, J. 
Kohlbrecher, cond-mat/0308424

\noindent
73. L. Landau, A. V. Sologubenko, H. R. Ott, \textit{Phys. Rev. B} \textbf{68}, 132506 (2003); 
cond-mat/0306094

\noindent
74. A. D. Klironomos, Alan T. Dorsey, cond-mat/0302422

\noindent
75. R. Gilardi et al., \textit{Phys. Rev. Lett.} \textbf{88}, 217003 (2002)

\noindent
76. R. Gilardi, S. Streule, A.J. Drew, et al. to be published in Int. J. Mod. Phys. B; cond-mat/0302413

\noindent
77. J. Hu, and A.H. MacDonald, \textit{Phys.Rev}. B \textbf{52}, 1286 (1995). 

\noindent
78. J. W. Lue, A. G. Montgomery, and R. R. Hake \textit{Phys.Rev. B} \textbf{11}, 3393 (1975).

\noindent
79. V.A. Marchenko and A.V. Nikulov, \textit{Physica} C \textbf{210}, 466 (1993). 

\noindent
80. D.J.Scalapino and M.Sears, \textit{Phys. Rev.} $B$ \textbf{6}, 3409 (1972).

\noindent
81. S. Hikami, A. Fujita, and A. I. Larkin, \textit{Phys. Rev. B} \textbf{44}, 10400 (1991).

\noindent
82. A. Houghton, R.A.Pelcovits, and A.Sudbo, \textit{Phys.Rev. B} \textbf{42}, 906 (1990)

\noindent
83. R.Ikeda, T.Ohmi, and T.Tsuneto, \textit{J.Phys.Soc.Japn.} \textbf{61}, 254 (1992).

\noindent
84. G. Sergeeva, \textit{Low Temp. Phys.} \textbf{20}, 3 (1994). 

\noindent
85. G. Sergeeva, \textit{Physica C} \textbf{235-240}, 1949 (1994).

\noindent
86. R. Lortz, C. Meingast, U. Welp, W. K. Kwok, G. W. Crabtree, \textit{Phys.Rev.Lett}. \textbf{90}, 
237002 (2003); cond-mat/0307110

\noindent
87. Mai Suan Li, Thomas Nattermann, \textit{Phys. Rev. B} \textbf{67}, 184520 (2003), 
cond-mat/0212229

\noindent
88. C. Dasgupta and O. T. Valls, \textit{Phys. Rev. Lett.} \textbf{91}, 127002 (2003)

\noindent
89. M. Menghini, Y. Fasano, F. de la Cruz, S. S. Banerjee, Y. Myasoedov, E. 
Zeldov, C. J. van der Beek, M. Konczykowski, and T. Tamegai, \textit{Phys. Rev. Lett.} \textbf{90}, 
147001 (2003)

\noindent
90. S. Savelersquolev, C. Cattuto, and F. Nori, \textit{Phys. Rev. B} \textbf{67}, 180509 (2003)

\noindent
91. Y. Radzyner, A. Shaulov, Y. Yeshurun, I. Felner, K. Kishio, and J. Shimoyama, \textit{Phys. Rev. B} \textbf{65}, 100503 (2002)

\noindent
92. C. Reichhardt, C. J. Olson, R. T. Scalettar, and G. T. Zimenyi, \textit{Phys. Rev. B} 
\textbf{64}, 144509 (2001).

\noindent
93. B. Lundqvist, O. Rapp, M. Andersson, and Yu. Eltsev, \textit{Phys. Rev. B} \textbf{64}, 060503 
(2001)

\noindent
94. Y. Nonomura and X. Hu, \textit{Phys. Rev. Lett.} \textbf{86}, 5140 (2001)

\noindent
95. M. Calame, S. E. Korshunov, Ch. Leemann, and P. Martinoli \textit{Phys. Rev. Lett.} \textbf{86}, 3630 
(2001)

\noindent
96. M. M. Mola, S. Hill, J. S. Brooks, and J. S. Qualls, \textit{Phys. Rev. Lett.} \textbf{86}, 2130 
(2001)

\noindent
97. S. N. Gordeev, A. A. Zhukov, P. A. J. de Groot, A. G. M. Jansen, R. Gagnon, 
and L. Taillefer \textit{Phys. Rev. Lett.} \textbf{85}, 4594 (2000).

\noindent
98. Chandan Dasgupta, Oriol T. Valls, to appear in Phys. Rev. Lett; 
cond-mat/0308084

\noindent
99. A. Zamora, cond-mat/0307598

\noindent
100. A.V. Nikulov, "The vortex lattice melting theory as example of science fiction" in{\it NATO Science Series: Symmetry and Pairing in Superconductors}, eds. M.Ausloos and S.Kruchinin, Kluwer Academic Publishers, p.131, 1999; cond-mat/9811051

\noindent
101. A.V. Nikulov, in NATO Science Series: Physics and Materials Science of Vortex States, Flux Pinning and Dynamics. R. Kossowski at al., eds. Kluwer Academic Publishers, p.609, 1999; physics/0202021

\noindent
102. A.I.Larkin and Yu.N.Ovchinnikov, \textit{J. Low Temp. Phys.} \textbf{73}, 109 (1979). 

\noindent
103. A.I.Larkin and Yu.N.Ovchinnikov, \textit{Pisma Zh.Eksp.Teor.Fiz.} \textbf{27}, 301 (1978)

\noindent
104. P.W.Anderson and Y.B. Kim, \textit{Rev.Mod.Phys.} \textbf{36}, 39 (1964).

\noindent
105. I.N. Goncharov and I.S.Kukhareva, \textit{Zh.Eksp.Teor.Fiz.} \textbf{62}, 627 (1972) (\textit{Sov.Phys.-JETP} \textbf{35}, 331 
(1972)). 

\noindent
106. P.Berghuis and P.H.Kes, \textit{Phys.Rev.B} \textbf{47}, 262 (1993).

\noindent
107. S. Bhattacharya and M.J. Higgins, \textit{Phys.Rev.Lett} \textbf{70}, 2617 (1993).

\noindent
108. S. Bhattacharya and M.J. Higgins, \textit{Phys.Rev.B} \textbf{49}, 10005 (1994).

\noindent
109. M.H. Theunissen and P.H. Kes, \textit{Phys.Rev.B} \textbf{55}, 15183 (1997).

\noindent
110. A.Pautrat, A.Daignere, C.Goupil, et al. cond-mat/0304488. 

\noindent
111. D. Li and B. Rosenstein, \textit{Phys. Rev. B} \textbf{65}, 220504 (2002)

\noindent
112. B. Rosenstein, \textit{Phys. Rev. B} \textbf{60}, 4268--4271 (1999)

\noindent
113. D. Li and B. Rosenstein, \textit{Phys. Rev. B} \textbf{65}, 024514 (2002).

\noindent
114. P.Berghuis, A.L.F. van der Slot, and P.H.Kes, \textit{Phys.Rev.Lett.} \textbf{65}, 2583 (1990). 

\noindent
115. A.V.Nikulov, E.Milani, G.Balestrino, and V.A.Oboznov, \textit{Supercond. Sci. Technol.} \textbf{12}, 442 (1999). 

\noindent
116. A.V. Nikulov, D.Yu. Remisov, and V.A. Oboznov, \textit{Pisma Zh.Eksp.Teor.Fiz.} \textbf{61}, 575 (1995) \textit{ (JETP Lett.} \textbf{61}, 588 (1995)).

\noindent
117. A.V. Nikulov, S.V. Dubonos, and Y.I. Koval, \textit{J. Low Temp.Phys.} \textbf{109}, 643 (1997).

\noindent
118. A.V.Nikulov, cond-mat/9812168

\noindent
119. W. A. Al-Saidi and D. Stroud \textit{Phys. Rev. B} \textbf{68}, 144511 (2003)

\noindent
120. B. J. Taylor, S. Li, M. B. Maple, and M. P. Maley \textit{Phys. Rev. B} \textbf{68}, 054523 
(2003)

\noindent
121. D. Li and B. Rosenstein \textit{Phys. Rev. Lett.} \textbf{90}, 167004 (2003)

\noindent
122. E. W. Carlson, A. H. Castro Neto, and D. K. Campbell, \textit{Phys. Rev. Lett.} \textbf{90}, 087001 
(2003) 

\noindent
123. N.P.Ong, Yayu Wang, cond-mat/0306399

\noindent
124. C.M. Aegerter, H. Keller, S.H. Lloyd, et al., cond-mat/0305593

\noindent
125. P. Matl and N. P. Ong \textit{Phys. Rev. B} \textbf{67}, 146502 (2003)

\noindent
126. A. Pautrat, C. Goupil, Ch. Simon, B. Plarais, and P. Mathieu \textit{Phys. Rev. B} \textbf{67}, 
146501 (2003)

\noindent
127. C. E. Creffield and J. P. Rodriguez \textit{Phys. Rev. B} \textbf{67}, 144510 (2003)

\noindent
128. M. Yasugaki, M. Tokunaga, N. Kameda, and T. Tamegai \textit{Phys. Rev. B} \textbf{67}, 104504 
(2003) 

\noindent
129. S. M. Ashimov, J. G. Chigvinadze, cond-mat/0306118. 

\noindent
130. L. Fruchter, cond-mat/0304523

\noindent
131. J.A.G. Koopmann, V.B. Geshkenbein, G. Blatter, cond-mat/0302225

\noindent
132. R. Besseling, N. Kokubo, P.H. Kes, cond-mat/0302187

\noindent
133. Sandeep Tyagi, Yadin Y. Goldschmidt, cond-mat/0302127

\noindent
134. A. B. Kolton, R. Exartier, L. F. Cugliandolo, D. Dominguez, and N. Grinbech-Jensen \textit{Phys. Rev. Lett.} \textbf{89}, 227001 (2002) 

\noindent
135. J. Sinova, C. B. Hanna, and A. H. MacDonald \textit{Phys. Rev. Lett.} \textbf{89}, 030403 (2002) 

\noindent
136. M. Tokunaga, M. Kishi, N. Kameda, K. Itaka, and T. Tamegai \textit{Phys. Rev. B} \textbf{66}, 
220501 (2002)

\noindent
137. K. Shibata, T. Nishizaki, T. Sasaki, and N. Kobayashi \textit{Phys. Rev. B} \textbf{66}, 214518 
(2002) 

\noindent
138. J. P. Rodriguez \textit{Phys. Rev. B} \textbf{66}, 214506 (2002)

\noindent
139. T. Joseph and C. Dasgupta \textit{Phys. Rev. B} \textbf{66}, 212506 (2002)

\noindent
140. C. Dasgupta and O. T. Valls \textit{Phys. Rev. B} \textbf{66}, 064518 (2002). 

\noindent
141. C. C. de Souza Silva and G. Carneiro \textit{Phys. Rev. B} \textbf{66}, 054514 (2002)

\noindent
142. V. Zhuravlev and T. Maniv \textit{Phys. Rev. B} \textbf{66}, 014529 (2002) 

\noindent
143. P. Matl, N. P. Ong, R. Gagnon, and L. Taillefer \textit{Phys. Rev. B} \textbf{65}, 214514 (2002) 

\noindent
144. M. Yasugaki, K. Itaka, M. Tokunaga, N. Kameda, and T. Tamegai \textit{Phys. Rev. B} \textbf{65}, 
212502 (2002) 

\noindent
145. M. Juneau, R. MacKenzie, M.-A. Vachon, and J. M. \textit{Phys. Rev. B} \textbf{65}, 140512 (2002) 

\noindent
146. L. Miu \textit{Phys. Rev. B} \textbf{65}, 096501 (2002)

\noindent
147. J. Figueras, T. Puig, X. Obradors, A. Erb, and E. Walker \textit{Phys. Rev. B} \textbf{65}, 
092505 (2002). 

\noindent
148. A. Schilling, U. Welp, W. K. Kwok, and G. W. Crabtree \textit{Phys. Rev. B} \textbf{65}, 054505 (2002) 

\noindent
149. C. Dasgupta and O. T. Valls \textit{Phys. Rev. Lett.} \textbf{87}, 257002 (2001) 

\noindent
150. J. P. Rodriguez \textit{Phys. Rev. Lett.} \textbf{87}, 207001 (2001) 

\noindent
151. A. Soibel, Y. Myasoedov, M. L. Rappaport, T. Tamegai, S. S. Banerjee, and E. Zeldov \textit{Phys. Rev. Lett}. \textbf{87}, 167001 (2001) 

\noindent
152. P. Olsson and S. Teitel \textit{Phys. Rev. Lett.} \textbf{87}, 137001 (2001)

\noindent
153. G. P. Mikitik and E. H. Brandt \textit{Phys. Rev. B} \textbf{64}, 184514 (2001)

\noindent
154. M. F. Laguna, C. A. Balseiro, D. Dominguez, and F. Nori \textit{Phys. Rev.} $B$ \textbf{64}, 104505 (2001) 

\noindent
155. S. E. Savelarsquotev, J. Mirkovic, and K. Kadowaki \textit{Phys. Rev. B} \textbf{64}, 094521 
(2001)

\noindent
156. J. Mirkovic, S. E. Savel'ev, E. Sugahara, and K. Kadowaki \textit{Phys. Rev. Lett.} \textbf{86}, 886-889 (2001) 

\noindent
157. C. D. Vaccarella and C. A. R. de Melo \textit{Phys. Rev. B} \textbf{63}, 180505 (2001) 

\noindent
158. V. I. Marconi and D. Dominguez \textit{Phys. Rev. B} \textbf{63}, 174509 (2001) 

\noindent
159. J. M. E. Geers, C. Attanasio, M. B. S. Hesselberth, J. Aarts, and P. H. Kes 
\textit{Phys. Rev. B} \textbf{63}, 094511 (2001) 

\noindent
160. T. Maniv, V. Zhuravlev, I. Vagner, and P. Wyder \textit{Rev. Mod. Phys.} \textbf{73}, 867-911 (2001) 

\noindent
161. G. Mohler and D. Stroud \textit{Phys. Rev. B} \textbf{62}, R14665 (2000) 

\noindent
162. J. P. Rodriguez \textit{Phys. Rev. B} \textbf{62}, 9117 (2000) 

\noindent
163. D. Pal, D. Dasgupta, B. K. Sarma, S. Bhattacharya, S. Ramakrishnan, and A. K. Grover \textit{Phys. Rev. B} \textbf{62}, 6699 (2000) 

\noindent
164. M. F. Laguna, D. Dominguez, and C. A. Balseiro \textit{Phys. Rev. B} \textbf{62}, 6692 (2000)

\noindent
165. M.-C. Miguel and M. Kardar \textit{Phys. Rev. B} \textbf{62}, 5942 (2000) 

\noindent
166. X. G. Qiu, V. V. Moshchalkov, and J. Karpinski \textit{Phys. Rev. B} \textbf{62}, 4119 (2000)

\noindent
167. J. Kierfeld and V. Vinokur \textit{Phys. Rev. B} \textbf{61}, R14928 (2000) 

\noindent
168. C. Reichhardt and G. T. Zimenyi \textit{Phys. Rev. B} \textbf{61}, 14354 (2000) 

\noindent
169. G. Dirsquo;Anna, V. Berseth, L. Forri, A. Erb, and E. Walker \textit{Phys. Rev. B} \textbf{61}, 4215 (2000) 

\noindent
170. T. Nishizaki, T. Naito, S. Okayasu, A. Iwase, and N. Kobayashi \textit{Phys. Rev. B} \textbf{61}, 3649 (2000)

\noindent
171. A. Schilling, M. Willemin, C. Rossel, et al. \textit{Phys. Rev. B} \textbf{61}, 3592 (2000) 

\noindent
172. H. M. Carruzzo and C. C. Yu \textit{Phys. Rev. B} \textbf{61}, 1521 (2000) 

\noindent
173. J. E. Sonier, J. H. Brewer, R. F. Kiefl, et al \textit{Phys. Rev. B} \textbf{61}, R890 (2000) 

\noindent
174. S. Grundberg and J. Rammer \textit{Phys. Rev. B} \textbf{61}, 699 (2000)

\noindent
175. T. Kita, cond-mat/0307067

\noindent
176. R.Sasik, D. Stroud and Z. Tesanovic\textit{, Phys. Rev. B} \textbf{51}, 3041 (1995).

\noindent
177. R. Ikeda, \textit{J. Phys.Soc.Jpn.} \textbf{65}, 3998 (1996).

\noindent
178. M.P.A. Fisher, \textit{Phys. Rev. Lett.} \textbf{62,} 1415 (1989).

\noindent
179. D.S. Fisher, M.P.A. Fisher and D.A. Huse, \textit{Phys. Rev. B} \textbf{43}, 130 (1991).

\noindent
180. T. Klein, I. Joumard, S. Blanchard, et al. Nature 413, 404 (2001)

\noindent
181. A. Hernandez, D. Dominguez, cond-mat/0308511

\noindent
182. M. Suzuki, I. S. Suzuki, J. Walter, Physica C (2003), in press; cond-mat/0308311

\noindent
183. S. L. Li, H. H. Wen, to appear in Physica C, cond-mat/0303111

\noindent
184. Hikaru Kawamura, cond-mat/0302284

\noindent
185. Jack Lidmar, \textit{Phys. Rev. Lett.} \textbf{91}, 097001 (2003); cond-mat/0302577

\noindent
186. S. Okuma, Y. Imamoto, and M. Morita \textit{Phys. Rev. Lett.} \textbf{86}, 3136-3139 (2001)

\end{document}